\documentclass[twocolumn,nofootinbib,amsmath,showpacs,amssymb,aps,prd,balancelastpage,superscriptaddress]{revtex4-1}

\pdfoutput=1 

\usepackage[T1]{fontenc} 
\usepackage{parskip}
\usepackage{bm}         
\usepackage{array}
\usepackage{multirow}   
\usepackage{booktabs}   

\usepackage{textcomp}
\usepackage{graphicx}
\usepackage{subcaption}
\usepackage{amssymb}

\usepackage{mleftright}
\usepackage[utf8x]{inputenc}
\usepackage{bbold}
\usepackage{color}
\usepackage[dvipsnames]{xcolor}
\usepackage{amsmath,amscd}
\usepackage{placeins}
\usepackage{soul}
\usepackage{braket}
\usepackage{tikz-cd}
\usepackage{empheq}
\usepackage[most]{tcolorbox}
\usetikzlibrary{arrows, shapes,shadows}
\usepackage{mathtools,slashed}
\usepackage{dsfont}
\usepackage[makeroom]{cancel}

\DeclareMathAlphabet{\mathpzc}{OT1}{pzc}{m}{it}

\newcommand{\ra}[1]{\renewcommand{\arraystretch}{#1}}
\newcommand*{\Scale}[2][4]{\scalebox{#1}{$#2$}}%

\newcommand{\del}{\partial}

\renewcommand{\i}{\mathrm i}
\renewcommand{\(}{\left(}
\renewcommand{\)}{\right)}
\renewcommand{\[}{\left[}
\renewcommand{\]}{\right]}

\newcommand{\mean}[1]{\left \langle #1 \right \rangle }
\newcommand{\vev}[1]{v_{\mathrm{#1}}}
\newcommand{\abs}[1]{\left| #1 \right| }

\newcommand{\U}[1]{\mathrm{U}(1)_{\mathrm{#1}}}			
\newcommand{\SU}[2]{\mathrm{SU}(#1)_{\mathrm{#2}}}		
\newcommand{\SO}[2]{\mathrm{SO}(#1)_{\mathrm{#2}}}		
\newcommand{\E}[1]{\mathrm{E}_{#1}}		
\newcommand{\T}[2]{T_{\mathrm{#1}}^{#2}}

\newcommand{\RN}[1]{%
  \textup{\uppercase\expandafter{\romannumeral#1}}%
}

\newcommand{\parityarrow}{\overset{\mathds{P}}{\rightarrow}}

\newcommand{\LLR}[3]{\big(\bm{L}^{ #1} \big)^{ #2 }{}_{ #3 }}
\newcommand{\QL}[3]{\big(\bm{Q}_{\mathrm{L}}^{ #1} \big)^{ #2 }{}_{ #3 }}
\newcommand{\QR}[3]{\big(\bm{Q}_{\mathrm{R}}^{ #1} \big)^{ #2 }{}_{ #3 }}
\newcommand{\LLRs}[3]{\big(\bm{L}^*_{ #1} \big)_{ #2 }{}^{ #3 }}

\newcommand{\DA}[2]{\bm{\Delta}_{\mathrm{#1}}^{ #2}}

\newcommand{\FLLR}[3]{\big(L^{ #1} \big)^{ #2 }{}_{ #3 }}

\newcommand{\FDA}[2]{\Delta_{\mathrm{#1}}^{ #2}}

\newcommand{\FLLRd}[3]{\big(L^{\dagger}_{ #1} \big)_{ #2 }{}^{ #3 }}

\newcommand{\SLLR}[3]{\big(\tilde{L}^{ #1} \big)^{ #2 }{}_{ #3 }}

\newcommand{\SQL}[3]{\big(\tilde{Q}_{\mathrm{L}}^{ #1} \big)^{ #2 }{}_{ #3 }}
\newcommand{\SQR}[3]{\big(\tilde{Q}_{\mathrm{R}}^{ #1} \big)^{ #2 }{}_{ #3 }}
\newcommand{\SDA}[2]{\tilde{\Delta}_{\mathrm{#1}}^{ #2}}

\newcommand{\SLLRs}[3]{\big(\tilde{L}^*_{ #1} \big)_ { #2 }{}^{ #3 }}

\newcommand{\SQRs}[3]{\big(\tilde{Q}_{\mathrm{R} #1}^* \big)_{ #2 }{}^{ #3 }}
\newcommand{\SDAs}[2]{\tilde{\Delta}_{\mathrm{#1}}^{\ast #2}}

\newcommand{\lam}[2]{\lambda_{\Scale[0.50]{#1 \-- #2}}}
\newcommand{\delt}[2]{\delta_{\Scale[0.5]{#1 \-- #2}}}
\newcommand{\la}[1]{\lambda_{\Scale[0.5]{#1}}}
\newcommand{\de}[1]{\delta_{\Scale[0.5]{#1}}}
\newcommand{\y}[1]{\mathrm{y}_{\Scale[0.5]{#1}}}
\newcommand{\g}[1]{g_{\Scale[0.5]{\mathrm{#1}}}}
\newcommand{\q}[1]{Y_{\Scale[0.5]{\mathrm{#1}}}}
\newcommand{\id}[1]{\mathbb{1}_{\Scale[0.6]{#1}}}
\newcommand{\ida}[1]{\mathbb{1}^{\Scale[0.6]{\rm adj}}_{\Scale[0.6]{#1}}}
\newcommand{\So}[1]{{\Scale[0.7]{#1}}}
\newcommand{\Sr}[2]{\Scale[#1]{\rm #2}}

\begin{document}

\pagestyle{plain}

\begin{flushright}
LU TP 17-37
\end{flushright}

\title{Scale hierarchies, symmetry breaking and particle spectra\\ 
in SU(3)-family extended SUSY trinification}

\author{Jos\'e~E.~Camargo-Molina}
\email{eliel@thep.lu.se}
\affiliation{Department of Physics, Imperial College London, \\
South Kensington Campus, London SW7 2AZ, UK}

\author{Ant\'onio~P.~Morais}
\email{aapmorais@ua.pt}
\affiliation{Departamento de F\'\i sica, Universidade de Aveiro and CIDMA, \\
Campus de Santiago, 3810-183 Aveiro, Portugal}

\author{Astrid~Ordell}
\email{astrid.ordell@thep.lu.se}
\affiliation{Department of Astronomy and Theoretical Physics,\\
Lund University, SE-223 62 Lund, Sweden}

\author{Roman~Pasechnik}
\email{roman.pasechnik@thep.lu.se}
\affiliation{Department of Astronomy and Theoretical Physics,\\
Lund University, SE-223 62 Lund, Sweden}

\author{Jonas~Wess\'en}
\email{jonas.wessen@thep.lu.se}
\affiliation{Department of Astronomy and Theoretical Physics,\\
Lund University, SE-223 62 Lund, Sweden}

\begin{abstract}
\noindent
A unification of left-right $\SU{3}{L}\times \SU{3}{R}$, colour $\SU{3}{C}$ and family $\SU{3}{F}$ symmetries in 
a maximal rank-8 subgroup of ${\rm{E}}_8$ is proposed as a landmark for future explorations beyond the Standard 
Model (SM). We discuss the implications of this scheme in a supersymmetric (SUSY) model based on the trinification 
gauge $\left[\SU{3}{}\right]^3$ and global $\SU{3}{F}$ family symmetries. Among the key properties 
of this model are the unification of SM Higgs and lepton sectors, a common Yukawa coupling for chiral fermions, 
the absence of the $\mu$-problem, gauge couplings unification and proton stability to all orders in perturbation theory. 
The minimal field content consistent with a SM-like effective theory at low energies is composed of one $\mathrm{E}_6$ 
$\bm{27}$-plet per generation as well as three gauge and one family $\SU{3}{}$ octets inspired by the fundamental 
sector of ${\rm{E}}_8$. The details of the corresponding (SUSY and gauge) symmetry breaking scheme, multi-scale 
gauge couplings' evolution, and resulting effective low-energy scenarios are discussed. 
\end{abstract}


\maketitle

\section{Introduction}

Finding successful candidate theories unifying the strong and electroweak interactions, leading to a detailed understanding of the SM origin, with all its parameters, hierarchies, symmetries and 
particle content remain a big challenge for the theoretical physics community. Some of the most popular SM extensions are based on supersymmetric (SUSY) GUTs where the SM gauge interactions are unified
under symmetry groups such as $\SU{5}{}$ and $\mbox{SO}(10)$ \cite{Georgi:1974sy,Fritzsch:1974nn,Chanowitz:1977ye,Georgi:1978fu,
Georgi:1979dq,Georgi:1979ga,Georgi:1982jb} as well as $\mathrm{E}_6$\footnote{The $\mathrm{E}_6$-based models are typically motivated by heterotic 
string theories where massless sectors consistent with the chiral structure of the SM are naturally described by an ${\rm E_8} \times {\rm E^{\prime}_8}$ 
gauge theory. For more details we refer the reader to Refs.~\cite{Gursey:1975ki,Gursey:1981kf,Achiman:1978vg}} and $\mathrm{E}_7$ \cite{Gursey:1976dn}. 
A particularly appealing scenario proposed by Glashow in 1984 \cite{original} is based upon the rank-6 trinification symmetry $[\SU{3}{}]^3 \equiv \SU{3}{L} \times  
\SU{3}{R} \times  \SU{3}{C} \rtimes \mathbb{Z}_3 \subset \mathrm{E}_6$ (T-GUT, in what follows) where all matter fields are embedded in bi-triplet representations 
and due to the cyclic permutation symmetry $\mathbb{Z}_3$, the corresponding gauge couplings unify at the T-GUT Spontaneous Symmetry Breaking (SSB) scale, 
or GUT scale in what follows. 

There have been many phenomenological and theoretical studies of T-GUTs, in both SUSY and non-SUSY formulations, motivated by their unique features 
(see e.g.~Refs.~\cite{Dias:2010vt,Reig:2016tuk,Babu:1985gi,He:1986cs,Greene:1986jb,Wang:1992hu,Lazarides:1993uw,Dvali:1994wj,Maekawa:2002qv,
Kim:2003cha,Kim:2004pe,Carone:2004rp,Carone:2005ha,Sayre:2006ma,DiNapoli:2006kq,Cauet:2010ng,Stech:2012zr,Stech:2014tla,Hetzel:2015bla,
Hetzel:2015cca,Pelaggi:2015kna,Pelaggi:2015knk,Rodriguez:2016cgr,Camargo-Molina:2016bwm}). For example, due to the fact that quarks and leptons 
belong to different gauge representations in T-GUT scenarios, the baryon number is naturally conserved by the gauge sector \cite{Babu:1985gi}, only allowing 
for proton decay via Yukawa and scalar interactions, if at all present. As was shown for a particular T-GUT realisation in Ref.~\cite{Sayre:2006ma}, the proton decay rates were consistent 
with experimental limits in the case of low-scale SUSY, or completely unobservable in the case of split SUSY. Many T-GUTs can also accommodate any quark 
and lepton masses and mixing angles \cite{Babu:1985gi,Stech:2014tla} whereas neutrino masses are generated by a see-saw mechanism \cite{Kim:2004pe} 
of radiative \cite{Sayre:2006ma} or inverse \cite{Cauet:2010ng} type.  

Despite a notable progress in exploring gauge coupling unification, neutrino masses, Dark Matter candidates, TeV-scale Higgs partners, collider and other 
phenomenological implications of GUTs, there are several yet unresolved problems. 
One of problems emerging in the case of SUSY T-GUT model building is the longstanding issue of avoiding GUT scale masses for the would-be SM leptons. To circumvent this, the usual solution is to 
add several $\bm{27}$-plets of ${\rm E_6}$ with scalar components responsible for SSB of gauge trinification \cite{Babu:1985gi,Wang:1992hu,Dvali:1994wj,Maekawa:2002qv,
Willenbrock:2003ca,Carone:2005ha,Sayre:2006ma,Cauet:2010ng,Stech:2012zr,Stech:2014tla,Hetzel:2015bla,Pelaggi:2015kna}, or to simply add higher dimensional operators \cite{Nath:1988xn,Dvali:1994wj,
Maekawa:2002qv,Carone:2005ha,Cauet:2010ng}. These approaches typically require a significant fine-tuning in high-scale parameter space (especially, 
in the Yukawa sector) \cite{Sayre:2006ma}. Otherwise, they exhibit phenomenological issues with proton 
stability \cite{Babu:1985gi,Maekawa:2002qv,Sayre:2006ma} and with a large amount of unobserved light states \cite{original,Nath:1988xn,Dvali:1994wj,
Stech:2014tla,Hetzel:2015bla,Pelaggi:2015knk}. Despite continuous progress, the SM-like EFTs originating from T-GUTs still remain underdeveloped 
in comparison to other GUT models such as $\SU{5}{}$, $\SO{10}{}$ or even ${\rm E_6}$ (see e.g.~Ref.~\cite{Hetzel:2015cca} 
and references therein).

In this paper, we explore in detail the SUSY T-GUT model proposed in \cite{Camargo-Molina:2016yqm} with a global $\SU{3}{F}$ family symmetry inspired 
by the embedding of ${\rm E_6} \times \SU{3}{}$ into ${\rm E_8}$. We will refer to 
this model as the SUSY Higgs-Unified Trinification (SHUT) model (for alternative ways of extending the SM by means of an $\SU{3}{F}$ symmetry see e.g.~Refs.~\citep{Berezhiani:1990wn,Berezhiani:1989fp,Berezhiani:1990jj,Sakharov:1994pr}). As we will see, the SHUT model offers solutions to some of the problems faced by previous T-GUTs. 
As the light Higgs and lepton sectors are unified, the model can be embedded into a single ${\rm E_8}$ representation. Furthermore, the embedding suggests
the introduction of adjoint scalars and a family $\SU{3}{F}$, where the former protects a sufficient amount of fermionic states from acquiring masses before EWSB to be in agreement with the SM.  
The interplay of the family $\SU{3}{F}$ also provides a unification 
of the high-scale Yukawa sector into a single coupling. 
This is in contrast to well-known $\mathrm{SO}(10)$ and Pati-Salam models where the Yukawa unification is constrained to the third family only 
(see e.g.~Refs.~\cite{Blazek:2001sb, Baer:2001yy, Anandakrishnan:2013cwa,Blazek:2002ta, Tobe:2003bc, Baer:2009ie, Badziak:2011wm, Anandakrishnan:2012tj, 
Joshipura:2012sr, Anandakrishnan:2013nca, Anandakrishnan:2014nea, Badziak:2013eda, Ajaib:2013zha}). 

The Yukawa and gauge couplings unification in the SHUT model largely reduces its parameter space, making a complete analysis of its low-energy EFT scenarios 
technically feasible. The model also has a particular feature in that no further spontaneous breaking of the symmetry towards the SM gauge group is provided by the SUSY conserving part of the model, and that the energy scales at which the symmetry is further broken are instead associated with the soft SUSY-breaking operators. As such, both the electro-weak scale and the scales of intermediate symmetry breaking are naturally suppressed relative to the GUT scale.

In Sect.~\ref{Sec:LRCF} we briefly discuss the key features of the SHUT model and its SSB scheme, and in Sect.~\ref{sec:SUSYHS} the high-scale SHUT model 
is introduced in its minimal setup in detail. In particular, we discuss its features and the details on how it solves the longstanding problems of previous T-GUT realizations and 
how the GUT scale SSB in this model leads to a Left-Right (LR) symmetric SUSY theory. In Sect.~\ref{sec:BreakingSUSY} we discuss the inclusion of soft SUSY-breaking 
interactions and how they lead to a breaking of the remaining gauge symmetries down to the SM gauge group, and in Sect~\ref{sec:masses2} we present a short overview of the low-energy limits of the SHUT model. Finally, Sect.~\ref{sec:EFTs} contains an analysis of RG evolution of gauge couplings at one loop and extraction of characteristic values of the GUT and soft scales, before concluding in Sect.~\ref{sec:conclusions}. 

\subsection*{A short note on notation}

In this article we adopt the following notations:
\begin{itemize}
\item Supermultiplets are always written in bold (e.g.~${\bf \Delta}$). As usual, the scalar components of chiral supermultiplets and fermionic components of vector 
supermultiplets carry a tilde (e.g.~$\widetilde{\Delta}$), except for the Higgs-Higgsino sector where the tilde serves to identify the fermion 
$\SU{2}{L} \times \SU{2}{R}$ bi-doublets (e.g.~$\widetilde{H}$).
\item Fundamental representations carry superscript indices while anti-fundamental representations carry subscript indices.
\item $\SU{3}{K}$ and $\SU{2}{K}$ (anti-)fundamental indices are denoted by $k,k',k_1,k_2 \dots$ for $K=L,R$, respectively, while colour indices are denoted by $x,x',x_1,x_2 \dots$.
\item Indices belonging to (anti-)fundamental representations of $\SU{3}{F}$ are denoted by $i,j,k \dots$.
\item If a field transforms both under gauge and global symmetry groups, the index corresponding to the global one is placed 
within the parenthesis around the field, while the indices corresponding to the gauge symmetries are placed outside.
\item Global symmetry groups will be indicated by $\{\dots\}$. 
\end{itemize} 

\section{Left-Right-Color-Family unification}
\label{Sec:LRCF}

In Glashow's formulation of the trinified  $[\SU{3}{L}\times \SU{3}{R}\times \SU{3}{C}]\rtimes \mathbb{Z}_3 \subset \mathrm{E}_6$ (LRC-symmetric)
gauge theory \cite{original}, three families of the fermion fields from the SM are arranged over three $\bm{27}$-plet copies
of the $\mathrm{E}_6$ group, namely, 
\begin{eqnarray*}
\bm{27}^i &\to& \LLR{i}{l}{r}\oplus\QL{i}{x}{l}\oplus\QR{i}{r}{x} \\ 
&\equiv& (\bm{3}^l,\bm{\bar{3}}_r,\bm{1})^i\oplus (\bm{\bar{3}}_l,\bm{1},
\bm{3}^x)^i\oplus (\bm{1},\bm{3}^r,\bm{\bar{3}}_x)^i \,,
\end{eqnarray*}
while the Higgs fields responsible for a high-scale SSB are typically introduced via e.g.~an additional $\bm{27}$-plet. Here and below, the left, right, and color 
$\SU{3}{}$ indices are denoted by $l,\,r,$ and $x$, respectively, while the fermion families are labelled by an index $i=1,2,3$.
%
%
\begin{figure}[hbt!]
\centering
\resizebox{8.2cm}{9.5cm}{
\includegraphics[scale=1]{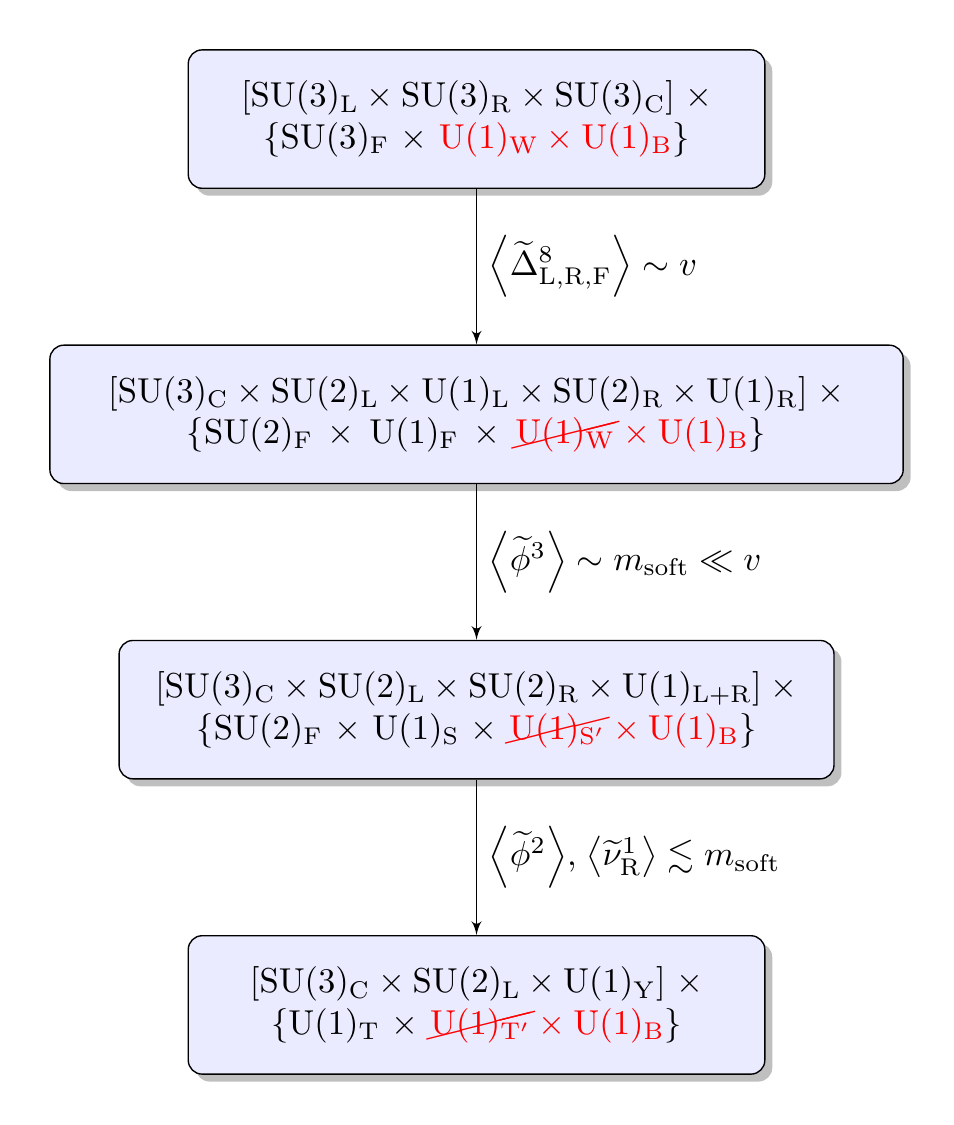}
}
\caption{\label{fig:1112abc}\emph{The symmetry breaking scheme in the SHUT model studied in this work. The symmetry groups in red correspond to 
the accidental symmetries of the high-scale theory. The global accidental $\U{W}$ and, consequently, its low-energy counterparts $\U{S',T'}$ discussed below 
are considered to be softly broken at low-energy scales and thus are shown as crossed-out symmetry groups.}}
\end{figure}

The SHUT model first presented in Ref.~\cite{Camargo-Molina:2016yqm}, in contrast to the Glashow's trinification, introduces the global family 
symmetry $\SU{3}{F}$ which acts in the generation-space. In this case, the light Higgs and lepton superfields, as well as quarks and colored scalars, 
all are unified into a single $(\bm{27},\bm{3})$-plet under $\mathrm{E}_6\times \SU{3}{F}$ symmetry, i.e.
\begin{eqnarray*}
(\bm{27},\bm{3}) &\to& \LLR{i}{l}{r}\oplus\QL{i}{x}{l}\oplus\QR{i}{r}{x} \\
&\equiv& (\bm{3}^l,\bm{\bar{3}}_r,\bm{1},\bm{3}^i)\oplus (\bm{\bar{3}}_l,\bm{1},
\bm{3}^x,\bm{3}^i)\oplus (\bm{1},\bm{3}^r,\bm{\bar{3}}_x,\bm{3}^i) \,.
\end{eqnarray*}
The leptonic tri-triplet superfield $\LLR{i}{l}{r}$ that unifies the SM left- and right-handed leptons and SM Higgs doublets can be conveniently represented as
\begin{eqnarray}
&& \Scale[0.95]{\LLR{i}{l}{r} =\begin{pmatrix}
\bm{H}_{11} & \bm{H}_{12} & \bm{e}_{\mathrm{L}}\\
\bm{H}_{21} & \bm{H}_{22} & \bm{\nu}_{\mathrm{L}}\\
\bm{e}_{\mathrm{R}}^{c} & \bm{\nu}_{\mathrm{R}}^{c} & \bm{\phi}
\end{pmatrix}^{i}\,,}
\label{eq:L-tri-triplet}
\end{eqnarray}
Besides, the left-quark $\QL{i}{x}{l}$ and right-quark $\QR{i}{r}{x}$ tri-triplets are
\begin{eqnarray}
\begin{aligned}
&\Scale[0.95]{\QL{i}{x}{l}=\begin{pmatrix}\bm{u}_{\mathrm{L}}^x & \bm{d}_{\mathrm{L}}^x & \bm{D}_{\mathrm{L}}^x
\end{pmatrix}^{i}}, \label{eq:Q-tri-triplets}
\\
&\Scale[0.95]{\QR{i}{r}{x}=\begin{pmatrix}\bm{u}_{\mathrm{R}x}^c & \bm{d}_{\mathrm{R}x}^c & \bm{D}_{\mathrm{R}x}^c
\end{pmatrix}^{\top\;\;i}}\,.
\end{aligned}
\end{eqnarray}
In addition, the SHUT model also incorporates the adjoint (namely, $\SU{3}{L,R,C,F}$ octet) superfields $\bm{\Delta}_{\mathrm{L,R,C,F}}$. The first SSB step 
in the SHUT model $\SU{3}{L,R,F}\to \SU{2}{L,R,F}\times \U{L,R,F}$ is triggered at the GUT scale by the SUSY-preserving vacuum expectation values (VEVs) 
in the scalar components of the corresponding octet superfields while all the subsequent low-scale SSB steps are triggered by VEVs in the leptonic tri-triplet 
$\LLR{i}{l}{r}$ through the soft SUSY-breaking operators.

Along this work, we will be focused on the symmetry breaking scheme shown in Fig.~\ref{fig:1112abc}. There it can be seen that an accidental global 
$\U{B} \times \U{W}$ symmetry (which is marked in red and will be discussed in detail in the next section) appears in the high-scale theory. 
As we will see, although alternative breaking schemes are possible, this is the one leading to the low energy SM-like scenarios we find most interesting.
As we shall see in Sec.~\ref{sec:masses2}, dimension-3 operators that softly break $\U{W}$, and consequently its low-energy descendants (that will be 
denoted below as $\U{S',T'}$), are needed for a phenomenologically viable low-scale fermion spectrum. Such interactions do not have a perturbative 
origin from the high-scale theory and are added to the effective theory that emerges once the heavy degrees of freedom of the SHUT model are integrated out. 

\section{Supersymmetric trinification with global $\SU{3}{F}$} 
\label{sec:SUSYHS}

This section contains a review of the SHUT model before and after the T-GUT symmetry is broken spontaneously by adjoint field VEVs. We here present the symmetries, particle content and interactions of the model at both stages, in addition to showing how it addresses the shortcomings of previous T-GUTs. 

\subsection{Tri-triplet sector}
\label{sec:tritri}

In the following, we consider the SHUT model -- a SUSY T-GUT theory based on the trinification gauge group with an accompanying global 
$\SU{3}{F}$ family symmetry, i.e.
\begin{eqnarray}
\nonumber
G_{333\{3\}} &\equiv& [\SU{3}{L} \times \SU{3}{R} \times \SU{3}{C} ] \\
&\rtimes& \mathbb{Z}_3^{(\mathrm{LRC})}  \times \{ \SU{3}{F} \} \,.
\label{eq:3333}
\end{eqnarray} 
Here and below, curly brackets indicate global (non-gauge) symmetries. The minimal chiral superfield content (shown in Tab.~\ref{table:ChiralSuperE6}) that 
can accommodate the SM (Higgs and fermion) fields, is comprised of three tri-triplet representations of $G_{333\{3\}}$ which we label as $\bm{L}$, $\bm{Q}_{\mathrm{L}}$ 
and $\bm{Q}_{\mathrm{R}}$ respectively (for their explicit relation to the SM field content up to a possible mixing, see Eqs.~(\ref{eq:L-tri-triplet}) and (\ref{eq:Q-tri-triplets})). 
\begin{table}[htb!]
\centering
\Scale[0.9]{  \begin{tabular}{|>{\centering}m{2.5cm}>{\centering}m{1.2cm}|>{\centering}m{1.2cm}|>{\centering}m{1.2cm}|>{\centering}m{1.2cm}|>{\centering}m{1.2cm}|}
\hline
\multicolumn{6}{|c|}{Chiral supermultiplet fields}\tabularnewline
\hline
\hline
\small{Superfield}&\small{}&\small{$\SU{3}{L}$}&\small{$\SU{3}{R}$}&\small{$\SU{3}{C}$}&\small{$\SU{3}{F}$}\tabularnewline
\hline
\small{Higgs-Lepton} & \small{ $\LLR{i}{l}{r}$}&\small{$\bm{3}^l$}&\small{$\bm{\bar{3}}_r$}&\small{$ \bm{1} $ }&\small{$\bm{3}^{i}$}\tabularnewline
\small{Left-Quark}& \small{ $\QL{i}{x}{l}$ }&\small{$\bm{\bar{3}}_l$}&\small{$\bm{1}$}&\small{$\bm{3}^x$ }&\small{$\bm{3}^{i}$}\tabularnewline
\small{Right-Quark} & \small{ $\QR{i}{r}{x}$}&\small{$\bm{1}$}&\small{$\bm{3}^r$}&\small{$ \bm{\bar{3}}_x$ }&\small{$\bm{3}^{i}$}\tabularnewline
\hline
\end{tabular} }
\caption[ ]{\emph{Tri-triplet chiral superfields in the SHUT model and their quantum numbers.}}
\label{table:ChiralSuperE6}
\end{table}
The $\mathbb{Z}_3^{(\mathrm{LRC})}$ in Eq.~\eqref{eq:3333} is realized on the chiral and vector superfields as the simultaneous cyclic permutation 
within $\{ \bm{L}, \bm{Q}_\mathrm{L}, \bm{Q}_\mathrm{R} \}$ and $\{\bm{V}_\mathrm{L},\bm{V}_\mathrm{C},\bm{V}_\mathrm{R}\}$ sets, respectively, 
where $\bm{V}_\mathrm{L,R,C}$ are the vector (super)fields for the respective gauge $\SU{3}{L,R,C}$ groups. The $\mathbb{Z}_3^{(\mathrm{LRC})}$ 
symmetry enforces the gauge couplings of the $\SU{3}{L,R,C}$ groups to unify, i.e.~$g_\mathrm{L}=g_\mathrm{R}=g_\mathrm{C}\equiv g_\mathrm{U}$.

As mentioned previously, all fields in Tab.~\ref{table:ChiralSuperE6} can be contained in a $(\bm{27},\bm{3})$ representation of $\mathrm{E}_6 \times \SU{3}{F}$. 
In turn, the group $\mathrm{E}_6 \times \SU{3}{F}$ is a maximal subgroup of $\mathrm{E}_8$,
\begin{equation}
{\rm E_8} \supset { \rm E_6} \times \SU{3}{F} \,,
\label{eq:E8maximal}
\end{equation}
where the $(\bm{27},\bm{3})$ fits neatly into the $\bm{248}$ irrep of $\mathrm{E}_8$ whose branching rule is given by
\begin{equation}
\bm{248} = \left(\bm{1}, \bm{8} \right) \oplus \left( \bm{78},\bm{1} \right) \oplus \left( \bm{27},\bm{3} \right) \oplus \left(\bm{\overline{27}},
\bm{\overline{3}} \right)\,. 
\label{eq:248}
\end{equation}

Note, for clarity, that we are only considering representations of the subgroup $\[\SU{3}{}\]^4$ of $\mathrm{E}_8$, which are chiral rather than vector-like, in agreement with the chiral fermion content of the SM. In this work, we treat $\SU{3}{F}$ as a global symmetry. While considerably simpler, the trinification model with global $\SU{3}{F}$ can be viewed as the principal part of the fully gauged version in the limit of a vanishingly small family-gauge coupling $g_{\mathrm{F}} \ll g_{\mathrm{U}}$. In that case, Goldstone bosons would become the longitudinal d.o.f of massive $\SU{3}{F}$ gauge bosons instead of remaining as massless scalars. Such a restricted model can thus be a first step towards the fully gauged $\mathrm{E}_8$-inspired version.

Considering only renormalizable interactions, the symmetry group $G_{333\{3\}}$ allows for just a single term in the superpotential with the tri-triplet superfields, 
\begin{equation} \label{eq:WE6}
W =  \lambda_{\bm{27}}\,  \varepsilon_{ijk} \LLR{i}{l}{r} \QL{j}{x}{l} \QR{k}{r}{x} \,.
\end{equation}
where $\lambda_{\bm{27}}$ can be taken to be real without any loss of generality, as any phase can be absorbed with a field redefinition. As the light Higgs and lepton sectors are fully contained 
in the single tri-triplet $\bm{L}$, this construction provides an exact unification of Yukawa interactions 
of the fundamental superchiral sector and the corresponding scalar quartic couplings to a common origin, $\lambda_{\bm{27}}$. 

The superpotential in Eq.~\eqref{eq:WE6} has an accidental $\U{W} \times \U{B} $ symmetry as we can perform independent 
phase rotations on two of the tri-triplets as long as we do a compensating phase rotation on the third. We can arrange the charges of the tri-triplets 
under $\U{W} \times \U{B} $ as shown in Tab.~\ref{tab:AccSym}, such that $\U{B}$ is identified as the symmetry responsible for baryon number conservation. With this, we have proton stability to all orders in perturbation theory. 
\begin{table}[htb!]
  \begin{center}
  \begin{tabular}{ccc}
     \toprule                     
                        				 						& $\U{W}$ & $\U{B}$  	\\    
      \midrule
       $\bm{L}$     			    							& $+1$		& $0$			\\
       $\bm{Q}_\mathrm{L}$  						& $-1/2$	& $+1/3$		\\
       $\bm{Q}_\mathrm{R}$ 	 					& $-1/2$	& $-1/3$ 		\\
     \bottomrule
    \end{tabular} 
    \caption{\emph{Charge assignment of the tri-triplets under the accidental symmetries.}}
    \label{tab:AccSym}  
  \end{center}
\end{table} 

The model with the superpotential in Eq.~\eqref{eq:WE6} also exhibits an accidental symmetry under LR-parity $\mathds{P}$. 
This is realized at the superspace level as
\begin{equation} 
\label{eq:SUSYParity}
\begin{aligned}
& \LLR{i}{s}{t} \parityarrow -\LLRs{i}{t}{s} \quad , \,  \left( \bm{Q}_{\mathrm{L,R}}^i \right)^x{}_s \parityarrow  \left( \bm{Q}^*_{\mathrm{R,L}}{}_i \right)_s{}^x \, \\
& \qquad \qquad \bm{V}_{\mathrm{L,R,C}}^a \parityarrow -\bm{V}_{\mathrm{R,L,C}}^a \quad \, , 
\end{aligned} 
\end{equation}
accompanied by
\begin{equation}
x^\mu \parityarrow x_\mu \quad , \quad \theta_\alpha \parityarrow \i \theta^{\dagger\;\alpha} \, . 
\end{equation}
Here, $\alpha$ is the spinor index on the Grassman valued superspace coordinate $\theta$. Note that $s$ and $t$ in Eq.~\eqref{eq:SUSYParity} 
label both $\SU{3}{L,R}$ indices as such representations are swapped under LR-parity. At the Lagrangian level, the LR-parity transformation 
rules become
\begin{equation} 
\label{eq:LagrParity}
\begin{aligned}
\SLLR{i}{s}{t} \parityarrow -\SLLRs{i}{t}{s}\, , & \quad \left[\FLLR{i}{s}{t}\right]_\alpha \parityarrow- \i \left[\FLLRd{i}{t}{s}\right]^\alpha \, ,	\\
\SQL{i}{x}{s} \parityarrow \SQRs{i}{s}{x} \, , & \quad \left[ \left( Q^i_{\mathrm{L,R}} \right)^x{}_s \right]_\alpha \parityarrow \i \left[ \left( Q_{\mathrm{R,L}}^{\dagger}{}_i \right)_s{}^x\right]^\alpha \, , \\
G_{\mathrm{L,R,C}}^a{}_\mu \parityarrow G_{\mathrm{R,L,C}}^{a}{}^\mu  \, , & 
\quad \left[ \widetilde{\lambda}^a_{\mathrm{L,R,C}} \right]_\alpha \parityarrow -\i \left[ \widetilde{\lambda}^a_{\mathrm{R,L,C}} {}^\dagger \right]^{\alpha} \, , 
\end{aligned} 
\end{equation}
which can be verified by expanding out the components of the superfields in Eq.~\eqref{eq:SUSYParity}. In this model, LR-parity exists already 
at the $\SU{3}{}$ level, unlike common $\SU{2}{L}\times\SU{2}{R}$ LR-symmetric realisations. Note also that there exist the corresponding 
accidental Right-Colour and Colour-Left parity symmetries due to the $\mathbb{Z}_3^{(\mathrm{LRC})}$ permutation symmetry imposed in the SHUT model.

As mentioned in the introduction, one of the main drawbacks of a SUSY T-GUT (as well as any SUSY GUT with very few free parameters) is the difficulty for spontaneous 
breaking of high-scale symmetries. For example, while the non-SUSY T-GUT in Ref.~\cite{Camargo-Molina:2016bwm} has no problem with SSB down to 
a LR-symmetric theory, when including SUSY the additional relations between potential and gauge couplings make it so that there is no minimum of 
the potential allowing for that breaking. Moreover, even when relaxing the family symmetry, any VEV in e.g.~$\widetilde{L}^i$ 
induces mass terms that mix the $L^i$ fermions with the gauginos $\widetilde{\lambda}^a_{\rm L,R}$ through $\mathcal{D}$-term interactions of the type 
\begin{equation}
\label{eq:FermiD}
\mathcal{L}_{\mathcal{D}} = -\sqrt{2} g_{\rm U}  \SLLRs{i}{l_1}{r} \left(T_a\right)^{l_1}{}_{l_2}\FLLR{i}{l_2}{r} \widetilde{\lambda}^a_{\mathrm{L}} \,.
\end{equation}
This is a common problem in the previous T-GUT realizations as the number of light fields would not be enough to accommodate the particle content of the SM 
at low energies. While it is possible to get around this issue by adding extra Higgs multiplets to the theory and making them responsible for the SSB, this significantly 
increases the amount of light exotic fields that might be present at low energies but are unobserved. Such theories typically contain a very large number of free 
parameters and a fair amount of fine tuning which significantly reduces their predictive power.

In the SHUT model, this issue is instead solved by the inclusion of adjoint $\SU{3}{L,R,C,F}$ chiral supermultiplets, 
$\bm{\Delta}_{\mathrm{L,R,C,F}}$.
By triggering the first SSB, while preserving SUSY, VEVs in scalar components of $\bm{\Delta}_{\mathrm{L,R,F}}$ 
do not lead to heavy would-be SM lepton fields. In addition, the scalar and fermion components of $\bm{\Delta}_{\mathrm{L,R,C}}$ are all automatically heavy 
after the breaking and thus do not remain in the low-energy theory.

\subsection{$\SU{3}{}$ adjoint superfields} 
\label{sec:SUSYGaugeAdjoints}

The addition of gauge adjoint superfields is the main feature preventing SM-like leptons from getting a GUT scale mass. As was briefly mentioned above, the gauge and family $\SU{3}{}$ adjoints are motivated by the $(\bm{78},\bm{1})$ and 
$(\bm{1},\bm{8})$ representations of $\mathrm{E}_6 \times \SU{3}{F}$ (which can be inspired by the branching rule of the $\bm{248}$-rep in its embedding into 
$\mathrm{E}_8$ as shown in Eq.~\eqref{eq:248}). Indeed, the $\bm{78}$-rep, in turn, branches as
\begin{equation}
\label{eq:78}
\Scale[0.9]{\bm{78} =  \left( \bm{8}, \bm{1}, \bm{1} \right) \oplus \left( \bm{1}, \bm{8}, \bm{1} \right) \oplus \left( \bm{1}, \bm{1}, \bm{8} \right) \oplus 
\left( \bm{3}, \bm{3}, \bm{\overline{3}} \right) \oplus \left( \bm{\overline{3}}, \bm{\overline{3}}, \bm{3} \right) } ,
\end{equation}
under $\mathrm{E}_6 \supset [\SU{3}{}]^3$. We include three gauge-adjoint chiral superfields $\bm{\Delta}_{\mathrm{L,R,C}}$ corresponding to 
$\left( \bm{8}, \bm{1}, \bm{1} \right)$, $\left( \bm{1}, \bm{8}, \bm{1} \right)$ and $\left( \bm{1}, \bm{1}, \bm{8} \right)$ in Eq.~\eqref{eq:78}, 
respectively, as well as the family $\SU{3}{F}$ adjoint, $\bm{\Delta}_{\mathrm{F}}$ (all listed in Table~\ref{table:ChiralSuperAdj}). 
The transformation rule for the $\mathbb{Z}_3^{(\mathrm{LRC})}$ symmetry in $G_{333\{3\}}$ of Eq.~\eqref{eq:3333} is now accompanied by the cyclic permutation of 
$\{ \bm{\Delta}_\mathrm{L}, \bm{\Delta}_\mathrm{C}, \bm{\Delta}_\mathrm{R} \}$ fields. 

In order to keep the minimal setup, in this work we will not consider the fields that correspond to $\left( \bm{3}, \bm{3}, \bm{\overline{3}} \right)$ and 
$ \left( \bm{\overline{3}}, \bm{\overline{3}}, \bm{3} \right)$ from Eq.~\eqref{eq:78}. In practice, they can be made very heavy and only couple to the tri-triplets via gauge interactions.
\begin{table}[htb!]
\centering
\Scale[0.9]{\begin{tabular}{|>{\centering}m{2.5cm}>{\centering}m{1.2cm}|>{\centering}m{1.2cm}|>{\centering}m{1.2cm}|>{\centering}m{1.2cm}|>{\centering}m{1.2cm}|}
\hline
\multicolumn{6}{|c|}{Chiral supermultiplet fields}\tabularnewline
\hline
\hline
\small{Superfield}&\small{}&\small{$\SU{3}{L}$}&\small{$\SU{3}{R}$}&\small{$\SU{3}{C}$}&\small{$\SU{3}{F}$}\tabularnewline
\hline
\small{Left-adjoint} & \small{ $\DA{L}{a}$}&\small{$\bm{8}^a$}&\small{$\bm{1}$}&\small{$ \bm{1}$ }&\small{$\bm{1}$}\tabularnewline
\small{Right-adjoint} & \small{ $\DA{R}{a}$}&\small{$\bm{1}$}&\small{$\bm{8}^a$}&\small{$ \bm{1}$ }&\small{$\bm{1}$}\tabularnewline
\small{Colour-adjoint} & \small{ $\DA{C}{a}$}&\small{$\bm{1}$}&\small{$\bm{1}$}&\small{$ \bm{8}^a$ }&\small{$\bm{1}$}\tabularnewline
\small{Family-adjoint} & \small{ $\DA{F}{a}$}&\small{$\bm{1}$}&\small{$\bm{1}$}&\small{$ \bm{1}$ }&\small{$\bm{8}^a$}\tabularnewline
\hline
\end{tabular} }
\caption[ ]{\it $\SU{3}{}$ adjoint chiral superfields in the SHUT model and their representations.}
\label{table:ChiralSuperAdj}
\end{table}

By introducing the adjoint chiral superfields, we have to add the following terms 
\begin{eqnarray} \nonumber
W \supset &\sum_{A=\mathrm{L,R,C}} \left[\frac{1}{2} \mu_\mathbf{78} \, 
\bm{\Delta}_A^a \bm{\Delta}_A^a + 
\frac{1}{3!} \lambda_\mathbf{78} \, d_{abc} \bm{\Delta}_A^a \bm{\Delta}_A^b 
\bm{\Delta}_A^c \right] \\
& + \frac{1}{2} \mu_{\bm{1}}  \DA{F}{a} \DA{F}{a}
+ \frac{1}{3!} \lambda_{\bm{1}} d_{abc} \DA{F}{a} \DA{F}{b} \DA{F}{c}  \,, 
\label{eq:SuperPotentialGaugeAdj}
\end{eqnarray}
to the superpotential in Eq.~\eqref{eq:WE6}. Here, $d_{abc} = 2 \mathrm{Tr}[\lbrace T_a, T_b \rbrace T_c]$ are the totally symmetric $\SU{3}{}$ coefficients. 

Note that bilinear terms are only present for the adjoint superfields and not for the fundamental ones, as they are forbidden by the T-GUT symmetry. Since the VEVs of the adjoint scalars set the first scale where the T-GUT symmetry is spontaneously broken, while all subsequent breaking steps occur at scales given by the soft parameters. In other words, the model is free of the so-called $\mu$-problem.

We can pick the phase of $\bm{\Delta}_{\mathrm{L,R,C,F}}$ to make $\mu_{\bm{78}}$ and $\mu_{\bm{1}}$ real, which makes $\lambda_{\bm{78}}$ and 
$\lambda_{\bm{1}}$ complex, in general. Notice that the superpotential provides no renormalisable interaction terms between the adjoint superfields 
and the tri-triplets. The accidental $\U{W} \times \U{B} $ symmetry of the tri-triplet sector is not affected by $\bm{\Delta}_{\mathrm{L,R,C,F}}$ as we 
can take these fields simply to not transform under this symmetry. The gauge interactions are parity-invariant with the following definitions for the transformation rules,
\begin{equation}
\tilde{\Delta}_{\mathrm{L,R,C,F}}^a \parityarrow \tilde{\Delta}_{\mathrm{R,L,C,F}}^{*a},
\quad \left[ \Delta_{\mathrm{L,R,C,F}}^a \right]_{\alpha} \parityarrow \i \left[ \Delta_{\mathrm{R,L,C,F}}^{\dagger a} \right]^{\alpha},
\label{eq:Delta-parities}
\end{equation}
or, equivalently, $\bm{\Delta}^a_{\mathrm{L,R,C,F}} \parityarrow \bm{\Delta}^{*a}_{\mathrm{R,L,C,F}}$ at the superfield level. However, LR-parity is not generally respected 
by the $\mathcal{F}$-term interactions unless $\lambda_{\bm{78}}$ and $\lambda_{\bm{1}}$ are real. In what follows, we assume a real $\lambda_{\bm{78}}$, whereas the accidental
LR-parity can be explicitly broken by the soft SUSY-breaking sector of the theory, at or below the GUT scale.

Now, for illustration, let us discuss briefly the first symmetry breaking step which determines the GUT scale in the SHUT model (see Fig.~\ref{fig:1112abc}). 
Eq.~\eqref{eq:SuperPotentialGaugeAdj} leads to a scalar potential containing several SUSY-preserving minima with VEVs that can be rotated to 
the eighth component of $\tilde{\Delta}^8_{\mathrm{L,R,F}}$. In particular, there is an $\SU{3}{C}$ and LR-parity preserving minimum with
\begin{equation}
\langle \tilde{\Delta}_{\mathrm{L,R}}^a \rangle = \frac{v_{\mathrm{L,R}}}{\sqrt{2}} \delta^a_8 \quad \mbox{with} \quad v_{\mathrm{L,R}} = v \equiv2 \sqrt{6} \, 
\frac{ \mu_\mathbf{78}}{\lambda_\mathbf{78}}, \quad v_{\mathrm{C}}=0\,,
\end{equation}
for the gauge-adjoints, and
\begin{equation}
\langle \tilde{\Delta}_\mathrm{F}^a \rangle = \frac{v_\mathrm{F}}{\sqrt{2}} \delta^a_8 \quad \mbox{with} \quad v_\mathrm{F} = 
2 \sqrt{6} \, \frac{ \mu_\mathbf{1}}{\lambda_\mathbf{1}} \,  \label{vF-VEV}\,,
\end{equation}
for the family-adjoint, setting the GUT scale $v\sim v_{\mathrm{F}}$. The vacuum structure $\langle \tilde{\Delta}_{\mathrm{L,R,F}}^8 \rangle \neq 0$ leads 
to the spontaneous breaking \mbox{$\mathrm{SU}(3)_{\mathrm{L,R,F}} \rightarrow \mathrm{SU}(2)_{\mathrm{L,R,F}} \times \mathrm{U}(1)_{\mathrm{L,R,F}}$} 
(see Appendix~\ref{sec:Symmetries} for the corresponding generators and $\U{}$ charges), resulting in the unbroken group
\begin{eqnarray}
\label{eq:322113}
G_{32211\lbrace 21 \rbrace} &\equiv& \SU{3}{C} \times [ \SU{2}{L} \times \SU{2}{R} 
\\ &\times&  
\U{L} \times \U{R} ] \times \lbrace \SU{2}{F} \times\U{F} \rbrace \, . 
\nonumber 
\end{eqnarray}
LR-parity also remains unbroken since $v_{\mathrm{L}} = v_{\mathrm{R}}^*$, which is true as long as $\lambda_{\bm{78}}$ is taken 
to be real.

By making the shift
\begin{equation} \label{eq:GaugeAdjShift}
\bm{\Delta}_{\mathrm{L,R}}^a \rightarrow \bm{\Delta}_{\mathrm{L,R}}^a + \frac{v}{\sqrt{2}} \delta^a_8,\;\; 
\bm{\Delta}_{\mathrm{F}}^a \rightarrow \bm{\Delta}_{\mathrm{F}}^a + \frac{v_\mathrm{F}}{\sqrt{2}} \delta^a_8
\end{equation}
and substituting $\mu_\mathbf{78} =\frac{\lambda_\mathbf{78} \, v}{2 \sqrt{6}}$, 
$\mu_\mathbf{1} =\frac{\lambda_\mathbf{1} \, v_{\mathrm{F}}}{2 \sqrt{6}}$ in the superpotential, we obtain

\onecolumngrid

\begin{equation}
\begin{aligned}
W&\supset \sum_{B=\mathrm{L,R}} \left[ \frac{ \lambda_\mathbf{78} \, v}{2\sqrt{2}} \left( d_{aa8} + 
\frac{1}{2\sqrt{3}} \right) \bm{\Delta}_B^a \bm{\Delta}_B^a + \frac{1}{3!} \lambda_\mathbf{78} \, 
d_{abc}  \bm{\Delta}_B^a \bm{\Delta}_B^b \bm{\Delta}_B^c \right] + \frac{ \lambda_\mathbf{1} \, v_\mathrm{F}}{2\sqrt{2}} \left( d_{aa8} + 
\frac{1}{2\sqrt{3}} \right) \bm{\Delta}_\mathrm{F}^a \bm{\Delta}_\mathrm{F}^a \\
&+\frac{1}{3!} \lambda_\mathbf{78} \, 
d_{abc}  \bm{\Delta}_\mathrm{F}^a \bm{\Delta}_\mathrm{F}^b \bm{\Delta}_\mathrm{F}^c + 
\frac{ \lambda_\mathbf{78} \, v }{4 \sqrt{6}}  \, \bm{\Delta}_{\mathrm{C}}^a 
\bm{\Delta}_{\mathrm{C}}^a + \frac{1}{3!} \lambda_\mathbf{78} \, 
d_{abc} \bm{\Delta}_{\mathrm{C}}^a \bm{\Delta}_{\mathrm{C}}^b 
\bm{\Delta}_{\mathrm{C}}^c + \mbox{const.}
\end{aligned}
\end{equation}

\twocolumngrid

The quadratic terms in the superpotential vanish for $\bm{\Delta}_{\mathrm{L,R,F}}^{4,5,6,7}$, since $d_{aa8}=-1/(2 \sqrt{3})$ for $a=4,5,6,7$,
meaning that these fields receive no $\mathcal{F}$-term contribution to their masses (contrary to the other components of $\bm{\Delta}_\mathrm{L,R}$ and 
$\bm{\Delta}_\mathrm{F}$ which receive GUT scale masses $m_{\Scale[0.7]{\Delta}}^2 \sim \lambda_{\bm{78}}^2 v^2$ and 
$\lambda_{\bm{1}}^2 v_{\mathrm{F}}^2$, respectively). While the global Goldstone bosons $\mathrm{Re} [\tilde{\Delta}_{\mathrm{F}}^{4,5,6,7}]$ 
are present in the physical spectrum, the gauge ones become the longitudinal polarisation states of 
the heavy gauge bosons related to the breaking $G_{333} \rightarrow G_{32211}$. 

The presence of massless scalar degrees of freedom can only be avoided in the extended model with the gauged family symmetry. It is clear, however, 
that even in the case of an approximately global $\SU{3}{F}$ with $g_{\rm F}\ll g_{\rm U}$ there are no massless Goldstones in the spectrum (provided 
that the accidental symmetries are softly broken at low energies) but a set of relatively light family gauge bosons very weakly interacting with the rest of 
the spectrum.

By performing the shifts in Eq.~\eqref{eq:GaugeAdjShift} in the $\mathcal{D}$-terms, we obtain
\begin{eqnarray*}
\mathcal{D}^a_B \supset -\i f^{abc} \tilde{\Delta}^b_B {}^\dagger \tilde{\Delta}^c_B 	&\rightarrow& -
\i \frac{v}{\sqrt{2}} f^{a8b} \left( \tilde{\Delta}^b_B -  \tilde{\Delta}^b_B {}^\dagger \right) \\ 
&-& \i f^{abc} \tilde{\Delta}^b_B {}^\dagger \tilde{\Delta}^c_B \, ,
\end{eqnarray*}
for $B=\mathrm{L,R}$, leading to the universal GUT scale mass term $m^2 = 3 g_{\rm U}^2 v^2/4$ for the gauge-adjoints 
$\mathrm{Im}[\tilde{\Delta}^{4,5,6,7}_\mathrm{L,R} ]$, while $\tilde{\Delta}_{\mathrm{F}}^{4,5,6,7}$ have no $\mathcal{D}$-term contributions 
(or a small one in the case of approximately global $\SU{3}{F}$ with $g_{\rm F}\ll g_{\rm U}$). Hence, all components of the gauge adjoints and 
$\tilde{\Delta}^{1,2,3,8}_\mathrm{F}$ receive masses of order GUT scale and are integrated out in the low-energy EFT. The remaining 
$\tilde{\Delta}_{\mathrm{F}}^{4,5,6,7}$, on the other hand, receive a much smaller mass from the soft SUSY-breaking sector (and strongly 
suppressed $\mathcal{D}$-terms) and stay in the physical spectrum of the EFT. In what follows, we shall denote by $\bm{\mathcal{H}}^i_F$ 
the superfields containing $\mathrm{Im}[\tilde{\Delta}^{4,5,6,7}_\mathrm{L,R} ]$, and by $\bm{\mathcal{G}}^i_F$ the superfields containing $\mathrm{Re}[\tilde{\Delta}^{4,5,6,7}_\mathrm{L,R} ]$.

\subsection{LR-symmetric SUSY theory}
\label{sec:SUSYLR}

In this section we describe the details of the supersymmetric theory left after the adjoint fields acquire VEVs.  
As shown in the previous section, all components of the gauge adjoint chiral superfields receive masses of the order of the GUT scale ($\mathcal{O}(v)$) 
in the vacuum given by Eq.~\eqref{eq:GaugeAdjShift}. This means that to study the low-energy predictions of the theory, we need to integrate out 
$\bm{\Delta}_\mathrm{L,R,C}$, as well as components 1, 2, 3 and 8 of $\bm{\Delta}_\mathrm{F}$. 

For the gauge sector of the SHUT model, $\langle \tilde{\Delta}_{\mathrm{L,R}} \rangle$ naturally triggers a $\SU{3}{\mathrm{L,R}}\rightarrow\SU{2}{\mathrm{L,R}}\times\U{\mathrm{L,R}}$ breaking also for the tri-triplets (whose interactions with $\tilde{\Delta}_{\mathrm{L,R}}$ are mediated via $V_{\mathrm{L,R}}^a$ gauge bosons). 
For the global $\SU{3}{F}$ sector, there is no coupling of $\tilde{\Delta}_{\mathrm{F}}$ to the tri-triplets and, thus, the $\SU{3}{F}$ symmetry remains intact 
(or approximate in the case of $g_{\rm F}\ll g_{\rm U}$) in the tri-triplet sector, resulting in $G_{32211\{3\}}$ rather than $G_{32211\{21\}}$. 
Integrating out $\bm{\Delta}_\mathrm{L,R,C}$, and components 1, 2, 3 and 8 of $\bm{\Delta}_\mathrm{F}$, therefore leaves us with a supersymmetric 
theory based on the symmetry group $G_{32211\{3\}}$, with a chiral superfield content given by $\bm{\Delta}^{4-7}_\mathrm{F}$ 
and by the branching of $\bm{L}$, $\bm{Q}_\mathrm{L}$ and $\bm{Q}_\mathrm{R}$. 

Writing the trinification tri-triplets in terms of $G_{32211\{3\}}$ representations one gets,
\begin{eqnarray}
\label{eq:tri-triplets}
\renewcommand\arraystretch{1.3}
&&\LLR{i}{l}{r} =
\mleft(
\begin{array}{cc|c}
\bm{H}_{11} & \bm{H}_{12} & \bm{e}_{\mathrm{L}}\\
\bm{H}_{21} & \bm{H}_{22} & \bm{\nu}_{\mathrm{L}}\\
\hline
\bm{e}_{\mathrm{R}}^{c} & \bm{\nu}_{\mathrm{R}}^{c} & \bm{\phi}
\end{array}
\mright)^{i}\;, \\
&&
\begin{aligned}
\renewcommand\arraystretch{1.3}
&\QL{i}{x}{l}=
\mleft(
\begin{array}{cc|c}
\bm{u}_{\mathrm{L}}^x & \bm{d}_{\mathrm{L}}^x & \bm{D}_{\mathrm{L}}^x
\end{array}
\mright)^{i}\;,
\\
\renewcommand\arraystretch{1.3}
&\QR{i}{r}{x}=
\mleft(
\begin{array}{cc|c}
\bm{u}_{\mathrm{R}x}^c & \bm{d}_{\mathrm{R}x}^c & \bm{D}_{\mathrm{R}x}^c
\end{array}
\mright)^{\top\;\;i}\;,
\end{aligned}
\end{eqnarray}
where the vertical and horizontal lines denote the separation of the original tri-triplets into $\SU{2}{}$-doublets and singlets
after the first SSB step. We will refer to the lepton and quark $\SU{2}{L,R}$ doublets as $\bm{E}_{\mathrm{L,R}}$ and $\bm{q}_{\mathrm{L,R}}$. 
With this, we find that the most general superpotential consistent with $G_{32211\{3\}}$ is
\begin{equation} \label{eq:WLR}
\begin{aligned}
W &= \varepsilon_{ijk} \left\lbrace  y_1 \bm{\phi}^i {\bm{D}_\mathrm{L}}^j {\bm{D}_\mathrm{R}}^k + 
y_2 (\bm{H}^i)^L{}_R  ({\bm{q}_\mathrm{L}}^j)_L ({\bm{q}_\mathrm{R}}^k)^R  \right. \\
& \left. + y_3 ({\bm{E}_\mathrm{L}}^i)^L ({\bm{q}_\mathrm{L}}^j)_L  {\bm{D}_\mathrm{R}}^k  +  y_4
({\bm{E}_\mathrm{R}}^i)_R  {\bm{D}_\mathrm{L}}^j ({\bm{q}_\mathrm{R}}^k)^R  \right\rbrace \,.
\end{aligned}
\end{equation}
Note, in this effective SUSY LR theory one could naively add a mass term like $\varepsilon_{i j} \tilde{\mu} \bm{\mathcal{H}}^i_F \bm{\mathcal{G}}^i_F$ 
(that is symmetric under $\SU{2}{F}\times \U{F}$ but not under full $\SU{3}{F}$) between the massless components of the family-adjoint superfield, 
$\bm{\mathcal{H}}^i_F$, and the massless superfield $\bm{\mathcal{G}}^i_F$ containing the Goldstone bosons. Such an effective $\mu$-term is matched 
to zero at tree level at the GUT scale. Due to SUSY non-renormalisation theorems \cite{Grisaru:1979wc}, in the exact SUSY limit this term cannot be 
regenerated radiatively at low energies so $\tilde{\mu}$ is identically zero and was not included in the superpotential given by Eq.~(\ref{eq:WLR}). So, the resulting 
superpotential contains only fundamental superfields coming from $\bm{L}$, $\bm{Q}_\mathrm{L}$ and $\bm{Q}_\mathrm{R}$ and is indeed invariant 
under $\SU{3}{F}$.

In the GUT scale theory, a complex $\lambda_{\bm{78}}$ would be the only source of LR-parity violation. In the low energy theory this should 
lead to $y_3 \neq y_4^*$. Otherwise, $y_3 = y_4^*$ and after the matching is performed we can always make any $y_{1,2,3,4}$ real by field redefinitions.
The same argument applies for the equality of the corresponding LR gauge couplings for $\SU{2}{L,R} \times \U{L,R}$ symmetries.

Since we now have an effective LR-symmetric SUSY model with a $\U{L,R}$ symmetry, there is a possibility of having gauge 
kinetic mixing. The $\U{L,R}$ $D$-term contribution to the Lagrangian is given by
\begin{equation}
\mathcal{L} \supset \frac{1}{2} (\chi\, \mathcal{D}_{\mathrm{L}}\mathcal{D}_{\mathrm{R}} + \mathcal{D}_{\mathrm{L}}^2 + 
\mathcal{D}_{\mathrm{R}}^2) - \kappa (\mathcal{D}_{\mathrm{L}} - \mathcal{D}_{\mathrm{R}}) + 
X_{\mathrm{L}} \mathcal{D}_{\mathrm{L}} + X_{\mathrm{R}} \mathcal{D}_{\mathrm{R}} \,,
\end{equation}
where the terms proportional to $\kappa$ are the Fayet-Iliopoulos terms, while the $D$-terms and the expressions for $X_{\mathrm{L,R}}$ 
are shown in Appendix~\ref{sec:ADterms}.

The values of the parameters $\{ y_{1,2,3,4}, \g{C}, \g{L,R}, \g{L,R}^{\prime}, \chi, \kappa \}$ in the LR-symmetric SUSY theory are determined by 
the values of the parameters $\{ \lambda_{\bm{27}}, \lambda_{\bm{78}}, g_{\mathrm{U}}, v \}$ in the high-scale trinification theory at the GUT scale boundary 
through a matching procedure\footnote{Before adding soft SUSY-breaking interactions, $\Delta_\mathrm{F}$ is completely decoupled from the fundamental sector when taking $\SU{3}{F}$ to be global, meaning that $\lambda_{\bm{1}}$ and $v_{\rm F}$ do not enter in the matching conditions.}. Regarding the RG evolution of the couplings, we note that the only dimensionful parameter in the effective theory 
is the Fayet-Iliopoulos parameter $\kappa$. This means that $\beta_\kappa \propto \kappa$ so that if $\kappa=0$ at the matching scale (which is true, at least, 
at tree level), then $\kappa$ will remain zero throughout the RG flow yielding no spontaneous SUSY-breaking. Thus, we stick to the concept of 
soft SUSY-breaking in what follows.\\

\section{Softly broken SUSY} 
\label{sec:BreakingSUSY}

In this section we describe the details of adding soft SUSY-breaking terms before the SHUT symmetry is broken spontaneously by adjoint field VEVs. One of the most important results is treated in Sec.~\ref{sec:SU3breakingSoft2}, where it is shown that the symmetry breakings below the GUT scale are triggered solely by the soft SUSY-breaking sector. This in turn allows for a strong hierarchy between the GUT scale and the scale of the following VEVs.\\

\subsection{The soft SUSY-breaking Lagrangian}
\label{sec:SSBSU3}

The soft SUSY-breaking scalar potential terms respecting the imposed $G_{333\{3\}}$ symmetry, are bilinear and trilinear interactions given by

\onecolumngrid

\begin{align}
\label{eq:Vsg}
\begin{aligned}
V_{{\rm{soft}}}^{\rm G} = &~\Bigg\{
~m^2_{\bm{27}}\,   \SLLR{i}{l}{r} \SLLRs{i}{l}{r} + m^2_{\bm{78}} \SDAs{L}{a} \SDA{L}{a}
+  \left[ \frac{1}{2} b_{\bm{78}} \SDA{L}{a} \SDA{L}{a} 
 + \mathrm{c.c} \right] \\ 					
& 
+ d_{a b c}\, \left[ \frac{1}{3!} A_{\bm{78}} \SDA{L}{a} \SDA{L}{b} \SDA{L}{c} 
+ \frac{1}{2} C_{\bm{78}} \SDAs{L}{a} \SDA{L}{b} \SDA{L}{c} + \mathrm{c.c.} \right] 
\\ 
&
+  \left[ A_{\mathrm{G}} \, \SDA{L}{a} \left( T_a \right)_{\phantom{l}l_2}^{l_1} \SLLRs{i}{l_1}{r} \SLLR{i}{l_2}{r} 
+ A_{\bar{\mathrm{G}}} \, \SDA{R}{a} \left( T_a \right)_{\phantom{l} r_1}^{r_2} \SLLRs{i}{l}{r_1} \SLLR{i}{l}{r_2}  + 
\mathrm{c.c.} \right]   \\ 
&
+ (\mathbb{Z}_3^{(\mathrm{LRC})} \mbox{ permutations}) \Bigg\} +\, \left[ A_{\bm{27}} \,  
\varepsilon_{ijk}\SQL{i}{x}{l} \SQR{j}{r}{x} \SLLR{k}{l}{r} +\mathrm{c.c.} \right]\,,
\end{aligned}
\end{align}

\twocolumngrid

for the gauge-adjoints and pure tri-triplet terms, and 

\onecolumngrid

\begin{equation}
\label{eq:SUSYbreakingNoDeltaHS}
\begin{aligned}
V_{{\rm{soft}}}^{\rm F}  = & ~ m^2_{\bm{1}} \SDAs{F}{a} \SDA{F}{a} + \left[ \frac{1}{2} b_{\bm{1}} \SDA{F}{a} \SDA{F}{a}  + 
\mathrm{c.c} \right] +  \,d_{a b c} \left[ \frac{1}{3!} A_{\bm{1}} \SDA{F}{a} \SDA{F}{b} \SDA{F}{c} + 
\frac{1}{2} C_{\bm{1}} \SDAs{F}{a} \SDA{F}{b} \SDA{F}{c} +  \mathrm{c.c.} \right]  \\ 
& +\left[A_{\mathrm{F}} \SDA{F}{a} \left( T_a \right)^i {}_j \SLLRs{i}{l}{r} \SLLR{j}{l}{r} + \mathrm{c.c.} + 
(\mathbb{Z}_3^{(\mathrm{LRC})} \mbox{ permutations})  \right] \,.
\end{aligned}
\end{equation}

\twocolumngrid

for the family adjoint. All parameters here are assumed to be real for simplicity. 
We note that although trilinear terms with the gauge singlets (such as $\SDAs{F}{} \SDA{F}{} \SDA{F}{} $ above) are not in general soft, 
due to the family symmetry and the fact that $\sum_a d_{aab} = 0 $, the dangerous tadpole diagrams do indeed cancel and do not lead 
to quadratic divergences.

The terms in Eq.~(\ref{eq:Vsg}) and (\ref{eq:SUSYbreakingNoDeltaHS}), which account for the most general soft SUSY-breaking scalar potential 
consistent with $G_{333\{3\}}$ and real parameters, also respect the accidental $\U{W} \times \U{B}$ symmetry of the original SUSY theory. However, accidental LR-parity is, 
in general, softly-broken as long as $A_{\mathrm{G}} \not= A_{\bar{\mathrm{G}}}$, and this breaking can then be transmitted to the other sectors of the effective theory 
radiatively (e.g.~via RG evolution and radiative corrections at the matching scale).

The only dimensionful parameters entering in the tree-level tri-triplet masses come from soft SUSY-breaking parameters, such that the corresponding scalar fields receive masses of the order of the soft SUSY-breaking scale. The full expressions are given in Appendix~\ref{sec:masses}, from which we notice that positive squared masses requires
\begin{eqnarray} \nonumber
\abs{A_{\mathrm{G},\bar{\mathrm{G}}}} v &\sim & \abs{A_{\rm F}} v_{\rm F} \sim \abs{A_{\bm 1}} 
v_{\rm F} \\  &\sim & m^2_{\rm soft} \Rightarrow 
 \begin{cases}
 \abs{A_{\mathrm{G},\bar{\mathrm{G}},\mathrm{F}}} & \lesssim \tfrac{m^2_{\bm{27}}}{v}\sim \frac{m_{\rm soft}^2}{v}\,,\\
 \abs{A_{\bm 1}}  & \lesssim \tfrac{m^2_{\bm{1}}}{v_{\rm F}} \,.
 \end{cases}
 \label{eq:softCond}
\end{eqnarray}
For more details, see Sect.~\ref{Sec:VacStab}.

Note that the $A_{\mathrm{F}}$-term in the soft sector introduces small $\SU{3}{F}$ violating (but $\SU{2}{F} \times \U{F}$ preserving) effects on 
the interactions in the effective theory once $\langle \tilde{\Delta}_{\rm F} \rangle \neq 0$. Consider, for example, effective quartic interactions 
between components of $\tilde{L}$  that come from two $A_{\mathrm{F}}$ tri-linear vertices connected by an internal $\tilde{\Delta}_{\mathrm{F}}^{1,2,3}$ or 
$\tilde{\Delta}_{\mathrm{F}}^{8}$ propagator. The value of this diagram is $\sim \i A_{\mathrm{F}}^2 / \lambda_{\bm{1}}^2 v_{\mathrm{F}}^2$ neglecting 
the external momentum in the propagator. Using Eq.~\eqref{eq:softCond}, we see that this diagram behaves as $[m_{\rm soft}/v]^4$. 

The possible fermion soft SUSY-breaking terms are the Majorana mass terms for the gauginos and the Dirac mass terms between the gauginos and the fermion 
components of $\bm{\Delta}_{\mathrm{L,R,C}}$, namely,
\begin{eqnarray}
\label{eq:Lsoft}
\mathcal{L}^{\rm{fermion}}_{\rm{soft}} &=&  \Big[ -\dfrac{1}{2} M_0 \widetilde{\lambda}^a_{\rm L} 
\widetilde{\lambda}^a_{\rm L} - M^{\prime}_0 \widetilde{\lambda}^a_{\rm L} \FDA{L}{a}  + 
\mathrm{c.c.} \\
&+& (\mathbb{Z}_3^{(\mathrm{LRC})} \mbox{ permutations})  \Big] \,, \nonumber
\end{eqnarray}
From the transformation rules in Eqs.~\eqref{eq:SUSYParity} and \eqref{eq:Delta-parities} it follows that LR-parity is not respected 
by $\mathcal{L}^{\rm{fermion}}_{\rm{soft}}$ unless $M^{\prime}_0 = 0$. \\

\subsection{Vacuum in the presence of soft SUSY-breaking terms} 
\label{sec:SU3breakingSoft2}

Here we show how the scalar potential changes in the presence of soft SUSY-breaking interactions. In particular, how soft SUSY-breaking terms trigger a VEV in $\SLLR{3}{3}{3} \equiv \widetilde{\phi}^3$ of the same order 
as the soft SUSY-breaking scale. 

With $\langle \Delta_{\mathrm{L,R,F}}^8 \rangle \equiv \frac{1}{\sqrt{2}} v_{\mathrm{L,R,F}}$ and 
$\langle \widetilde{\phi}^3\rangle \equiv \frac{1}{\sqrt{2}} v_{\varphi}$ being the VEVs present, our potential evaluated in the vacuum is given by

\onecolumngrid

\begin{equation}
\begin{aligned}
V_{\rm vac} = &    \left[ \frac{1}{2} m_{\bm{27}}^2   - \frac{1}{\sqrt{6}} \left( A_\mathrm{G} v_{\mathrm{L}}  +  A_{\bar{\mathrm{G}}} 
v_{\mathrm{R}}+A_\mathrm{F} v_\mathrm{F} \right) \right]  v_\varphi^2 +  \frac{1}{12} g_{\mathrm{U}}^2 v_{\varphi}^4 + 
\Big\{ \frac{1}{2} \left( m_{\bm{78}}^2 + b_{\bm{78}} \right) v_{\mathrm{L}}^2		
\\
&  - \frac{1}{\sqrt{6}} 
\left( \frac{1}{3!} A_{\bm{78}} + \frac{1}{2} C_{\bm{78}} \right) v_{\mathrm{L}}^3 + \frac{1}{2} 
v_{\mathrm{L}}^2 \, \left( \frac{1}{2 \sqrt{6}} \lambda_{\bm{78}}  v_{\mathrm{L}} - 
\mu_{\bm{78}} \right)^2 +(v_{\mathrm{L}} \rightarrow v_{\mathrm{R}} ) \Big\} 
\\
& +\frac{1}{2} \left( m_{\bm{1}}^2 + b_{\bm{1}} \right) v_{\mathrm{F}}^2 - \frac{1}{\sqrt{6}} 
\left( \frac{1}{3!} A_{\bm{1}} + \frac{1}{2} C_{\bm{1}} \right) v_{\mathrm{F}}^3 \,.
\end{aligned}
\end{equation}

\twocolumngrid

As all other fields (that do not acquire VEVs) only enter in bi-linear combinations, it suffices to consider the above terms to solve the conditions for vanishing 
first derivatives of the scalar potential. We retain the notation $v=2 \sqrt{6} \mu_{\bm{78}}/\lambda_{\bm{78}}$ for 
the VEVs of $\tilde{\Delta}_{\mathrm{L,R}}^8$ in the absence of soft terms. Assuming that the soft terms are much smaller than the GUT scale, 
i.e.~$m_{\rm soft}\ll v$, we can approximately solve the extremum conditions for $v_{\mathrm{L,R},\varphi}$ by Taylor expanding them to the leading 
order in soft terms. Doing so we find
\begin{equation}
\begin{aligned}
v_\varphi^2 \approx & \,  \frac{3}{g^2_{\mathrm{U}} } \left[ - m_{\bm{27}}^2 + 
\sqrt{\frac{2}{3}} \left( A_\mathrm{G} + A_{\bar{\mathrm{G}}} \right) v + 
\sqrt{\frac{2}{3}} A_\mathrm{F} v_\mathrm{F} \right] \, 	,			\\
v_{\mathrm{L,R}}  \approx & \, v + \frac{24}{\lambda_{\bm{78}}} \Big[ - \frac{ m_{\bm{78}}^2 + b_{\bm{78}} }{v} + 
\sqrt{\frac{3}{2}} \left( \frac{1}{3!} A_{\bm{78}} + \frac{1}{2} C_{\bm{78}} \right) \\
+& 
\frac{1}{\sqrt{6}} A_{\mathrm{G},\bar{\mathrm{G}}} \left(\frac{v_\varphi}{v} \right)^2 \Big],
\end{aligned}
\end{equation}
where in the top equation we see that the $\widetilde{\phi}^3$ VEV is of the order of the soft SUSY-breaking scale. In other words, the $\widetilde{\phi}^3$ VEV cannot be triggered unless soft terms are introduced. As is described in Sec.~\ref{sec:SSBSU3}, the soft tri-linear couplings $A_{{\rm G},\bar{\rm G}}$, $A_{\bm{78}}$ 
and $C_{\bm{78}}$ need to be $\lesssim m_{\bm{27}}^2 / v$ for having positive squared masses. 

Adding the soft terms 
shifts the values of the VEVs $v_{\mathrm{L,R}}$ described in Sec.~\ref{sec:SUSYGaugeAdjoints} by a relative amount behaving as 
\vspace{-2mm}
\begin{eqnarray}
\sim \Big[ \frac{m_{\rm soft}}{v} \Big]^2\,.
\end{eqnarray}
Furthermore, we note that the presence of $v_{\varphi}$ slightly affects the equality of $v_{\mathrm{L,R}}$,
\begin{equation}
v_{\mathrm{L}} - v_{\mathrm{R}} \approx \frac{4 \sqrt{6} }{\lambda_{\bm{78}}} 
\left( \frac{v_\varphi}{v} \right)^2 \, \left( A_\mathrm{G} - A_{\bar{\mathrm{G}}} \right) \,,
\end{equation}
as long as $A_\mathrm{G}\neq A_{\bar{\mathrm{G}}}$. The relative difference between $v_{\mathrm{L,R}}$, therefore, behaves as 
\begin{eqnarray}
\sim \Big[ \frac{m_{\rm soft}}{v} \Big]^4\,.
\end{eqnarray}
That is, although the VEVs of $\tilde{\Delta}_{\mathrm{L,R}}$ are shifted by the soft terms, the effect is very small, if not negligible, for $m_{\rm soft}\ll v$.

With a non-zero $v_\varphi\sim m_{\rm soft}\ll v$, the symmetry is further broken as
\begin{eqnarray}
\label{eq:SymmetriesAfterPhiVev}
&& \U{L} \times \U{R} \times \{ \U{F} \times \U{W} \} \\
&& \overset{\langle \widetilde{\phi}^3 \rangle}{\rightarrow} \U{L+R} \times 
\{ \U{S} \times \U{S^\prime} \} \,, \nonumber
\end{eqnarray} 
where $\U{L+R}$ consists of simultaneous $\U{L,R}$ phase rotations by the same phase. $\U{S}$ and $\U{S'}$ are also simultaneous 
$\U{L,R}$ phase rotations, but with opposite phase, which is compensated by an appropriate $\U{F}$ and $\U{W}$ transformation, respectively. 
All generators are presented in Appendix~\ref{sec:Symmetries}.

In the limit of vanishingly small $A_{\mathrm F}\to 0$ in Eq.~\eqref{eq:SUSYbreakingNoDeltaHS}, the model exhibits an exact global $\SU{3}{F'} \times \SU{3}{F''}$ 
symmetry as we could then perform independent $\SU{3}{}$ family rotations on $(\bm{L},\bm{Q}_{\mathrm{L,R}})$ and $\bm{\Delta}_{\mathrm{F}}$. With non-zero 
$v_\varphi$ and $v_{\mathrm{F}}$, we would in this case end up with Goldstone fields built up out of $\widetilde{\phi}^{1,2}$ and $\mathrm{Re} 
[ \tilde{\Delta}_{\mathrm{F}}^{4,5,6,7}]$ from the spontaneous breaking of $\SU{3}{F'}$ and $\SU{3}{F''}$, respectively. With $A_\mathrm{F} \neq 0$ the 
$\SU{3}{F'} \times \SU{3}{F''}$ symmetry softly breaks to the familiar $\SU{3}{F}$. This causes $\widetilde{\phi}^{1,2}$ and 
$\mathrm{Re}[ \tilde{\Delta}_{\mathrm{F}}^{4,5,6,7}]$ to arrange themselves into one pure Goldstone and one pseudo-Goldstone $\SU{2}{F}$ doublet 
(the mass of the latter is proportional to $A_\mathrm{F}$). Since $v_\varphi \ll v_{\mathrm{F}}$, the pure Goldstone is mostly 
$\mathrm{Re} [ \tilde{\Delta}_{\mathrm{F}}^{4,5,6,7}]$ (it has a small $\mathcal{O}(v_\varphi/v)$ admixture of $\widetilde{\phi}^{1,2}$, 
while the pseudo-Goldstone mode is mostly $\widetilde{\phi}^{1,2}$ containing an $\mathcal{O}(v_\varphi/v)$ amount of 
$\mathrm{Re} [ \tilde{\Delta}_{\mathrm{F}}^{4,5,6,7}] $). 

\subsection{Masses in presence of soft SUSY-breaking terms}\label{sec:massespres}

The inclusion of soft SUSY-breaking interactions results in non-zero masses for the fundamental scalars contained in 
the $\bm{L}$, $\bm{Q}_{\rm L}$ and $\bm{Q}_{\rm R}$ superfields as well as for the gauginos. By construction, the soft SUSY-breaking parameters 
are small in comparison to the GUT scale, i.e.~$m_{\rm soft} \ll v$, which means that the heavy states in the SUSY theory discussed in Sect.~\ref{sec:SUSYHS} 
will remain heavy and only those that were massless will receive contributions whose size is relevant for the low-energy EFT.

The masses of the fundamental scalars are purely generated in the soft SUSY-breaking sector. Furthermore, for a vacuum where only adjoint scalars 
acquire VEVs as in Eq.~\eqref{eq:GaugeAdjShift}, there is no mixing among the components of the fundamental scalars corresponding to the physical 
eigenstates at the first breaking stage shown in Fig.~\ref{fig:1112abc}. 

The Higgs-slepton masses (no summation over the indices is implied) read
\begin{eqnarray} \nonumber
m^2_{\SLLR{i}{l}{r}}  &=&  m^2_{\bm{27}} + 2 \Big[ A_{\rm G} v \(\T{}{8}\)^l_l \\ 
&+&  A_{\bar{\rm G}}\(\T{}{8}\)^r_r +  A_{\rm F} v_{\rm F} \(\T{}{8}\)^i_i\Big]\,,
\end{eqnarray}
while the corresponding squark masses are given by
\begin{align}
\begin{aligned}
m^2_{\SQL{i}{ }{l}} &= m^2_{\bm{27}} + 2 \[ A_{\rm G} v \(\T{}{8}\)^l_l +   A_{\rm F} v_{\rm F} \(\T{}{8}\)^i_i\]\,, \\
m^2_{\SQR{i}{r}{ }} &= m^2_{\bm{27}} + 2 \[A_{\bar{\rm G}}  v \(\T{}{8}\)^r_r +   A_{\rm F} v_{\rm F} \(\T{}{8}\)^i_i\]\,.
\end{aligned}
\end{align}
In Tab.~\ref{table:SpecFund} of Appendix~\ref{sec:masses} we show the masses for each fundamental scalar component in the LR-parity 
symmetric limit corresponding to $ A_{\rm G} = A_{\bar{\rm G}}$, for simplicity.

Moreover, the $\widetilde{\mathcal{H}}_{\rm F}$ mass is given by 
\begin{align}
m^2_{\widetilde{\mathcal{H}}_{\rm F}} \simeq 2m^2_{\bm{1}} + \mathcal{O}\( m^4_\mathrm{soft}/v^2_\mathrm{F}\)\,,
\end{align}
The exact expressions for scalar fields' squared masses can be found in Tab.~\ref{table:SpecAdj} of Appendix~\ref{sec:masses}.

The massless superpartners of the gauge bosons associated with the unbroken symmetries also acquire soft-scale masses. In particular, they mix with 
the chiral adjoint fermions via Dirac-terms whose strength, $M_0^{\prime}$ in Eq.~\eqref{eq:Lsoft}, is also of the order $m_{\rm soft}$. Typically, 
for minimal Dirac-gaugino models, the ad-hoc introduction of adjoint chiral superfields has the undesirable side effect of spoiling the gauge 
couplings' unification. However, in the model studied in Refs.~\cite{Benakli:2014cia,Benakli:2016ybe}, this problem is resolved by evoking trinification 
as the natural embedding for the required adjoint chiral scalars needed to form Dirac mass terms with gauginos. With this point in mind, we want 
to note that the SHUT model, with softly broken SUSY at the GUT scale, is on its own a Dirac-gaugino model and a possible high-scale framework 
for such a class of models.

The mass matrix for the adjoint fermions in the basis $\{\widetilde{\lambda}^{1,2,3}_{\rm L,R}, \Delta^{1,2,3}_{\rm L,R}, 
\widetilde{\lambda}^{8}_{\rm L,R}, \Delta^{8}_{\rm L,R}\}$ is then
\begin{align}
\label{eq:Mgauginos}
\renewcommand\arraystretch{1.3}
\mathcal{M}_{\Scale[0.6]{\widetilde{\lambda},\Delta}} = \mleft(
\begin{array}{ccccccc}
M_0 & & M_0^{\prime} & & 0 & & 0  \\
M_0^{\prime} & & \frac{v \lambda_{\bm{78}} }{\sqrt{6}} + \mu_{\bm{78}}  & & 0 & & 0 \\
0 & & 0 & & M_0 & & M_0^{\prime}  \\
0 & & 0 & & M_0^{\prime} & & \frac{v \lambda_{\bm{78}} }{\sqrt{6}} - \mu_{\bm{78}}
\end{array}
\mright)\;.\;
\end{align}
We denote the resulting mass eigenstates as $\{\mathcal{T}_{\mathrm{L,R}},\mathcal{T}^\perp_{\mathrm{L,R}}, \mathcal{S}_{\mathrm{L,R}}, \mathcal{S}^\perp_{\mathrm{L,R}}\}$ 
where $\mathcal{S}_{\mathrm{L,R}}$ and $\mathcal{T}_{\mathrm{L,R}}$ are the light (soft-scale) adjoint fermions while 
$\mathcal{S}^\perp_{\mathrm{L,R}}$ and $\mathcal{T}^\perp_{\mathrm{L,R}}$ denote the heavy (GUT scale) ones. Note that, due to a small mixing, 
both the low- and high-scale gauginos are essentially Majorana-like. Indeed, the mass of the former ones are approximately given by $M_0$, while the high-scale adjoint fermions $\mathcal{T}^\perp_{\mathrm{L,R}}$ 
and $\mathcal{S}^\perp_{\mathrm{L,R}}$ get their masses from $\mathcal{F}$-terms being approximately equal to 
$\big(\mathcal{M}_{\Scale[0.6]{\widetilde{\lambda},\widetilde{\Delta}}}\big)^2_2$ and 
$\big(\mathcal{M}_{\Scale[0.6]{\widetilde{\lambda},\widetilde{\Delta}}}\big)^4_4$, respectively. 

The same effect is observed for the gluinos $\tilde{g}^a$ whose masses, in the limit $M_0 \sim M_0^\prime \ll v \sim \mu_{78}$, are equal to $M_0$, for the light states, and $\mu_{\bm{78}}$, for the heavy states. There is also an $\SU{2}{F}$-doublet fermion $\mathcal{H}_{\rm F}$ that acquires a mass of the order of soft SUSY-breaking 
scale $m_{\rm soft}$. Note that $\mathcal{H}_{\rm F}$ as well as its superpartner $\widetilde{\mathcal{H}}_{\rm F}$ receive $\mathcal{D}$-term 
contributions if $\SU{3}{F}$ is gauged. Finally, the chiral fundamental fermions are massless at this stage.

\section{Particle masses at lower scales -- a qualitative analysis} 
\label{sec:masses2}

In this section we give a short overview of the low-energy limits of the SHUT model, i.e.~the spectrum after $\widetilde{\phi}^3$, $\widetilde{\phi}^2$ 
and $\widetilde{\nu}_\mathrm{R}^1$ acquire VEVs. 
In particular, we investigate whether the SM-extended symmetry, $G_{\rm SM} \times \U{T} \times \U{T'}$ as represented at the bottom 
of Fig.~\ref{fig:1112abc}, leaves enough freedom to realise the SM particle spectrum. Note that $\mathrm{SU}(2)$ (anti-)fundamental indices 
are denoted with lowercase letters for the remainder of the text, rather than with uppercase letters. 

\subsection{Colour-neutral fermions}

Once the $\SU{2}{R} \times \SU{2}{F}$ symmetries are broken, the tri-doublet $\widetilde{H}^{f\,l}_r$ and the bi-doublet $\widetilde{h}^l_r$ 
are split into three distinct generations of $\SU{2}{L}$ doublets. We will then rename them as $\widetilde{H}_{r=1}^{f\,l} \equiv \widetilde{H}_{\rm u}^{f\,l}$, 
$\widetilde{h}_{r=1}^{l} \equiv \widetilde{h}_{\rm u}^{l}$, $\widetilde{H}_{r=2}^{f\,l} \equiv \widetilde{H}_{\rm d}^{f\,l}$ and 
$\widetilde{h}_{r=2}^{l} \equiv \widetilde{h}_{\rm d}^{l}$, such that 
\begin{align}
\begin{aligned}
\widetilde{H}^{i}_{\rm u} &= \( 
\begin{array}{c}
       \widetilde{H}_\mathrm{u}^{i\,0}    \\
         \widetilde{H}_\mathrm{u}^{i\,+}   
\end{array}
\)
\qquad
\widetilde{H}^{i}_{\rm d} = \( 
\begin{array}{c}
       \widetilde{H}_\mathrm{d}^{i\,-}    \\
         \widetilde{H}_\mathrm{d}^{i\,0}   
\end{array}
\)
\qquad
E^{i}_{\rm L} = \( 
\begin{array}{c}
      e_\mathrm{L}^i    \\
         \nu_\mathrm{L}^i   
\end{array}
\) 
\\
\widetilde{h}_{\rm u} &= \( 
\begin{array}{c}
       \widetilde{H}_\mathrm{u}^{3\,0}    \\
         \widetilde{H}_\mathrm{u}^{3\,+}   
\end{array}
\)
\qquad
\widetilde{h}_{\rm d} = \( 
\begin{array}{c}
       \widetilde{H}_\mathrm{d}^{3\,-}    \\
         \widetilde{H}_\mathrm{d}^{3\,0}   
\end{array}
\)
\qquad
\mathcal{E}_{\rm L} = \( 
\begin{array}{c}
      e_\mathrm{L}^3    \\
         \nu_\mathrm{L}^3   
\end{array}
\)
\end{aligned}
\end{align}
where $i=1,2$, and where their scalar counterparts follow the same notation but without and with tildes, respectively. From this we can 
build mass terms for the charged lepton and charged Higgsinos as
\begin{eqnarray}
\label{eq:DiracM}
\mathcal{L}_{C} &=& 
\begin{pmatrix} e_\mathrm{L}^1 & e_\mathrm{L}^2 & e_\mathrm{L}^3 & \widetilde{H}_{\rm d}^{1\,-} & 
\widetilde{H}_{\rm d}^{2\,-} & \widetilde{H}_{\rm d}^{3\,-} \end{pmatrix} 
\mathcal{M}^C \\
&\times & \begin{pmatrix} e_\mathrm{R}^1 & e_\mathrm{R}^2 & e_\mathrm{R}^3 & \widetilde{H}_{\rm u}^{1\,+} & 
\widetilde{H}_{\rm u}^{2\,+} & \widetilde{H}_{\rm u}^{3\,+} \end{pmatrix}^{\mathrm{T}}
+{\rm c.c.}\,. \nonumber
\end{eqnarray}
Let us start by classifying all possible EW Higgs doublet and complex-singlet bosons, whose VEVs may have a role in the SM-like fermion 
mass spectrum. There are three types of Higgs doublets distinguished in terms of their $\U{Y} \times \U{T}$ charges and one possibility 
for complex singlets (and their complex conjugates). In particular, we can have
\begin{enumerate}
\item $( 1,\,1 )$: $H_{\rm u}^{2}$, $h_{\rm u}$, $H_{\rm d}^{*1}$, $\widetilde{E}_\mathrm{L}^{*2}$, $\widetilde{\mathcal{E}}_{\rm L}^*$, with VEVs denoted as $\bullet$-type.
\item $( 1,\,5 )$: $H_{\rm u}^{1}$, $H_{\rm d}^{* 2}$, $h_{\rm d}^{*}$, with VEVs denoted as $\star$-type.
\item $( 1,\,-3 )$: $\widetilde{E}_\mathrm{L}^{* 1}$, with VEVs denoted as $\ast$-type.
\item $( 0,\,-4 )$: $\widetilde{S}_{1,2}$, with VEVs denoted as $\diamond$-type.
\end{enumerate}
Note that the doublets in each line can mix, in particular, in the last line the two complex singlets emerge from the mixing $\big(\widetilde{\phi}^{*1},\, 
\widetilde{\nu}_\mathrm{R}^2,\, \widetilde{\nu}_\mathrm{R}^3\big) \mapsto 
\big( \widetilde{S}_{1}, \, \widetilde{S}_{2}, \, \mathcal{G}_s \big)$ induced by the third breaking step in Fig.~\ref{fig:1112abc}, with 
$\mathcal{G}_s$ being a complex Goldstone boson\footnote{The breaking $\SU{2}{R} \times \SU{2}{F} \times \U{L+R} \times 
\U{S} \to \U{Y} \times \U{T}$ gives rise to six Goldstone bosons, three gauge and three global ones, where the former are 
$\rm{Im}\[ \widetilde{\nu}_\mathrm{R}^1 \],\, \rm{Re}\[ \widetilde{e}_\mathrm{R}^1 \]$ and 
$\rm{Im}\[ \widetilde{e}_\mathrm{R}^1 \]$ while the latter ones are $\rm{Im}\[\widetilde{\phi}^2\]$ and $\mathcal{G}_s$.}. 

According to the quantum numbers shown in Tab.~\ref{Table:EFT-content-snu1Phi} of Appendix~\ref{sec:Symmetries}, 
the matrix $\mathcal{M}^C$ has the structure
\begin{align}
\label{eq:MLTp}
\mathcal{M}^{C} & \sim  \(
\begin{array}{ccc|ccc}
       0  \; & \; \star \; & \; \star \; & \; 0\; & \; 0\; &\; 0 \\
       \star  \;      &\; \bullet\; &\; \bullet\; &\; 0\; &\; 0\; &\; 0 \\
       \star \;      &\; \bullet\; &\; \bullet\; &\; 0\; &\; 0\; &\; 0 \\
\hline
       \star \;      &\; \bullet\; &\; \bullet\; &\; 0\; &\; 0\; &\; 0 \\
       \bullet \;   &\; 0\; &\; 0\; &\; 0\; &\; 0\; &\; 0 \\
      \bullet       & \; 0\;  & \; 0\;  & \; 0\;  & \; 0\;  & \; 0
\end{array}
\)\,,
\end{align}
where the symbols denote the type of VEVs contributing to the entry. In this case, the rank of the matrix $\mathcal{M}^{C}$ is at most three, 
which means that while we may be able to identify the correct patterns for the masses of the charged leptons in the SM, there will be massless 
charged Higgsinos remaining in the spectrum after EWSB, which is in conflict with phenomenology. The mass terms are forbidden by the $\U{T'}$ symmetry, 
which remains unbroken after EWSB, and the latter is independent on the number of Higgs doublets involved. 

In order to get a particle content consistent with the SM, one needs to break the $\U{W}$ symmetry, thus avoiding the remnant $\U{T'}$ symmetry. 
The most general $\U{W}$ violating terms after $\langle\Delta^8_\mathrm{L,R,F}\rangle$ (obeying all other symmetries) are 
\begin{align} 
\begin{aligned}
V^{\mathrm{\slashed{\Scale[0.5]{W}}}}_{\rm soft}= & 
\varepsilon_{f f'} \varepsilon_{l l'} \varepsilon^{r r'}\( A_{Hh\phi} H^{l\,f}_{r} h^{l'}_{r'} \widetilde{\phi}^{f'}
+ A_{hEE} h^{l}_{r} \widetilde{E}_{\rm L}^{f \,l'} \widetilde{E}_{{\rm R}\,r'}^{f'}
\right. \\
&
\left.
+ A_{hE\mathcal{E}} H^{f\,l}_{r} \widetilde{E}_{\rm L}^{f' l'} \widetilde{\mathcal{E}}_{{\rm R}\,r'}
+ \bar{A}_{hE\mathcal{E}} H^{f\,l}_{r} \widetilde{E}_{{\rm R}\,r'}^{f'} \widetilde{\mathcal{E}}^{l'}_{{\rm L}\,} \) 
 + \mathrm{c.c.} \,,
\label{eq:accbrk}
\end{aligned}
\end{align}
with $A_{ijk} \ll v$. The charged lepton mass matrix now reads
\begin{align}
\label{eq:MLT}
\mathcal{M}^{C} & \sim  \(
\begin{array}{ccc|ccc}
       0   \;    & \;\star\; &\; \star\; &\; 0\; &\; \diamond\; &\; \diamond \\
       \star  \;     &\; \bullet\; &\; \bullet\; &\; \diamond\; &\; \blacklozenge\; &\; \blacklozenge \\
      \star  \;     &\; \bullet\; &\; \bullet\; &\; \diamond\; &\; \blacklozenge\; &\; \blacklozenge \\
\hline
       \star  \;     &\; \bullet\; &\; \bullet\; &\; \diamond\; &\; \blacklozenge\; &\; \blacklozenge \\
       \bullet  \;     &\; \ast\; &\; \ast\; &\; \blacklozenge\; &\; \diamond\; &\; \diamond \\
       \bullet  \;     &\; \ast\; &\; \ast\; &\; \blacklozenge\; &\; \diamond\; &\; \diamond \\
\end{array}
\)\,,
\end{align}
where $\blacklozenge$ labels entries related to the $\widetilde{\nu}_{\rm{R}}^1$ VEV and 
can thus be well above the EW scale. We now have a mass matrix of rank-6 which means that no charged leptons and Higgsinos are left massless after EWSB. 
Note that before the EW symmetry is broken there are three massless lepton doublets, as the matrix in \eqref{eq:MLT} with only $\blacklozenge$-type entires has rank 3, 
in accordance with the SM. Furthermore, due to large $\blacklozenge$-type entries, the structure of $\mathcal{M}^{C}$ allows for three exotic lepton eigenstates heavier 
than the EW scale. Similarly, in the neutrino sector, no massless states remain after EWSB.

We see from the structure of Eq.~\eqref{eq:MLT} that, while the maximal amount of light $\SU{2}{L}$ Higgs doublets is nine, the minimal low-scale model needs at 
least two Higgs doublets, one of the $\star$-type and one of the $\bullet$-type, for the rank of the matrix to remain at 6. Note also that the low-scale remnant of 
the family symmetry, $\U{T}$, is non-universal in the space of fermion generations. As such, the various generations of Higgs bosons couple differently to different 
families of the SM-like fermions, offering a starting point for a mechanism explaining the mass and mixing hierarchies among the charged leptons. In addition, with 
the only tree-level interaction among fundamental multiplets arising from the high scale term $\bm{L}^i\bm{Q}_{\rm L}^j\bm{Q}_{\rm R}^k\epsilon_{ijk}$, the masses 
for all leptons must be generated at loop-level, providing a possible explanation for the lightness of the charged leptons observed in nature. 

To see this, we write the allowed lepton Yukawa terms (omitting the heavy vector-like lepton contributions)
\begin{eqnarray}
-\mathcal{L}_\mathrm{Y} = \Pi^a_{ij} \overline{\ell_{\mathrm{L}i}} H_a e_{\mathrm{R}j} + \mathrm{c.c.}.
\end{eqnarray}
Note that the equation above is written in terms of Dirac spinors rather than left-handed Weyl spinors (such that the charges for all right-handed spinors 
in Table VI should be conjugated). Also, to match conventional notation, the left-handed spinor $E_\mathrm{L}$ is here denoted as $\ell_\mathrm{L}$. 

For the case of the three Higgs doublets being $H_u^1$, $H_u^2$ and $H_d^2*$ (which is one of the possible scenarios enabling the Cabbibo mixing 
at tree-level, as shown in the following subsection), the charged lepton mass form reads 
\begin{eqnarray*}
\Scale[0.9]{M_e = \dfrac{1}{\sqrt{2}}\mleft(
	\begin{array}{ccc}
	0 & v_1 \Pi^1_{12} + v_d \Pi^3_{12} & v_1 \Pi^1_{13} + v_d \Pi^3_{13} \\
	v_1 \Pi^1_{21} + v_d \Pi^3_{21} & v_2 \Pi^2_{22} & v_2 \Pi^2_{23} \\
	v_1 \Pi^1_{31} + v_d \Pi^3_{31} & v_2 \Pi^2_{32} & v_2 \Pi^2_{33}
	\end{array}
	\mright) } \,,
\end{eqnarray*}
where $v_1,\; v_2$ and $v_d$ is the VEV of $H_u^1,\;H_u^2$ and $H_d^2*$, respectively. The Yukawa couplings $\Pi^a_{ij}$ are generated radiatively, 
by a higher-order sequential matching of the EFT to the high-scale SHUT theory at each of the breaking steps (tree-level matching yields $\Pi^a_{ij}=0$). 

With this form, and with $\Pi^a_{ij}$ as free parameters, there is enough freedom to reproduce the pattern of charged SM-like lepton masses. 
However, whether or not it can be derived in terms of the high-scale SHUT parameters remains to be seen after the RG evolution and the calculations of 
the radiative threshold corrections have been carried out.

Finally, consider the neutrino sector of the model composed of 15 neutral leptons emerging from the leptonic tri-triplet $\LLR{i}{l}{r}$ after the EWSB,
\begin{eqnarray*}
\Scale[0.85]{\Psi_N = \{\phi^1\, \phi^2\, \phi^3\, \nu^1_\mathrm{R}\, \nu^2_\mathrm{R}\, \nu^3_\mathrm{R}\, \nu^1_\mathrm{L}\, \nu^2_\mathrm{L}\, \nu^3_\mathrm{L}\, 
\widetilde{H}^{1\,0}_\mathrm{d}\, \widetilde{H}^{2\,0}_\mathrm{d}\, \widetilde{H}^{3\,0}_\mathrm{d}\, \widetilde{H}^{1\,0}_\mathrm{u}\, 
\widetilde{H}^{2\,0}_\mathrm{u}\, \widetilde{H}^{3\,0}_\mathrm{u}\} } \,.
\end{eqnarray*}
Note, in this first consideration we ignore the adjoint (chiral superfields $\DA{L,R,F}{a}$ and neutral gaugino $\widetilde{\lambda}^a_{\rm L,R}$) sectors 
for the sake of simplicity, while they should be included in a complete analysis of the neutrino sector involving the RG running and the radiative threshold 
corrections at every symmetry breaking scale. The corresponding 15$\times$15 mass form with all the Dirac and Majorana terms allowed after the EWSB
\begin{eqnarray}
\mathcal{L}_\mathrm{N} = \Psi_N \mathcal{M}^N \Psi_N^{\top}  \,,
\end{eqnarray}
has the following generic structure
\begin{align*}
\Scale[0.95]{\mathcal{M}^{N}  =
    \(
\begin{array}{cccccccccccccccccc}
0         & 0         & 0         & 0          & \otimes & \otimes   & 0        & 0         & 0          & 0         & \vee    & \vee     & 0         & \vee    & \vee     \\
0        & \times & \times & \times &         0 & 0          & 0         & \vee    & \vee     & \vee    & \vee    & \vee     & \vee    & \vee    & \vee     \\
0         & \times & \cup & \times &         0 & 0          & 0         & \vee    & \vee     & \vee    & \vee    & \vee     & \vee    & \vee    & \vee     \\
0         & \times & \times & \times &         0 & 0          & 0         & \vee    & \vee     & \vee    & \vee    & \vee     & \vee    & \vee    & \vee     \\
\otimes  & 0         & 0         & 0         &         0 & 0          & \vee    & \vee    & \vee     & \vee    & 0         & 0          & \vee    & 0         & 0         \\
\otimes     & 0         & 0         & 0         &         0 & 0          & \vee    & \vee    & \vee     & \vee    & 0         & 0          & \vee    & 0         & 0         \\
0         & 0         & 0         & 0         & \vee    & \vee    & 0          & 0         & 0          & 0         & 0         & 0          & 0         & 0         & 0         \\
0         & \vee    & \vee    & \vee    & \vee    & \vee    & 0          & 0         & 0          & 0         & 0         & 0          & 0         & \otimes &\otimes  \\
0         & \vee    & \vee    & \vee    & \vee    & \vee    & 0          & 0         & 0          & 0         & 0         & 0          & 0         &\otimes & \otimes  \\
0         & \vee    & \vee    & \vee    & \vee    & \vee    & 0          & 0         & 0          & 0         & 0         & 0          & 0         &\otimes  & \otimes  \\
\vee    & \vee    & \vee    & \vee    & 0         & 0         & 0          & 0         & 0          & 0         & 0         & 0          &\otimes  & 0         & 0         \\
\vee    & \vee    & \vee    & \vee    & 0         & 0         & 0          & 0         & 0          & 0         & 0         & 0          & \otimes & 0         & 0         \\
0         & \vee    & \vee    & \vee    & \vee    & \vee    & 0          & 0         & 0          & 0         & \otimes  &\otimes & 0         & 0         & 0         \\
\vee    & \vee    & \vee    & \vee    & 0         & 0         & 0          & \otimes  & \otimes   & \otimes & 0         & 0          & 0         & 0         & 0         \\
\vee    & \vee    & \vee    & \vee    & 0         & 0         & 0          & \otimes  & \otimes  & \otimes  & 0         & 0          & 0         & 0         & 0         \\
\end{array}
\) }
\end{align*}
where the symbol $\cup$ denotes the only Majorana bilinear below the $\langle\widetilde{\phi}^3\rangle$ scale, $\times$ the Majorana bilinears 
below the $\langle\widetilde{\phi}^2\rangle$ and $\langle\widetilde{\nu}_\mathrm{R}^1\rangle$ scales, $\otimes$ the Dirac bilinears below 
$\langle\widetilde{\phi}^2\rangle$ and $\langle\widetilde{\nu}_\mathrm{R}^1\rangle$ scales, and $\vee$ the Dirac bilinears at the lowest EWSB scale. 
For mass terms receiving contributions from more than one symmetry breaking scale, only the highest scale is displayed in the matrix above. 
Note that all bilinears with both fields having zero charge under all U(1) groups are referred to as Majorana bilinears, and not just combinations 
consisting of a field with itself. 

Despite of the absence of tree-level Yukawa interaction for the leptonic tri-triplet $\LLR{i}{l}{r}$ 
at the GUT scale, the Majorana mass terms in the upper-left 3$\times$3 block of the mass form are generated at tree-level 
at the intermediate matching $\langle\widetilde{\phi}^3\rangle$, $\langle\widetilde{\phi}^2\rangle$ and $\langle\widetilde{\nu}_\mathrm{R}^1\rangle$ scales due to interactions with gauginos, while all other Majorana and Dirac terms are generated radiatively, either at one- or two-loop level. 
With this structure, and with the hierarchy of scales presented in Sec.~\ref{sec:EFTs}, there are solutions with three sub-eV neutrino states. Two 
of these states are present for a wide range of parameter values, while a third light state, in the considered simplistic approach, typically requires a tuned suppression of one or more entries in the lower right 8$\times$8 block. Whether this can be obtained with less fine-tuning when including the full set of neutral states coming from the adjoint superfields, remains to be seen once the full RG evolution and matching has been carried out.

\subsection{Quark sector}\label{quarkquark}

In the absence of the accidental $\U{T'}$ symmetry, the low-energy limit of the SHUT model also offers good candidates for SM quarks without massless 
states after EWSB. To see this we first note that once $\widetilde{\phi}^3$ develops a VEV at the second SSB stage shown in Fig.~\ref{fig:1112abc}, two generations of $D$-quarks 
mix and acquire mass terms of the form $ m_D D_{\rm L}^f D_{\rm R}^{f'} \varepsilon_{f f'}$, with $m_D = \mathcal{O} \big(m_{\rm soft}\big) \gg M_{\rm EW}$. Then, 
at the third breaking stage, the $\widetilde{\nu}_{\rm R}^1$  and $\widetilde{\phi}^2$ VEVs trigger a mixing between the R-type quarks $D_{\rm R}^i$ and $d_{\rm R}^i$
\begin{align}
\begin{pmatrix}
\vspace{0.5mm}
{d}_{\rm R}^1
\\
\vspace{0.5mm}
{D}_{\rm R}^2
\\
\vspace{0.5mm}
{D}_{\rm R}^3
\\
\vspace{0.5mm}
{D}_{\rm R}^1
\\
\vspace{0.5mm}
{d}_{\rm R}^2
\\
\vspace{0.5mm}
{d}_{\rm R}^3
\\
\end{pmatrix}
=
\begin{pmatrix}
\vspace{0.5mm}
1 &0 & 0 & 0 & 0 & 0
\\
\vspace{0.5mm}
0 & a_1 & a_2 & 0 & 0 & 0
\\
\vspace{0.5mm}
0 & a_3 & a_4 & 0 & 0 & 0
\\
\vspace{0.5mm}
0 & 0 & 0 & a_5 & a_6 & a_7
\\
\vspace{0.5mm}
0 & 0 & 0 & a_8 & a_9 & a_{10} 
\\
\vspace{0.5mm}
0  & 0 & 0 & a_{11} & a_{12} & a_{13}
\end{pmatrix}
\begin{pmatrix}
\vspace{0.5mm}
\mathpzc{d}_{\rm R}^1
\\
\vspace{0.5mm}
\mathpzc{d}_{\rm R}^2
\\
\vspace{0.5mm}
\mathpzc{D}_{\rm R}^1
\\
\vspace{0.5mm}
\mathpzc{d}_{\rm R}^3
\\
\vspace{0.5mm}
\mathpzc{D}_{\rm R}^2
\\
\vspace{0.5mm}
\mathpzc{D}_{\rm R}^3
\\
\end{pmatrix},
\label{eq:changeOfBasis}
\end{align}
where the parameters $a_{1}$ through $a_{13}$ are not all independent as the matrix is unitary.
At the classical level, and with $\langle\widetilde{\phi}^3\rangle=\langle\widetilde{\phi}^2\rangle=\langle\widetilde{\nu}_{\rm R}^1\rangle$, the parameters are given by
\begin{eqnarray*}
&& a_{1,3,4,12}=-a_{2,9}=\frac{1}{\sqrt{2}},\;\;a_{5,8,11}=\frac{1}{\sqrt{3}}\,, \\
&& a_6=0,\;\;a_7=-\sqrt{\frac{2}{3}},\;\;a_{10,13}=\frac{1}{\sqrt{6}},
\end{eqnarray*}
while the corresponding expressions for general $\langle\widetilde{\phi}^3\rangle$, $\langle\widetilde{\phi}^2\rangle$, $\langle\widetilde{\nu}_{\rm R}^1\rangle$ are too extensive to be presented here.

Defining the components of the $\SU{2}{L}$ quark doublets as $Q_\mathrm{L}^{1,2} \equiv \big(u_\mathrm{L}^{1,2}, d_\mathrm{L}^{1,2} \big)^{\mathrm{T}}$ 
and $q_\mathrm{L} \equiv \big(u_\mathrm{L}^3, d_\mathrm{L}^3 \big)^{\mathrm{T}}$, we can construct the Lagrangian for the SM-like quarks as
\begin{eqnarray} \nonumber
\mathcal{L}_{\rm quarks} &=& 
\begin{pmatrix} {u_{\rm L}^1} & {u_{\rm L}^2} & {u_{\rm L}^3} \end{pmatrix} 
\mathcal{M}^{\rm u} 
\begin{pmatrix} u_{\rm R}^1 \\\\ u_{\rm R}^2 \\\\ u_{\rm R}^3 \end{pmatrix} \\
&+&
\begin{pmatrix} {d_{\rm L}^1} & {d_{\rm L}^2} & {d_{\rm L}^3} \end{pmatrix} 
\mathcal{M}^{\rm d}
\begin{pmatrix} \mathpzc{d}_{\rm R}^1 \\\\ \mathpzc{d}_{\rm R}^2 \\\\ \mathpzc{d}_{\rm R}^3 \end{pmatrix}
+{\rm c.c.}
\label{eq:quarks}
\end{eqnarray}
With the different possibilities found for the Higgs sector, the most generic structure for $\mathcal{M}^{\rm u}$ and $\mathcal{M}^{\rm d}$ matrices 
obey the following patterns:
\begin{equation}
\mathcal{M}^{\rm u}  \sim 
\begin{pmatrix}
       \ast  \;     &\; \bullet\; &\; \bullet \\
\bullet\; &\; \star\; &\; \star \\
\bullet\; &\; \star\; &\; \star \\
\end{pmatrix}
, \;\;\;\;
\mathcal{M}^{\rm d}  \sim
\begin{pmatrix}
	   0\; 	   &\; \bullet\; &\; \star  \\
\star \; & \;     \ast \;		  &\; \bullet  \\
\star\; & \;    \ast	\;      &\; \bullet
\end{pmatrix}\label{eq:Md}
\,.
\end{equation}
In order for all quarks to gain a mass after EWSB, the matrices in Eq.~\eqref{eq:Md} must be of rank-3. As such, the low-scale limit of the SHUT model requires, at least, two Higgs doublets, where both $\bullet$-type and $\star$-type ones are present. In contrast to charged leptons, for which the contributions arise solely from effective Yukawa couplings, in Eq.~\eqref{eq:Md} there are allowed tree-level bilinears for the SM-like quarks.

Next, let us consider the possible flavour structure in the low-scale limit. At the classical level, we have Cabbibo mixing with a minimum of three Higgs doublets. For a realistic mass spectrum, it is also required to incorporate RG effects as well as loop-induced threshold corrections, which make the Yukawa couplings different from each other. Take for example the 3HDM with two up-type Higgs doublets, $H_{\rm u}^{1}$ and $H_{\rm u}^{2}$ and a down type Higgs doublet $H_\mathrm{d}^2$. In the classical limit of the theory, this corresponds to
\begin{eqnarray} \nonumber
\mathcal{M}^{\rm u}  &= \dfrac{\lambda_{27}}{\sqrt{2}}\(
\begin{array}{ccc}
0                   & 0                                & -v_2 \\
0                   & 0                                & v_1 \\
v_2 & -v_1 & 0 \\
\end{array}
\) \,,\,  \\
\mathcal{M}^{\rm d}  &= \dfrac{\lambda_{27}}{\sqrt{2}}\(
\begin{array}{ccc}
	   0\; 	   &\; 0\; &\; -\frac{1}{\sqrt{3}}v_d  \\
0 \; & \;     0 \;		  &\; 0  \\
 v_d\; & \;    0	\;      &\; 0
\end{array}\label{eq:MuMd}
\)\,,
\end{eqnarray}

where $v_{1,2,d}$ are the corresponding Higgs VEVs and where $\lambda_{27}$ is the high-scale Yukawa coupling. With this, the Cabbibo angle satisfies $\tan \theta_C=\frac{v_1}{v_2}$ and results in the quark mass spectrum
\begin{align}
\begin{split}
\label{eq:degenerateQuarkMasses}
&m^2_{\rm{c,t}}=\frac{1}{2}\lambda_{27}^2({v_1}^2 + {v_2}^2),\\ 
&m^2_{\rm{b}}=3m^2_{\rm{s}}=\frac{1}{2}\lambda_{27}^2v_{\rm{d}}^2,\;\;\;\; m^2_{\rm{u,d}}=0,
\end{split}
\end{align}
i.e.~the lowest order contributions to the particle spectrum imply a degeneracy of charm and top quark masses, while strange and bottom
quark masses squared are related with a factor three. 
\begin{figure}[!h]
\includegraphics[scale=1]{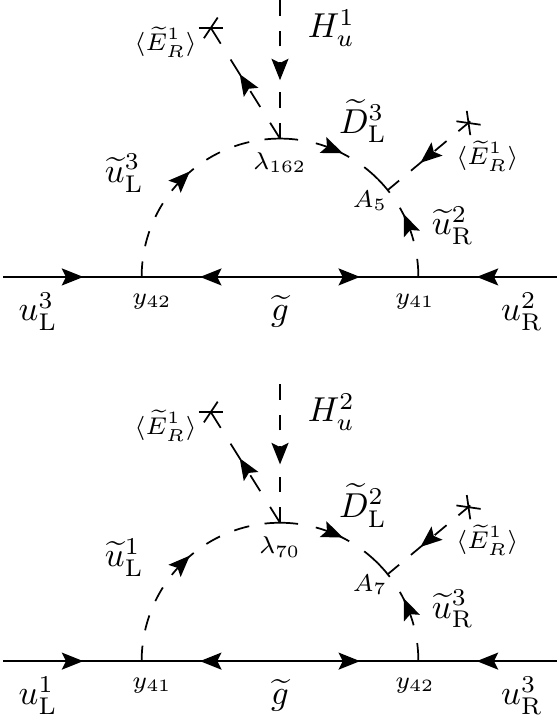}
\caption{Diagrams contributing to the one-loop matching conditions for Yukawa interactions with the upper diagram representing the dominant contributing to the top quark mass and the lower one a correction to the charm mass.}
\label{fig:loop-Yuk}
\end{figure}	

When radiative corrections are considered as well, the mass forms become more involved. 
Indeed, for an effective quark Yukawa Lagrangian the allowed terms (omitting, for simplicity, the heavy vector-like quark Yukawa terms)
	\begin{align*}
	-\mathcal{L}^q_\mathrm{Y} = \Gamma^a_{ij} \overline{q_{\mathrm{L}i}} H_a d_{\mathrm{R}j}
		+ \Delta^a_{ij} \overline{q_{\mathrm{L}i}} \widetilde{H}_a u_{\mathrm{R}j} + \mathrm{c.c.}
	\end{align*}
written again in terms of Dirac fermions rather than left-handed Weyl fermions, and where the tilde on the Higgs doublet refers to $\widetilde{H}^l=\varepsilon^{ll'}H^*_{l'}$ and not it being a Higgsino, as in the other parts of the paper. 
With the three Higgs doublets again being $H_u^{1,2}$ and $H_d^{2*}$, we have the mass forms
\begin{eqnarray} \nonumber
		&&\mathcal{M}^{\rm u} \approx \dfrac{1}{\sqrt{2}}\mleft(
		\begin{array}{ccc}
		0 & v_2 \Delta^2_{12} & v_2 \Delta^2_{13} \\
		v_2 \Delta^2_{21} & v_1 \Delta^1_{22} + v_d \Delta^3_{22} & v_1 \Delta^1_{23} + v_d \Delta^3_{23} \\
		v_2 \Delta^2_{31} & v_1 \Delta^1_{32} + v_d \Delta^3_{32} & v_1 \Delta^1_{33} + v_d \Delta^3_{33}
		\end{array}
		\mright) , \\
		&&\mathcal{M}^{\rm d} \approx \dfrac{1}{\sqrt{2}}\mleft(
		\begin{array}{ccc}
		0 & v_2 \Gamma^2_{12} & v_1 \Gamma^1_{13} + v_d \Gamma^3_{13} \\
		v_1 \Gamma^1_{21} + v_d \Gamma^3_{21} & 0 & v_2 \Gamma^2_{23} \\
		v_1 \Gamma^1_{31} + v_d \Gamma^3_{31} & 0 & v_2 \Gamma^2_{33}
		\end{array}
		\mright) .  \label{eq:MuMd-full}
\end{eqnarray}
where the zeros are put in as a good approximation since the corresponding Yukawa terms come from
higher-loop contributions which are generated only at the $\U{T}$ breaking scale. 

Let us estimate whether the radiative corrections can be sufficiently large to correct for the degeneracy in Eq.~\eqref{eq:degenerateQuarkMasses}. As a demonstration, we will consider the largest mass discrepancy, namely the degeneracy between the top and charm mass whose tree level value is proportional to $\lambda_{\bm{27}}$. The key idea here is that $\lambda_{\bm{27}} \sim \mathcal{O}\(10^{-2}\)$ which readily generates a viable charm mass but leaves the top quark two orders of magnitude lighter than its measured value. To lift such a degeneracy, one needs an order $\mathcal{O}\(1\)$ correction to $\Delta_{32}^1$ while leaving $\Delta_{13}^2 \lesssim \mathcal{O}\(10^{-2}\)$. To have an estimate for these radiative corrections, we can start with an instance of the mass forms ${\mathcal{M}}^{\mathrm{u},\mathrm{d}}$, with textures as in Eq.~\eqref{eq:MuMd-full}, that reproduce measured quark masses and mixing angles \cite{Bora:2012tx,Tanabashi:2018oca}, e.g.
\begingroup
\small
\begin{equation}
\begin{aligned}
&\mathcal{M}^{\rm u} = \mleft(
\begin{array}{ccc}
0 & -7.287 & 0.636 \\
-0.0013 & -0.159 - i 0.521 & -0.0016 - i 0.005 \\
0.124 & -171.944 & 0.00011
\end{array}
\mright) \mathrm{GeV}
\\
&\;\;\;\;\;\;\;\;\;\;\;\;\;\;\;\;\;\;\;\;\;\;\;\;\;\;\;\;\;\;\;\;\;\;\;\;\;\;\;\;\;\;\;\;\;\;\;\;\;\;\;\;\;\;\;\; \;\;\;\;\;\;\;\;\;\;\;\;\;\;\;\;\;\;\;\;\;\;\;
\\
&\mathcal{M}^{\rm d} = \mleft(
\begin{array}{ccc}
0 & -0.013 & 0.055 \\
-0.0006 & 0 & 0.013 \\
2.814 & 0 & 0.188
\end{array}
\mright) \mathrm{GeV} \,.
\end{aligned}\label{massNum}
\end{equation}
\endgroup

Keeping in mind that $v_1^2 + v_2^2 + v_d^2=\left(246\;\mathrm{GeV}\right)^2$, we then get an idea of what the values for $\Delta_{32}^1$ and $\Delta_{13}^2$ need to be. In particular, we see that the magnitude of $\Delta_{32}^1$ has to be larger than 0.7. 

The one-loop dominating contributions\footnote{Which diagrams that dominate depends on the specific parameter point and the details of the RG evolution. However, the gauge coupling for $\mathrm{SU}(3)_{\mathrm{C}}$ is larger than any other gauge coupling in the model at all scales, and as such the diagram with the gluino propagator dominates over diagrams with other gauginos, unless the gluino would be significantly heavier.} for the Yukawa couplings $\Delta_{32}^1$ and $\Delta_{13}^2$ are illustrated in Fig.~\ref{fig:loop-Yuk}. When a propagator in the loop becomes heavier than the renormalization scale, thus integrated out, we generate a threshold correction. For illustration purposes we will choose this scale to be either the gluino or the squark mass.

At this scale, the squark (gluino) propagators are resummed such that the masses are given by their $\overline{\mathrm{MS}}$ values at the gluino (squark) mass scale, which should be some function of quartic couplings, soft parameters and VEVs. We also have that $y_{41,42}$ are approximately equal to $\sqrt{2}g_\mathrm{S}$, with $\alpha_\mathrm{S} \sim 0.03$ at the $\langle\widetilde{\phi}^3\rangle$ scale, such that the two diagrams only differ when it comes to one of the couplings, and possibly by a mass difference for the squarks in the loop. The analytic expression for both diagrams, in the zero external momentum limit, is given by
\begin{align}
\begin{split}
\frac{ 2i\alpha_{\mathrm{S}} Z m_3}{ 3\pi\left(m_2^2-m_3^2\right) }\left( \frac{m_3^2\log\left(\frac{m_1^2}{m_3^2}\right)}{m_3^2-m_1^2} - 
\frac{m_2^2\log\left(\frac{m_1^2}{m_2^2}\right)}{m_2^2-m_1^2} \right) \,,
\end{split}\label{loop}
\end{align}
with 
\begin{equation}
Z=\lambda_{162}\langle\widetilde{\nu}_\mathrm{R}^1\rangle, \;\;\{m_1,m_2,m_3\}=\{m_{\widetilde{u}_{\mathrm{L}}^3},m_{\widetilde{u}_{\mathrm{R}}^2\widetilde{D}_{\mathrm{L}}^3},m_{\widetilde{g}}\}
\label{eq:pars-top}
\end{equation}
for the top diagram in Fig.~\ref{fig:loop-Yuk}, and with
\begin{equation}
Z=\lambda_{70}\langle\widetilde{\nu}_\mathrm{R}^1\rangle, \;\;\{m_1,m_2,m_3\}=\{m_{\widetilde{u}_{\mathrm{L}}^1},m_{\widetilde{u}_{\mathrm{R}}^3\widetilde{D}_{\mathrm{L}}^1},m_{\widetilde{g}}\}
\label{eq:pars-charm}
\end{equation}
for the bottom diagram. Note that the result is finite also in the limit of degenerate masses and has the form
\begin{equation}
\dfrac{i \alpha_S Z}{3 \pi m_{\tilde{g}} }.
\label{loopDeg}
\end{equation} 
In what follows we will consider the case where the intermediate symmetries are simultaneously broken by the VEVs 
\begin{equation}
\langle\widetilde{\phi}^3\rangle\sim\langle\widetilde{\phi}^2\rangle\sim\langle\widetilde{\nu}_\mathrm{R}^1\rangle\sim 8.8\cdot 10^{10}\; \mathrm{GeV},
\end{equation}
consistent with Sec.~\ref{sec:EFTs}, and with couplings as specified in Appendix~\ref{sec:FullEffectiveLs}. 

The magnitude of the dominant contributions to the top and charm Yukawa couplings are shown for a selection of gluino and squark masses in tabs.~\ref{tab:Vals} and \ref{tab:Vals2} respectively. Here we have, for example, a scenario with squark masses at the TeV scale, offering an interesting phenomenological probe to be studied in the context of LHC searches, or alternatively a scenario where both the gluino and squark masses in the top diagram are closely degenerate. Interestingly enough, we see that radiative corrections to the charm quark are sub-leading if at least one squark propagator is heavy enough and close to the $\langle \widetilde{\phi}^3 \rangle$ scale. With the examples provided we see that a hierarchy in the squark sector is reflected as a hierarchy in the radiative Yukawa couplings, necessary for the phenomenological viability of the model.
\begin{table}[htb!]
	\begin{center}
		\begin{tabular}{c|cccc}
			\toprule                     
			$\Delta^1_{32}$ \;&\; $\lambda_{162}$ \;&\; $m_{\tilde{g}}$ \;&\; $m_{\tilde{u}_\mathrm{L}^3}$ \;&\; $m_{\tilde{u}_\mathrm{R}^2,\,D_\mathrm{L}^3}$  	\\    
			\midrule
			$1$  						\;&\; $10^{-2}$	\;&\; $10^8$	\;&\; $10^3$ \;&\; $10^3$		\\
			$1$ 	 					\;&\; $10^{-2}$	\;&\; $10^6$ 	\;&\; $10^6$ \;&\; $ 10^6$	\\
			\bottomrule
		\end{tabular} 
		\caption{\it{Order of magnitude of the radiative correction to the top-quark Yukawa coupling (first column) and of the parameters contributing to the one-loop function \eqref{loop} (second to fifth columns). Masses are expressed in $\mathrm{GeV}$}}
		\label{tab:Vals}  
	\end{center}
\end{table}
\begin{table}[htb!]
	\begin{center}
		\begin{tabular}{c|cccc}
			\toprule                     
			$\Delta^2_{13}$ \;&\; $\lambda_{70}$ \;&\; $m_{\tilde{g}}$ \;&\; $m_{\tilde{u}_\mathrm{L}^1}$ \;&\; $m_{\tilde{u}_\mathrm{R}^3,\,D_\mathrm{L}^2}$  	\\    
			\midrule
			$10^{-5}$  						\;&\; $10^{-2}$	\;&\; $10^8$	\;&\; $10^{10}$ \;&\; $10^3$		\\
			$10^{-6}$ 	 					\;&\; $10^{-2}$	\;&\; $10^6$ 	\;&\; $10^{10}$ \;&\; $ 10^6$	\\
			\bottomrule
		\end{tabular} 
		\caption{\it{Order of magnitude of the radiative correction to the charm-quark Yukawa coupling (first column) and of the parameters contributing to the one-loop function \eqref{loop} (second to fifth columns). Masses are expressed in $\mathrm{GeV}$}}
		\label{tab:Vals2}  
	\end{center}
\end{table}
Note that for the degenerate scenario $\Delta^1_{32} = 2.8\times 10^8~\mathrm{GeV} \(\lambda_{162}/m_{\tilde{g}}\)$, which means that a viable correction to the top quark mass requires the ratio $\lambda_{162}/m_{\tilde{g}} \sim \mathcal{O}\(10^{-8}~\mathrm{GeV^{-1}}\)$. This means that, depending on the details of the renormalization procedure that may enhance or suppress the quartic coupling $\lambda_{162}$, an appropriate choice of the free gluino mass parameter will in principle make it possible to naturally lift the top-charm mass degeneracy in the right direction.

The required parameter values for compatible couplings at the EW scale remains unknown until the full RG evolution and sequential matching of all couplings in the model has been carried out, which is a subject of a further much more involved and dedicated study. What we can say at this point is that there do exist parameter space points with a potential of reproducing the correct hierarchy between the top and charm masses. 
%
%
\section{Estimating the scales of the theory} 
\label{sec:EFTs}

In this section we estimate the symmetry breaking scales of the model, i.e. the GUT scale $\langle\widetilde{\Delta}^8_\mathrm{L,R,F}\rangle\sim v$, and the intermediate scales $\langle\widetilde{\phi}^3\rangle$, $\langle\widetilde{\phi}^2\rangle$ and $\langle\widetilde{\nu}_\mathrm{R}^1\rangle$, by forcing the unified gauge coupling at the GUT scale to evolve such that it reproduces the measured values of the $\mathrm{SU}(3)_\mathrm{C} \times \mathrm{SU}(2)_\mathrm{L} \times \mathrm{U}(1)_\mathrm{Y}$ gauge couplings at the EW scale. This is done through a matching and running procedure, where the gauge couplings are matched at tree-level accuracy and evolved with one-loop RGEs, as a first step before matching at one-loop in future work. At each breaking scale, fermions obtaining a mass from the associated VEV are integrated out, giving rise to four intermediate energy ranges of RG evolution with different $\beta$-functions. We will refer to these regions as
%
%
\begin{equation}
\begin{aligned}
&\mathrm{Region\;\RN{1}:}\;\mu\in\[\langle\widetilde{\phi}^3\rangle,v\], \\
&\mathrm{Region\;\RN{2}:}\;\mu\in\[\langle\widetilde{\phi}^2\rangle,\langle\widetilde{\phi}^3\rangle\], 
\\
&\mathrm{Region\;\RN{3}:}\;\mu\in\[\langle\widetilde{\nu}_\mathrm{R}^1\rangle,\langle\widetilde{\phi}^2\rangle\], 
\\
&\mathrm{Region\;\RN{4}:}\;\mu\in\[m_{Z},\langle\widetilde{\nu}_\mathrm{R}^1\rangle\].
\end{aligned}
\end{equation}
The symmetry alone does not dictate the structure of the scalar mass spectrum, and we will therefore have to make assumptions about what scalars are to be integrated out at each matching scale. However, by studying the extreme cases we will show that the soft SUSY-breaking scale (which we associate with the scale of the largest tri-triplet VEV, $\langle\widetilde{\phi}^3\rangle$) is bounded from below by roughly $10^{11}$ GeV, independently of the scalar content.   

With the $\beta$-functions and matching conditions presented in Appendix~\ref{sec:betabeta}, we may set up a system of equations with three known values, the SM couplings at the $Z$-mass scale, and five unknown quantities, $\alpha^{-1}_g(v)$, $\log(\langle\widetilde{\phi}^3\rangle/v)$, $\log(m_Z/\langle\widetilde{\phi}^2\rangle)$, $\log(\langle\widetilde{\phi}^2\rangle/\langle\widetilde{\phi}^3\rangle)$  and $\log(\langle\widetilde{\nu}_\mathrm{R}^1\rangle/\langle\widetilde{\phi}^2\rangle)$: 
%
%
%
\begin{align}
\begin{split}
\label{eq:evolutionequations}
{\alpha}^{-1}_{g_\mathrm{C}}(m_Z)&={\alpha}^{-1}_{g}(v) - \frac{b^{\RN{2}}_{g_\mathrm{C}}}{2\pi} \log\left(\frac{\langle\widetilde{\phi}^2\rangle}{\langle\widetilde{\phi}^3\rangle}\right) \\
& - \frac{b^{\RN{3}}_{g_\mathrm{C}}}{2\pi} \log\left(\frac{\langle\widetilde{\nu}_\mathrm{R}^1\rangle}{\langle\widetilde{\phi}^2\rangle}\right)
 - \frac{b^{\RN{4}}_{g_\mathrm{C}}}{2\pi} \log\left(\frac{m_{{Z}}}{\langle\widetilde{\nu}_\mathrm{R}^1\rangle}\right),
\end{split}
\end{align}
%
%
\begin{align}
\begin{split}
\label{eq:evolutionequations2}
{\alpha}^{-1}_{g_\mathrm{L}}(m_{{Z}})&={\alpha}^{-1}_{g}(v) - \frac{b^{\RN{1}}_{g_\mathrm{L,R}}}{2\pi}\log\left(\frac{\langle\widetilde{\phi}^3\rangle}{v}\right) 
\\
& - \frac{b^{\RN{2}}_{g_\mathrm{L}}}{2\pi}\log\left(\frac{\langle\widetilde{\phi}^2\rangle}{\langle\widetilde{\phi}^3\rangle}\right) - \frac{b^{\RN{3}}_{g_\mathrm{L}}}{2\pi}\log\left(\frac{\langle\widetilde{\nu}_\mathrm{R}^1\rangle}{\langle\widetilde{\phi}^2\rangle}\right) 
\\
& - \frac{b^{\RN{4}}_{g_\mathrm{L}}}{2\pi}\log\left(\frac{m_{{Z}}}{\langle\widetilde{\nu}_\mathrm{R}^1\rangle}\right),
\end{split}
\end{align}
%
%
\begin{align}
\begin{split}
\label{eq:evolutionequations3}
{\alpha}^{-1}_{\widetilde{g}_\mathrm{Y}}(m_{{Z}}) &= \frac{5}{3}{\alpha}^{-1}_{g}(v) + \frac{b^{\RN{4}}_{\widetilde{g}_\mathrm{Y}}}{2\pi}\log\left(\frac{\langle\widetilde{\nu}_\mathrm{R}^1\rangle}{m_{{Z}}}\right) 
\\
&-\frac{1}{2\pi} \log\left(\frac{\langle\widetilde{\phi}^3\rangle}{v}\right)\left[{b^{\RN{1}}_{g_\mathrm{L,R}}} + \frac{2}{3}{b^{\RN{1}}_{\widetilde{g}_\mathrm{L,R}}}\right]   
\\
&- \frac{1}{2\pi}\log\left(\frac{\langle\widetilde{\phi}^2\rangle}{\langle\widetilde{\phi}^3\rangle} \right) \left[{b^{\RN{2}}_{g_\mathrm{R}}} + \frac{1}{3}{b^{\RN{2}}_{\widetilde{g}_\mathrm{L+R}}}\right]
\\
&
- \frac{1}{2\pi}\log\left(\frac{\langle\widetilde{\nu}_\mathrm{R}^1\rangle}{\langle\widetilde{\phi}^2\rangle} \right) \left[{b^{\RN{3}}_{g_\mathrm{R}}} + \frac{1}{3}{b^{\RN{3}}_{\widetilde{g}_\mathrm{L+R}}}\right],
\end{split}
\end{align}
%
%
%
with the following known parameters at the $m_Z$ scale ($\sim 91.2$ GeV) \cite{Patrignani:2016xqp} 
\begin{align}
\begin{split}
\label{betti2}
\alpha_{g_\mathrm{C}}^{-1}(m_{{Z}})&\sim 8.5, 
\\
\alpha_{g_\mathrm{L}}^{-1}(m_{{Z}})&=\sin^2(\theta_W)\cdot 128\sim 29.6, 
\\
\alpha_{\widetilde{g}_\mathrm{Y}}^{-1}(m_{{Z}})&=\cos^2(\theta_W)\cdot 128\sim 98.4. 
\end{split}
\end{align}
As we have more than three unknowns, the scales cannot be solved for uniquely, but are functions of $\log(\langle\widetilde{\phi}^2\rangle/\langle\widetilde{\phi}^3\rangle)$ and $\log(\langle\widetilde{\nu}_\mathrm{R}^1\rangle/\langle\widetilde{\phi}^2\rangle)$. If we take, for example, the scenario of having no hierarchies between these three scales,
\begin{equation}
\label{scenario11}
\langle\widetilde{\phi}^3\rangle \sim \langle\widetilde{\phi}^2\rangle \sim \langle\widetilde{\nu}_\mathrm{R}^1\rangle \sim m_{\mathrm{soft}},
\end{equation}
we end up with the following values
\begin{align}
\begin{split}
\label{eq:finalscales}
m_{\mathrm{soft}} & \sim 8.8\cdot 10^{10}\;\mathrm{GeV}, 
\\
v &  \sim 4.9\cdot 10^{17}\;\mathrm{GeV}, 
\\
\alpha^{-1}_{g}(v)& \sim  31.5, 
\end{split}
\end{align}
where hence the unified gauge coupling satisfies the perturbativity constraint, the GUT scale is below $\mathrm{M}_\mathrm{Planck}$ and the soft scale is well separated from both the GUT scale and the EW scale. Note that while the hierarchy between the GUT scale and the soft SUSY-breaking scale is stable with respect to radiative corrections, the hierarchy between the EW scale and the soft SUSY-breaking scale needs to be finely tuned.

\pagebreak
\onecolumngrid

\begin{figure}[!h]
\begin{minipage}{0.42\textwidth}
 \centerline{\includegraphics[width=1.05\linewidth]{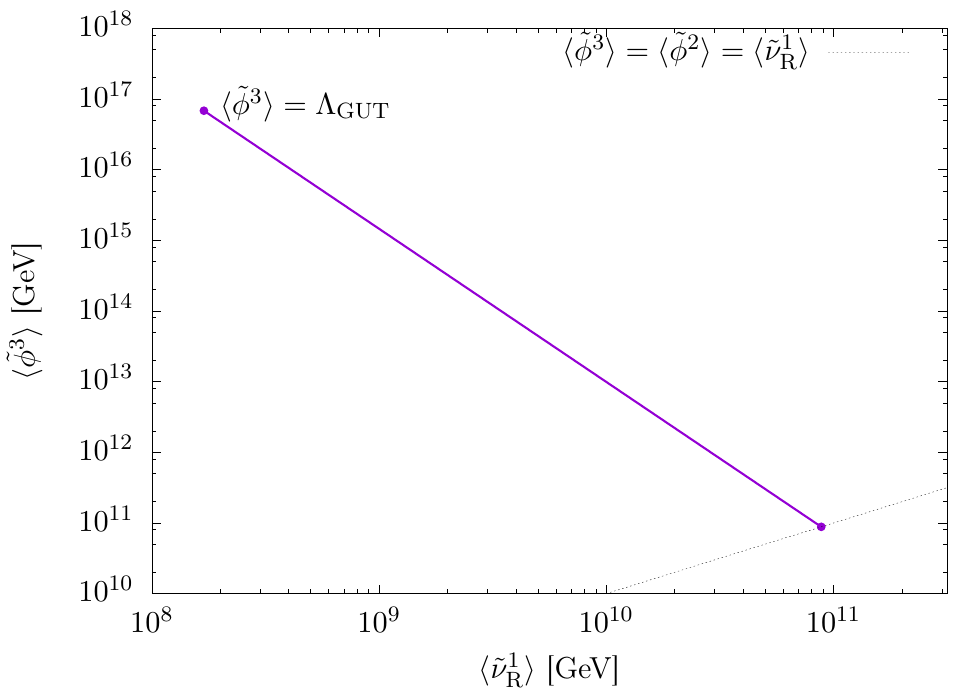}}
\end{minipage}
\hspace{0.8cm}
\begin{minipage}{0.42\textwidth}
 \centerline{\includegraphics[width=1.2\linewidth]{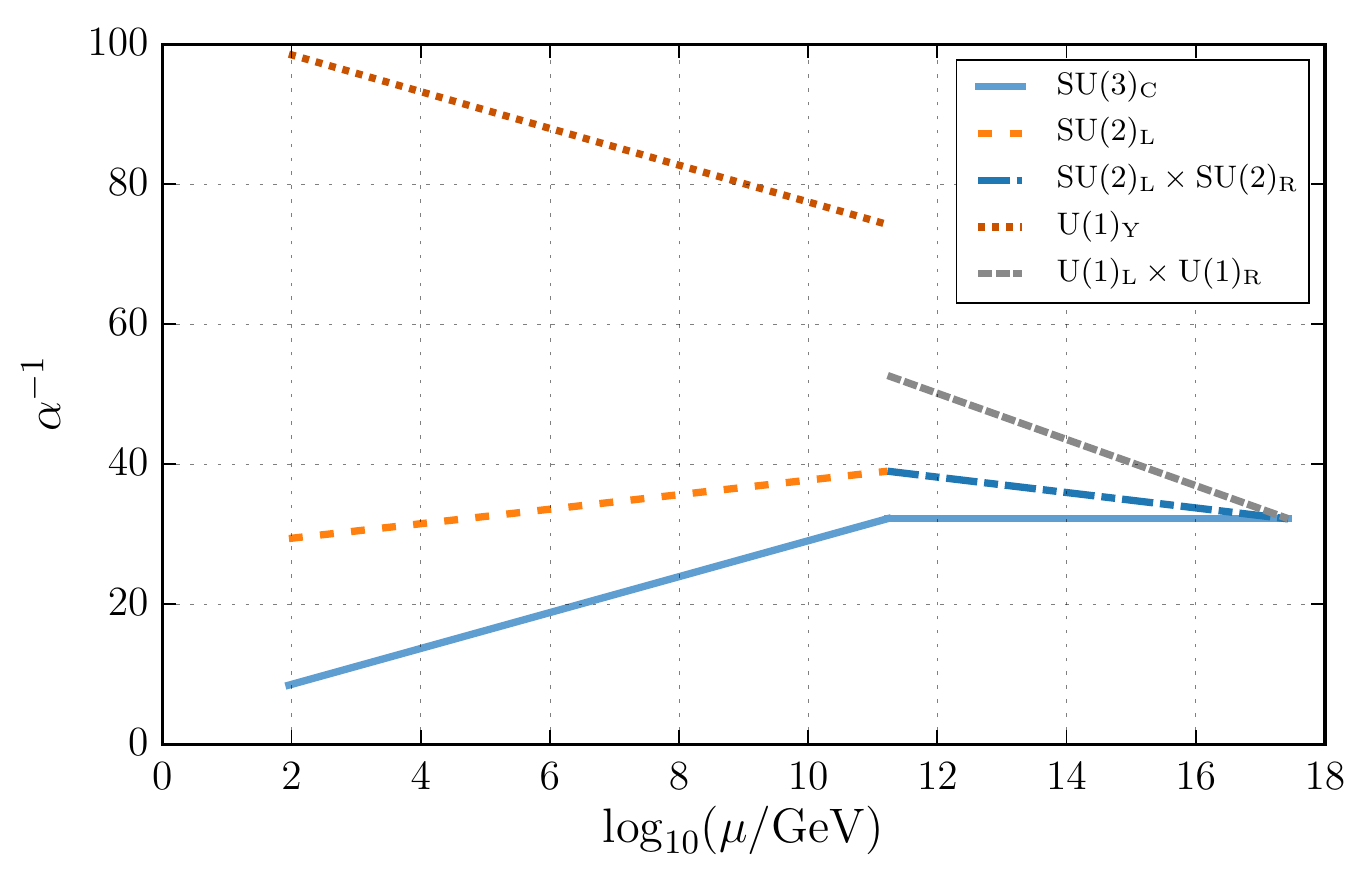}}
\end{minipage}
\caption{
{\bf Left panel:} \emph{Figure showing that the $\langle\widetilde{\phi}^3\rangle$ scale is minimised when there is no hierarchy between the soft scales, i.e.~where both lines meet in the lower right corner. The purple (solid) line corresponds to VEVs for which the gauge couplings run down to the measured standard model values.
The grey (dashed) line corresponds to the case of no hierarchy between the VEVs. Hence, the optimal choice corresponds to $\langle\widetilde{\phi}^3\rangle\sim \langle\widetilde{\phi}^2\rangle\sim \langle\widetilde{\nu}_\mathrm{R}^1\rangle$ and as such the scalar content in the intermediate regions will not affect the running of the gauge couplings.} \\
{\bf Right panel:} \emph{RG evolution of the gauge couplings for the scenario where there is no hierarchy between the three intermediate scales. To match the gauge couplings measured at the EW scale, the soft scale ends up at $8.8\cdot10^{10}$ GeV and the GUT scale at $4.9\cdot10^{17}$ GeV, i.e. we end up with a distinct hierarchy between all three scales.}
}
\label{fig:RGE}
\end{figure}

\twocolumngrid

Let us investigate whether the introduction of a hierarchy between $\langle\widetilde{\phi}^3\rangle$, $\langle\widetilde{\phi}^2\rangle$ and $\langle\widetilde{\nu}_\mathrm{R}^1\rangle$ can lower the soft scale $\langle\widetilde{\phi}^3\rangle$. By solving for $\langle\widetilde{\nu}_\mathrm{R}^1\rangle$ in Eq.~\eqref{eq:evolutionequations} and inserting all known values, we have the equation
\begin{eqnarray} \nonumber
\langle\widetilde{\nu}_\mathrm{R}^1\rangle=&m_{Z}\exp\Bigg\{20.69 - \frac{1}{19}\log\left(\frac{\langle\widetilde{\phi}^3\rangle}{\langle\widetilde{\phi}^2\rangle}\right)\Big[4b^{\RN{2}}_{g_\mathrm{C}} - 9b^{\RN{2}}_{g_\mathrm{L}} 
\\&
+ {3}{b^{\RN{2}}_{g_\mathrm{R}}}+ {b^{\RN{2}}_{\widetilde{g}_\mathrm{L+R}}} \Big]
- \frac{1}{19}\log\left(\frac{\langle\widetilde{\phi}^3\rangle}{\langle\widetilde{\phi}^2\rangle}\right)
\nonumber
\\&
\left[4b^{\RN{3}}_{g_\mathrm{C}} - 9b^{\RN{3}}_{g_\mathrm{L}} + {3}{b^{\RN{3}}_{g_\mathrm{R}}}+ {b^{\RN{3}}_{\widetilde{g}_\mathrm{L+R}}} \right]
\Bigg\}. 
\label{equationForM3}
\end{eqnarray}

The $b$-values will vary depending on the scalar field content with the extreme values presented in Appendix~\ref{sec:betabeta}. To minimise the argument of the exponential (and thereby minimising the value of $\langle\widetilde{\nu}_\mathrm{R}^1\rangle$), we should maximise the values of $b^{\RN{2},\RN{3}}_{g_\mathrm{C}}$, $b^{\RN{2},\RN{3}}_{g_\mathrm{R}}$ and $b^{\RN{2},\RN{3}}_{\widetilde{g}_\mathrm{L+R}}$, while minimising $b^{\RN{2},\RN{3}}_{g_\mathrm{L}}$. This occurs when including all scalars apart from the left-handed doublets $Q_\mathrm{L}^{1,2,3}$, $E_\mathrm{L}^{1,2,3}$ and $H^3$, in both region $\RN{2}$ and $\RN{3}$. In that case the values are
\begin{eqnarray} \nonumber
b^{\RN{2},\RN{3}}_{g_\mathrm{C}}=-\frac{13}{3},\;\;\;\;
b^{\RN{2},\RN{3}}_{g_\mathrm{L}}=-\frac{2}{3},\\
b^{\RN{2},\RN{3}}_{g_\mathrm{R}}=\frac{4}{3},\;\;\;\;
b^{\RN{2},\RN{3}}_{\widetilde{g}_\mathrm{L+R}}=\frac{40}{3}.
\label{eq:bvalues}
\end{eqnarray}
When ranging over various hierarchies using the $b$-values in~\eqref{eq:bvalues}, we see that the scale of $\langle\widetilde{\nu}_\mathrm{R}^1\rangle$ decreases as the hierarchy between $\langle\widetilde{\phi}^2\rangle$ and $\langle\widetilde{\phi}^3\rangle$ increases. The soft scale $\langle\widetilde{\phi}^3\rangle$, on the other hand, is minimised when it is equal to $\langle\widetilde{\nu}_\mathrm{R}^1\rangle$, i.e.~when there are no hierarchies, as shown in Fig.~\ref{fig:RGE} (left), by which we conclude that Eq.~\eqref{eq:finalscales} is in fact the optimal scenario in the sense that it provides the strongest hierarchy between the GUT scale and the soft SUSY-breaking scale. In Fig.~\ref{fig:RGE} (right) we show the evolution of the gauge couplings for this scenario. 

It is important to mention that these scales are obtained from gauge couplings evolved to one-loop accuracy but matched at tree-level, where one-loop matching conditions could introduce significant corrections, due to the many fields involved, as indicated in Ref.~\cite{Braathen:2018}. As the resulting scales could be sensitive to potentially significant threshold corrections, we are careful not to draw any strong conclusions at this point.   

Furthermore, there is a possibility for lowering the soft scale by relaxing the $\mathbb{Z}_3$ symmetry at the GUT scale, with gauge unification instead happening at the $\mathrm{E}_6$ level. In fact, as was demonstrated in \citep{Chakrabortty:2008zk}, a non-universal gauge coupling at the GUT breaking scale may arise from corrections to the gauge kinetic terms induced by dimension 5 operators, emerging due to higher dimensional $\mathrm{E}_6$ representations. This would
also open up the possibility for the emergence of new gauge bosons at, or at least close to, the TeV scale. We leave the question about a significance of such effects and its phenomenological implications for a further study. 
\section{Summary}
\label{sec:conclusions}

Here, we would like to summarise the basic features of the LRCF-symmetric SHUT theory considered in this paper:

\begin{itemize}
\item In contrast to previous GUT scale formulations based on gauge trinification, all three fermion generations are unified into 
a single $(\bm{27},\bm{3})$-plet of $\SU{3}{F}\times \mathrm{E}_6$, and no copies of any fundamental $\mathrm{E}_6$ reps are required 
for its consistent breaking down to the gauge symmetry of the SM. The considered $\SU{3}{F}\times \mathrm{E}_6$ symmetry can be embedded
into $\mathrm{E}_8$, motivating the addition of $(\bm{1},\bm{8})$ and $(\bm{78},\bm{1})$ multiplets corresponding to four $\SU{3}{}$-octet reps.
The gauge couplings are enforced to unify by means of a cyclic permutation symmetry $\mathbb{Z}_3$ acting on the trinification subgroup of the LRCF-symmetry 
in the same way as in the Glashow's formulation.
\item The chiral-adjoint sector $\bm{\Delta}_{\rm F}^a=(\bm{1},\bm{8})$ and $\bm{\Delta}_{\rm L,R,C}^a\subset(\bm{78},\bm{1})$ is 
necessary for a consistent breaking of the LRCF-symmetry down to the SM gauge symmetry in the softly-broken SUSY formulation of 
the theory while none of the adjoint fields remain at the EW scale. In our model, the fields developing VEVs at lower energies (the tri-triplets) 
happen to have the mass terms of $\mathcal{O}(\mathrm{m}_\mathrm{soft})$, while the fields whose VEVs spontaneously break the high-scale SHUT LRCF-symmetry 
(the adjoints) have their GUT scale mass term in the superpotential. Hence, our model does not exhibit an analogue of the $\mu$-problem in the MSSM. 
\item With the first symmetry breaking being triggered at the GUT scale by VEVs in the adjoint (octet) scalars, 
mass terms in the fundamential ($\bm{L},\,\bm{Q}_{\rm L},\,\bm{Q}_{\rm R}$ tri-triplet) sector are forbidden. This means that the SM-like quarks and leptons
remain massless until EWSB.
\item In the SHUT model, all possible tree-level masses for fermions come from a single term in the superpotential, $\bm{L}^i\bm{Q}_{\rm L}^j\bm{Q}_{\rm R}^k\epsilon_{ijk}$. As we have seen, only two generations of would-be SM quarks get such contributions to their masses. As such, the model offers a starting point for a mechanism explaining the mass hierarchies of the SM, where, for example, the charged leptons are all light as they have no allowed tree-level masses and instead attain their masses radiatively (i.e.~via loop-induced threshold corrections). Also, with three Higgs doublets at low energies, the model has Cabbibo quark mixing at tree-level, while radiatively generated (and RG evolved) Yukawa interactions open the possibility of reproducing the complete structure.
\item The symmetry breaking scales below the GUT scale (including the EW scale) are fully determined by the dynamics of the soft SUSY-breaking interactions and 
are thus naturally protected from the GUT scale radiative corrections. A particularly relevant multi-stage symmetry breaking scheme in the SHUT theory 
down to the SM-like gauge effective theory has been shown in Fig.~\ref{fig:1112abc}. 
\item The LRCF-symmetric theory contains an accidential $\U{B}$ baryon symmetry, by which the proton remains stable to all orders in perturbation theory. Other accidental $\U{W}$ and LR-parity symmetries can be 
(softly) broken in the low-energy EFT ensuring there being no massless charged leptons below the EWSB scale, and allowing the breaking of $\SU{2}{R}$ and $\SU{2}{L}$ 
symmetries at different energy scales, respectively. 
\item The smallest possible hierarchy between the EW scale and the soft scale, and the largest possible hierarchy between the soft scale and the GUT scale, occurs as the VEVs of $\widetilde{\phi}^3$, $\widetilde{\phi}^2$ and $\widetilde{\nu}_\mathrm{R}^1$ are all put at the same scale. For this scenario, the soft scale ends up at $\sim9\cdot10^{10}$ GeV and the GUT scale at $\sim5\cdot10^{17}$ GeV. However, these numbers do not take into account potentially large one-loop threshold corrections.
\item While our estimates have shown a potential agreement with the SM particle spectrum, and in particular the possibility to lift the top-charm mass degeneracy via quantum effects, it is not less true that the large $\langle \widetilde{\phi}^3 \rangle$, $\langle \widetilde{\phi}^2 \rangle$ and $\langle \widetilde{\nu}_\mathrm{R}^1 \rangle$ VEVs introduce fine-tuning in the scalar sector in order to satisfy the requirement of light Higgs doublets and possibly light squarks. We have pointed out that to solve this issue we need to relax the $\mathbb{Z}_3$ symmetry and transfer the unification of gauge interactions to the $\E{6}$ level, which is left for a future work.
\end{itemize}

Given the above properties, the SHUT model offers interesting new possibilities for deriving the structure and parameters of the SM from the GUT scale physics.
This is a good motivation for investigations of this model, its multi-scale symmetry breaking patterns,
loop-level matching and RG flow. Among the first natural steps would be to uncover some of the features of the simplest SM-like 
low-energy EFTs in a symmetry-based study without invoking the full-fledged radiative analysis of the SHUT theory. 
The EFT scenarios studied in this work pave the ground for further phenomenological studies of trinification based GUTs and move beyond 
the most common issues of such theories in the past. 


\acknowledgments
The authors would like to thank N.-E.~Bomark, A.~E.~C\'arcamo Hern\'andez, C.~Herdeiro, S.~Kovalenko, 
W.~Porod, J.~Rathsman, H.~Serodio and F.~Staub for insightful discussions in the various stages of this 
work.~J. E. C.-M., J.W. and R.P. thank Prof.~C.~Herdeiro for support of the project and hospitality during their 
visits at Aveiro University.~J. E. C.-M. was partially supported by the Crafoord Foundation and Lund 
University.~A.P.M.~was initially funded by FCT grant SFRH/BPD/97126/2013.~A.P.M.~is supported by
Funda\c{c}\~ao para a Ci\^encia e a Tecnologia (FCT),
within project UID/MAT/04106/2019 (CIDMA) and by national funds (OE), through FCT, I.P., in the scope
of the framework contract foreseen in the numbers 4, 5 and 6
of the article 23, of the Decree-Law 57/2016, of August 29,
changed by Law 57/2017, of July 19.~A.P.M.~is also supported by the \textit{Enabling Green E-science for the Square Kilometer Array Research Infrastructure} (ENGAGESKA), POCI-01-0145-FEDER-022217, and by the project \textit{From Higgs Phenomenology to the Unification of Fundamental Interactions}, PTDC/FIS-PAR/31000/2017.~R.P.,~A.O. and J.W.~were partially 
supported by the Swedish Research Council, contract numbers 621-2013-428 and 2016-05996.~R.P.~was 
also partially supported by CONICYT grants PIA ACT1406 and MEC80170112. The work by R.P.~was partially supported 
by the Ministry of Education, Youth and Sports of the Czech Republic project LT17018, as well as by the COST Action 
CA15213. This project has received funding from the European Research Council (ERC) under the European Union's 
Horizon 2020 research and innovation programme (grant agreement No 668679). This work is also supported by 
the CIDMA project UID/MAT/04106/2013.

\appendix


\section{Symmetry breaking schemes and charges} 
\label{sec:Symmetries}

In this appendix we provide a summary of the SSB scheme from 
the high-scale GUT symmetry down to that of the SM.

\subsection{Breaking path and generators}

The breaking path from the GUT symmetry down to a LR-symmetric effective theory reads
%

\onecolumngrid

\begin{eqnarray}
&&\left[ \SU{3}{C} \times \SU{3}{L} \times \SU{3}{R}\right] \rtimes \mathbb{Z}^{\rm (LRC)}_3 
\times \{\SU{3}{F} \times  \U{W} \times \U{B} \}  \nonumber \\
&& \overset{v,v_{\rm F}}{\to} \quad \; \SU{3}{C}\times \left[ \SU{2}{L} \times \SU{2}{R} \times 
\U{L} \times \U{R} \right]  \times \{\SU{2}{F} \times \U{F} \times  \U{W} \times \U{B} \} \nonumber \\
&& \overset{\langle \widetilde{\phi}^{3}\rangle}{\to} \quad \SU{3}{C} \times [\SU{2}{L} \times 
 \SU{2}{R}]  \times \mathrm{U}(1)_{\mathrm{L}+\mathrm{R}} \times 
 \{ \SU{2}{F}\times\mathrm{U}(1)_{\mathrm{S}} 
\;\times\; \U{S'} \times \U{B}\} \equiv G_{3221\{21\}}\,, \label{eq:brk2}
\end{eqnarray}

\twocolumngrid

where global symmetries (including the accidental ones) are indicated by $\{ \cdots \}$. 
The generators of the $\U{}$ groups after the GUT SSB are
\begin{equation}
\label{generators0}
\begin{aligned}
\Scale[0.95]{
 \T{L}{8}\,, \qquad  \T{R}{8}\,, \qquad \T{F}{8}\,, \qquad \T{W}{} \,, \qquad \T{B}{}\,,
}
\end{aligned}
\end{equation}
whereas after the $\langle \widetilde{\phi}^{3}\rangle$ VEV we have
\begin{equation}
\label{generators1}
\begin{aligned}
\T{L+R}{} = \T{L}{8} &+ \T{R}{8} \,, \quad \T{S}{} =   \T{L}{8} - \T{R}{8} - 2 \T{F}{8} \,, \\
&\T{S'}{} = \T{L}{8} - \T{R}{8} + \tfrac{2}{\sqrt{3}} \T{W}{} \,.
\end{aligned}
\end{equation}
with normalization factors conveniently chosen to provide integer charges for leptons and scalar bosons. 

Note that, according to 
the discussion in Sect.~\ref{sec:SSBSU3} the LR-parity can be explicitly broken in the soft SUSY-breaking sector and is therefore
absent in the effective theory.

We may also place a VEV in $\widetilde{\phi}^{2}$ and $\widetilde{\nu}_{\mathrm{R}}^{1}$. In such a case the breaking scheme takes the form

\begin{equation}
\label{eq:brknuphi}
\begin{aligned}
G_{3221\{21\}}&\overset{\langle \widetilde{\nu}_{\mathrm{R}}^{1}\rangle\,, \langle \widetilde{\phi}^{2} \rangle}{\longrightarrow} \, 
\SU{3}{C}\times  \SU{2}{L} \times  \mathrm{U}(1)_{\mathrm{Y}} \, \\
&\times\, \{\mathrm{U}(1)_{\mathrm{T}} \,\times\, 
\mathrm{U}(1)_{\mathrm{T'}} \times \U{B}\} \,,
\end{aligned}
\end{equation}
where the generators of $\U{Y}$, $\U{T}$ and $\U{T'}$ read
\begin{eqnarray}
\nonumber
&&\T{Y}{} = -\tfrac{1}{\sqrt{3}}\(  \T{L+R}{} + \sqrt{3} \T{R}{3}  \) \,, \quad
\T{T}{} = \T{R}{3} + \tfrac{1}{3\sqrt{3}}\T{S}{} - \tfrac{2}{3}\T{F}{3} \,, \\
&&\qquad \qquad \T{T'}{} =  \T{S'}{} + \tfrac{1}{3}\T{S}{} -\tfrac{2}{\sqrt{3}} \T{F}{3} \,.
\label{generators3}
\end{eqnarray}

\subsection{Quantum numbers}

In this section we present the representations and charges of the light states after each breaking step. 
We consider as light states all fields that are decoupled from the GUT scale after the first SSB step. 

In what follows, the Higgs bi-doublets are referred to as ${H}^{1,2,3}$, the singlet Higgs-lepton fields denoted as $\phi^{1,2,3}$ and the lepton doublets as $E_{\mathrm{L,R}}^{1,2,3}$, while the quark multiplets split up into $\mathcal{Q}_{\mathrm{L,R}}^{1,2,3}$ and $D_{\rm L,R}^{1,2,3}$, 
where $\mathcal{Q}$ are the $3 \times 2$ blocks and $D$ the $3 \times 1$ blocks. The superscript $1,2,3$ is the generation number. 
Whenever convenient we will adopt a simplifying notation according to
\begin{equation}
\label{1}
\begin{aligned}
H^{3} \rightarrow h\,, \\
E_{\rm{L,R}}^{3} \rightarrow \mathcal{E}_{\rm{L,R}}\,, \\
\mathcal{Q}_{\rm{L,R}}^{3} \rightarrow q_{\rm{L,R}}\,,\\
\end{aligned} \qquad
\begin{aligned}
\phi^{3} \rightarrow \varphi\,, \\
D_{\rm{L,R}}^{3} \rightarrow \mathcal{B}_{\rm{L,R}}\,,\\
X^{1,2} \rightarrow X^f\,,
\end{aligned}
\end{equation}
where $f$ is a family index running over the first two generations with $X$ representing any of such $\SU{2}{F}$ doublets.
\onecolumngrid

\begin{table}[!h]
\ra{1.4}
\centering
\scalebox{1.0}{
\begin{tabular}{@{}ccccccccccc@{}}
\hline
\small{Fermion}&\small{Boson}&\small{$\SU{3}{C}$}&\small{$\SU{2}{L}$}&\small{$\SU{2}{R}$}&\small{$\{\SU{2}{F}\}$}&\small{$\U{L}$}&\small{$\U{R}$}&
\small{$\{\U{F}\}$}&\small{$\{\U{B}\}^{\Scale[0.7]{\rm acc}}$}&\small{$\{\U{W}\}^{\Scale[0.7]{\rm acc}}$}\tabularnewline
\hline
$\varphi$ & $\widetilde{\varphi}$               	& $\bm{1}$      & $\bm{1}$  & $\bm{1}$ & $\bm{1}$ & $-2$ & $2$ & $-2$ & $0$ & $1$ \tabularnewline
$\phi^f$ & $\widetilde{\phi}^f$               	& $\bm{1}$      & $\bm{1}$  & $\bm{1}$ & $\bm{2}^f$ & $-2$ & $2$ & $1$ & $0$ & $1$ \tabularnewline
$ \mathcal{E}_{\rm L}^{l}$ & $\widetilde{\mathcal{E}}_{\rm L}^{l}$               	& $\bm{1}$      & $\bm{2}^l$  & $\bm{1}$ & $\bm{1}$ & $1$ & $2$ & $-2$ & $0$ & $1$ \tabularnewline
$E_{\rm L}^{f\,l}$ & $\widetilde{E}_{\rm L}^{f\,l}$               	& $\bm{1}$      & $\bm{2}^l$  & $\bm{1}$ & $\bm{2}^f$ & $1$ & $2$ & $1$ & $0$ & $1$ \tabularnewline
$ \mathcal{E}_{{\rm R}\,r}$ & $\widetilde{\mathcal{E}}_{{\rm R}\,r}$               	& $\bm{1}$      & $\bm{1}$  & $\bm{2}_r$ & $\bm{1}$ & $-2$ & $-1$ & $-2$ & $0$ 
& $1$ \tabularnewline
$E_{{\rm R}\,r}^{f}$ & $\widetilde{E}_{{\rm R}\,r}^{f}$               	& $\bm{1}$      & $\bm{1}$  & $\bm{2}_r$ & $\bm{2}^f$ & $-2$ & $-1$ & $1$ & $0$ & $1$ \tabularnewline
$\widetilde{h}^l_{r}$ & $h^l_{r}$               	& $\bm{1}$      & $\bm{2}^l$  & $\bm{2}_r$ & $\bm{1}$ & $1$ & $-1$ & $-2$ & $0$ & $1$ \tabularnewline
$\widetilde{H}^{f\,l}_{r}$ & $H^{f\,l}_{r}$               	& $\bm{1}$      & $\bm{2}^l$  & $\bm{2}_r$ & $\bm{2}^f$ & $1$ & $-1$ & $1$ & $0$ & $1$ \tabularnewline
$q_{{\rm L}\,l}^{x}$ & $\widetilde{q}_{{\rm L}\,l}^{x}$               	& $\bm{3}^x$      & $\bm{2}_l$  & $\bm{1}$ & $\bm{1}$ & $-1$ & $0$ & $-2$ & $1/3$ & $-1/2$ \tabularnewline
$\mathcal{Q}_{{\rm L}\,l}^{x\,f}$ & $\widetilde{\mathcal{Q}}_{{\rm L}\,l}^{x\,f}$               	& $\bm{3}^x$      & $\bm{2}_l$  & $\bm{1}$ & $\bm{2}^f$ & $-1$ & $0$ & $1$ 
& $1/3$ & $-1/2$ \tabularnewline
$q_{{\rm R}\,x}^{r}$ & $\widetilde{q}_{{\rm R}\,x}^{r}$               	& $\bm{3}_x$      & $\bm{1}$  & $\bm{2}^r$ & $\bm{1}$ & $0$ & $1$ & $-2$ & $-1/3$ & $-1/2$ \tabularnewline
$\mathcal{Q}_{{\rm R}\,x}^{f\,r}$ & $\widetilde{\mathcal{Q}}_{{\rm R}\,x}^{f\,r}$               	& $\bm{3}_x$      & $\bm{1}$  & $\bm{2}^r$ & $\bm{2}^f$ & $0$ & $1$ 
& $1$ & $-1/3$ & $-1/2$ \tabularnewline
$\mathcal{B}_{\rm L}^{x}$ & $\widetilde{\mathcal{B}}_{\rm L}^{x}$               	& $\bm{3}^x$      & $\bm{1}$  & $\bm{1}$ & $\bm{1}$ & $2$ & $0$ & $-2$ & $1/3$ 
& $-1/2$ \tabularnewline
$\mathcal{B}_{{\rm R}\,x}$ & $\widetilde{\mathcal{B}}_{{\rm R}\,x}$               	& $\bm{3}_x$      & $\bm{1}$  & $\bm{1}$ & $\bm{1}$ & $0$ & $-2$ & $-2$ 
& $-1/3$ & $-1/2$ \tabularnewline
$D_{\rm L}^{x\,f}$ & $\widetilde{D}_{\rm L}^{x\,f}$               	& $\bm{3}^x$      & $\bm{1}$  & $\bm{1}$ & $\bm{2}^f$ & $2$ & $0$ & $1$ & $1/3$ & $-1/2$ \tabularnewline
$D_{{\rm R}\,x}^{f}$ & $\widetilde{D}_{{\rm R}\,x}^{f}$               	& $\bm{3}_x$      & $\bm{1}$  & $\bm{1}$ & $\bm{2}^f$ & $0$ & $-2$ & $1$ & $-1/3$ & $-1/2$ \tabularnewline
\hline
$\widetilde{g}^a$ & $G_{\rm C}^{\mu\,a}$   & $\bm{8}^a$      & $\bm{1}$  & $\bm{1}$ & $\bm{1}$ & $0$ & $0$ & $0$ & $0$ & $0$ \tabularnewline
$\mathcal{T}_{\rm L}^i$ & $G_{\rm L}^{\mu\,i}$   & $\bm{1}$      & $\bm{3}^i$  & $\bm{1}$ & $\bm{1}$ & $0$ & $0$ & $0$ & $0$ & $0$ \tabularnewline
$\mathcal{T}_{\rm R}^i$ & $G_{\rm R}^{\mu\,i}$   & $\bm{1}$      & $\bm{1}$  & $\bm{3}^i$ & $\bm{1}$ & $0$ & $0$ & $0$ & $0$ & $0$ \tabularnewline
$\mathcal{S}_{\rm L,R}$ & $G_{\rm L,R}^{\mu\,8}$   & $\bm{1}$      & $\bm{1}$  & $\bm{1}$ & $\bm{1}$ & $0$ & $0$ & $0$ & $0$ & $0$ \tabularnewline
\hline
$\mathcal{H}_{\rm F}^f$ & $\widetilde{\mathcal{H}}_{\rm F}^f$ & $\bm{1}$      & $\bm{1}$  & $\bm{1}$ & $\bm{2}^f$ & $0$ & $0$ & $-1$ & $0$ & $0$ \tabularnewline
\hline
\end{tabular}
}
\caption[ ]{\it Field content and quantum numbers of the LR-symmetric EFT after $\widetilde{\Delta}_{\rm L,R,F}$ VEVs in Eq.~\eqref{eq:brk2}. Here and below, 
$\{\dots\}^{\rm acc}$ denote the accidental symmetries. The charges for $\U{L}$, $\U{R}$ and $\U{F}$ are to be rescaled with a factor $1/(2\sqrt{3})$.}
\label{Table:EFT-content}
\end{table}

\twocolumngrid

\onecolumngrid

\begin{table}[tbh]
\ra{1.4}
\centering
\scalebox{1.0}{
\begin{tabular}{@{}ccccccccccc@{}}
\hline
\small{Fermion}&\small{Boson}&\small{$\SU{3}{C}$}&\small{$\SU{2}{L}$}&\small{$\SU{2}{R}$}&\small{$\{\SU{2}{F}\}$}&\small{$\U{L+R}$}&\small{$\{\U{S}\}$}
&\small{$\{\U{S^{\prime}}\}^{\Scale[0.7]{\rm acc}}$}&\small{$\{\U{B}\}^{\Scale[0.7]{\rm acc}}$}\tabularnewline
\hline
$\varphi$ & $\widetilde{\varphi}$               	& $\bm{1}$      & $\bm{1}$  & $\bm{1}$ & $\bm{1}$ & $0$ & $0$ & $0$ & $0$  \tabularnewline
$\phi^f$ & $\widetilde{\phi}^f$               	& $\bm{1}$      & $\bm{1}$  & $\bm{1}$ & $\bm{2}^f$ & $0$ & $-2$ & $0$ & $0$ \tabularnewline
$ \mathcal{E}_{\rm L}^{l}$ & $\widetilde{\mathcal{E}}_{\rm L}^{l}$               	& $\bm{1}$      & $\bm{2}^l$  & $\bm{1}$ & $\bm{1}$ & $1$ & $1$ & $1$ & $0$\tabularnewline
$E_{\rm L}^{f\,l}$ & $\widetilde{E}_{\rm L}^{f\,l}$               	& $\bm{1}$      & $\bm{2}^l$  & $\bm{1}$ & $\bm{2}^f$ & $1$ & $-1$ & $1$ & $0$ \tabularnewline
$ \mathcal{E}_{{\rm R}\,r}$ & $\widetilde{\mathcal{E}}_{{\rm R}\,r}$               	& $\bm{1}$      & $\bm{1}$  & $\bm{2}_r$ & $\bm{1}$ & $-1$ & $1$ & $1$ & $0$ \tabularnewline
$E_{{\rm R}\,r}^{f}$ & $\widetilde{E}_{{\rm R}\,r}^{f}$               	& $\bm{1}$      & $\bm{1}$  & $\bm{2}_r$ & $\bm{2}^f$ & $-1$ & $-1$ & $1$ & $0$  \tabularnewline
$\widetilde{h}^l_{r}$ & $h^l_{r}$               	& $\bm{1}$      & $\bm{2}^l$  & $\bm{2}_r$ & $\bm{1}$ & $0$ & $2$ & $2$ & $0$ \tabularnewline
$\widetilde{H}^{f\,l}_{r}$ & $H^{f\,l}_{r}$               	& $\bm{1}$      & $\bm{2}^l$  & $\bm{2}_r$ & $\bm{2}^f$ & $0$ & $0$ & $2$ & $0$ \tabularnewline
$q_{{\rm L}\,l}^{x}$ & $\widetilde{q}_{{\rm L}\,l}^{x}$               	& $\bm{3}^x$      & $\bm{2}_l$  & $\bm{1}$ & $\bm{1}$ & $-1/3$ & $1$ & $-1$ & $1/3$ \tabularnewline
$\mathcal{Q}_{{\rm L}\,l}^{x\,f}$ & $\widetilde{\mathcal{Q}}_{{\rm L}\,l}^{x\,f}$               	& $\bm{3}^x$      & $\bm{2}_l$  & $\bm{1}$ & $\bm{2}^f$ & $-1/3$ & $-1$ 
& $-1$ & $1/3$ \tabularnewline
$q_{{\rm R}\,x}^{r}$ & $\widetilde{q}_{{\rm R}\,x}^{r}$               	& $\bm{3}_x$      & $\bm{1}$  & $\bm{2}^r$ & $\bm{1}$ & $1/3$ & $1$ & $-1$ & $-1/3$ \tabularnewline
$\mathcal{Q}_{{\rm R}\,x}^{f\,r}$ & $\widetilde{\mathcal{Q}}_{{\rm R}\,x}^{f\,r}$               	& $\bm{3}_x$      & $\bm{1}$  & $\bm{2}^r$ & $\bm{2}^f$ & $1/3$ & $-1$ 
& $-1$ & $-1/3$ \tabularnewline
$\mathcal{B}_{\rm L}^{x}$ & $\widetilde{\mathcal{B}}_{\rm L}^{x}$               	& $\bm{3}^x$      & $\bm{1}$  & $\bm{1}$ & $\bm{1}$ & $2/3$ & $2$ & $0$ & $1/3$ \tabularnewline
$\mathcal{B}_{{\rm R}\,x}$ & $\widetilde{\mathcal{B}}_{{\rm R}\,x}$               	& $\bm{3}_x$      & $\bm{1}$  & $\bm{1}$ & $\bm{1}$ & $-2/3$ & $2$ & $0$ 
& $-1/3$ \tabularnewline
$D_{\rm L}^{x\,f}$ & $\widetilde{D}_{\rm L}^{x\,f}$               	& $\bm{3}^x$      & $\bm{1}$  & $\bm{1}$ & $\bm{2}^f$ & $2/3$ & $0$ & $0$ & $1/3$ \tabularnewline
$D_{{\rm R}\,x}^{f}$ & $\widetilde{D}_{{\rm R}\,x}^{f}$               	& $\bm{3}_x$      & $\bm{1}$  & $\bm{1}$ & $\bm{2}_f$ & $-2/3$ & $0$ & $0$ & $-1/3$ \tabularnewline
\hline
$\widetilde{g}^a$ & $G_{\rm C}^{\mu\,a}$               	& $\bm{8}^a$      & $\bm{1}$  & $\bm{1}$ & $\bm{1}$ & $0$ & $0$ & $0$ & $0$  \tabularnewline
$\mathcal{T}_{\rm L}^i$ & $G_{\rm L}^{\mu\,i}$               	& $\bm{1}$      & $\bm{3}^i$  & $\bm{1}$ & $\bm{1}$ & $0$ & $0$ & $0$ & $0$ \tabularnewline
$\mathcal{T}_{\rm R}^i$ & $G_{\rm R}^{\mu\,i}$               	& $\bm{1}$      & $\bm{1}$  & $\bm{3}^i$ & $\bm{1}$ & $0$ & $0$ & $0$ & $0$ \tabularnewline
$\mathcal{S}_{\rm L,R}$ & $G_{\rm L,R}^{\mu\,8}$                 	& $\bm{1}$      & $\bm{1}$  & $\bm{1}$ & $\bm{1}$ & $0$ & $0$ & $0$ & $0$ \tabularnewline
\hline
$\mathcal{H}_{\rm F}^f$ & $\widetilde{\mathcal{H}}_{\rm F}^f$ & $\bm{1}$      & $\bm{1}$  & $\bm{1}$ & $\bm{2}^f$ & $0$ & $-2$ & $0$ & $0$ \tabularnewline
\hline
\end{tabular}
}
\caption[ ]{\it Field content and quantum numbers of the LR-symmetric EFT after $\widetilde{\varphi}$ VEV as in Eq.~\eqref{eq:brk2}. The charges for $\U{L+R}$, $\U{S}$ and $\U{S'}$ are to be rescaled with a factor $\sqrt{3}/2$.}
\label{Table:EFT-content-phi}
\end{table}

\twocolumngrid

The quantum numbers of the light eigenstates after the $v$ and $v_{\rm F}$ VEVs are given in Tab.~\ref{Table:EFT-content} while 
those of the model after $\widetilde{\phi}^{3}$ VEV are shown in Tab.~\ref{Table:EFT-content-phi}. In Tab.~\ref{Table:EFT-content-snu1Phi} we show the charges of the SM-like EFT after the $\widetilde{\nu}_{\rm R}^{1}$ and $\widetilde{\phi}^{2}$ VEVs 
which may either occur simultaneously or at separate scales. Note that the $\mean{\varphi}$ VEV enables mixing between the first and second generations of singlet (s)quarks. For example, it allows fermion mass 
terms of the form $ m_D D_{\rm L}^f D_{\rm R}^{f'} \varepsilon_{f f'}$.


\section{Particle masses in the high-scale theory} 
\label{sec:masses}

\subsection{Scalar spectra and minimisation conditions}

The extremizing conditions obtained after taking the first derivatives of the scalar potential of the SHUT model can be solved, 
e.g.~w.r.t. the soft parameters $m^2_{\bm{78}}$ and $m^2_{\bm{1}}$ from where we obtain
\begin{align} 
\nonumber
m^2_{\bm{78}} = &  - b_{\bm{78}} + \tfrac{v}{12} \left( \sqrt{6} A_{\bm{78}} + 3 \sqrt{6} C_{\bm{78}} - 
v \lambda^2_{\bm{78}} \right) \\ &+ 
\tfrac{\sqrt{6}}{4} v \lambda_{\bm{78}} \mu_{\bm{78}} - \mu_{\bm{78}}^2 \,, \label{eq:tadE8} \\
m^2_{\bm{1}} = &  - b_{\bm{1}} + \tfrac{\vev{F}}{12} \left( \sqrt{6} A_{\bm{1}} - \vev{F} \lambda^2_{\bm{1}} \right)  + 
\tfrac{\sqrt{6}}{4} \vev{F} \lambda_{\bm{1}} \mu_{\bm{1}} - \mu_{\bm{1}}^2\,.\nonumber
\end{align}
The minimisation conditions are then used in the Hessian matrix whose eigenvalues corresponding to the fundamental and adjoint scalar sectors are 
shown in Tabs.~\ref{table:SpecFund} and \ref{table:SpecAdj}, respectively. Note that, for simplicity, we use the LR-symmetric 
case with $A_{\bar{\mathrm{G}}} = A_\mathrm{G}$.

The branching rule for a fundamental representation of $\SU{3}{A}$, ${\rm A = L,R,F}$ when it is broken down to \mbox{$\SU{2}{A} \times \U{A}$} reads
\begin{equation}
\label{eq:3to21}
\bm{3} \rightarrow \bm{2}_1 \oplus \bm{1}_{-2}\,,
\end{equation} 
where, up to an overall normalization factor, the subscripts represent the $\U{A}$ charge. Therefore, after the SSB, the eigenstates shown 
in Tab.~\ref{table:SpecFund} form representations of the $G_{32211\lbrace 21 \rbrace}$ symmetry given in Eq.~\eqref{eq:322113}
and transform as singlets, doublets, bi-doublets and tri-doublets under the $\SU{2}{L,R,F}$ symmetries, as schematically represented by 
the blocks in Eq.~\eqref{eq:tri-triplets}\footnote{The family $\SU{3}{F}$ triplets are also split up into $\SU{2}{F}$ doublets, containing the first and second 
generations, and singlets corresponding to the third generation.}. The LR-parity discussed in Sect.~\ref{sec:tritri} yields 
identical masses for the $\SU{2}{L}$ and $\SU{2}{R}$ eigenstates at the trinification SSB scale.

The adjoint scalars $\widetilde{\Delta}^a_{\rm{A=L,R,F}}$ are complex octets whose branching rule is given by
\begin{equation}
\label{eq:8to3221}
\bm{8} \rightarrow \bm{3}_0 \oplus \bm{2}_1 \oplus \bm{2}_{-1} \oplus \bm{1}_0\,,
\end{equation} 
where the complex octet is a reducible representation while its real and imaginary parts are the irreducible representations. As such, we end up with two real triplets, two real singlets and two complex doublets and their complex conjugates after the SSB. Each broken symmetry provides four Goldstone degrees of freedom 
out of which eight correspond to breaking of the local symmetries whereas four of them -- to the global ones. While the triplet mass eigenstates, $\bm{3}_0$, 
can be written as
\begin{align}
\nonumber
\renewcommand\arraystretch{1.3}
&\widetilde{\mathcal{T}}_{\rm A} \equiv \dfrac{1}{\sqrt{2}} \mleft(
\begin{array}{ccc}
\rm{Re}[\widetilde{\Delta}^1_{\rm A}] - \i \rm{Re}[\widetilde{\Delta}^2_{\rm A}]\\
\sqrt{2} \rm{Re}[\widetilde{\Delta}^3_{\rm A}]\\
\rm{Re}[\widetilde{\Delta}^1_{\rm A}] + \i \rm{Re}[\widetilde{\Delta}^2_{\rm A}]
\end{array}
\mright)\;,
\end{align}
\begin{align}
\renewcommand\arraystretch{1.3}
&\widetilde{\mathcal{T}}_{\rm A}^{\prime} \equiv \dfrac{1}{\sqrt{2}} \mleft(
\begin{array}{ccc}
\rm{Im}[\widetilde{\Delta}^1_{\rm A}] - \i \rm{Im}[\widetilde{\Delta}^2_{\rm A}]\\
\sqrt{2} \rm{Im}[\widetilde{\Delta}^3_{\rm A}]\\
\rm{Im}[\widetilde{\Delta}^1_{\rm A}] + \i \rm{Im}[\widetilde{\Delta}^2_{\rm A}]
\end{array}
\mright)\;,\label{eq:Triplets}
\end{align}
the two real singlets $\bm{1}_0$ read
\begin{align}
\label{eq:Singlets}
\widetilde{\mathcal{S}}_{\rm A} \equiv \rm{Re}[\widetilde{\Delta}^8_{\rm A}]\;,\; \widetilde{\mathcal{S}}_{\rm A}^{\prime} \equiv 
\rm{Im}[\widetilde{\Delta}^8_{\rm A}]\,.
\end{align}

\onecolumngrid

\begin{table}[tbh]
\ra{1.4}
\centering
\scalebox{1.0}{
\begin{tabular}{@{}ccccccccc@{}}
\hline
\small{Fermion}&\small{Boson}&\small{$\SU{3}{C}$}&\small{$\SU{2}{L}$}&\small{$\U{Y}$}&\small{$\{\U{T}\}$}&\small{$\{\U{T^{\prime}}\}^{\Scale[0.7]{\rm acc}}$}
&\small{$\{\U{B}\}^{\Scale[0.7]{\rm acc}}$}\tabularnewline
\hline
$\phi^1$ & $\widetilde{\phi}^1$   & $\bm{1}$ & $\bm{1}$ & $0$ & $4$ & $2$ & $0$ \tabularnewline
$\phi^{2},\varphi$ & $\widetilde{\phi}^{2},\, \widetilde{\varphi}$   & $\bm{1}$ & $\bm{1}$ & $0$ & $0$ & $0$ & $0$ \tabularnewline
$E_{\rm L}^{1\,l}$ & $\widetilde{E}_{\rm L}^{1\,l}$ & $\bm{1}$ & $\bm{2}^l$ & $-1/2$ & $3$ & $0$ & $0$ \tabularnewline
$E_{\rm L}^{2\,l},\,\mathcal{E}_{\rm L}^{l}$ & $\widetilde{E}_{\rm L}^{2\,l},\, \widetilde{\mathcal{E}}_{\rm L}^{l}$ & $\bm{1}$ & $\bm{2}^l$ & $-1/2$ & $-1$ 
& $-2$ & $0$ \tabularnewline
$ e^1_{\rm R}$ & $\widetilde{e}^1_{\rm R}$ & $\bm{1}$ & $\bm{1}$ & $1$ & $6$ & $0$ & $0$  \tabularnewline
$ \nu^1_{\rm R}$ & $\widetilde{\nu}^1_{\rm R}$ & $\bm{1}$ & $\bm{1}$ & $0$ & $0$ & $0$ & $0$  \tabularnewline
$ e^{2,3}_{\rm R}$ & $\widetilde{e}^{2,3}_{\rm R}$ & $\bm{1}$ & $\bm{1}$ & $1$ & $2$ & $-2$ & $0$  \tabularnewline
$ \nu^{2,3}_{\rm R}$ & $\widetilde{\nu}^{2,3}_{\rm R}$ & $\bm{1}$ & $\bm{1}$ & $0$ & $-4$ & $-2$ & $0$  \tabularnewline
$\widetilde{H}^{1\,l}_{\rm u}$ & $H^{1\,l}_{\rm u}$  & $\bm{1}$ & $\bm{2}^l$ & $1/2$ & $5$ & $-2$ & $0$ \tabularnewline
$\widetilde{H}^{1\,l}_{\rm d}$ & $H^{1\,l}_{\rm d}$  & $\bm{1}$ & $\bm{2}^l$ & $-1/2$ & $-1$ & $-2$ & $0$ \tabularnewline
$\widetilde{H}^{2\,l}_{\rm u},\, \widetilde{h}^{l}_{\rm u}$ & $H^{2\,l}_{\rm u},\, h^l_{\rm u}$  & $\bm{1}$ & $\bm{2}^l$ & $1/2$ & $1$ & $-4$ & $0$ \tabularnewline
$\widetilde{H}^{2\,l}_{\rm d},\, \widetilde{h}^{l}_{\rm d}$ & $H^{2\,l}_{\rm d},\, h^l_{\rm d}$  & $\bm{1}$ & $\bm{2}^l$ & $-1/2$ & $-5$ & $-4$ & $0$ \tabularnewline
$\mathcal{Q}_{{\rm L}\,l}^{x\,1}$ & $\widetilde{\mathcal{Q}}_{{\rm L}\,l}^{x\,1}$ & $\bm{3}^x$ & $\bm{2}_l$ & $1/6$ & $3$ & $3$ & $1/3$ \tabularnewline
$\mathcal{Q}_{{\rm L}\,l}^{x\,2},\, q_{{\rm L}\,l}^{x}$ & $\widetilde{\mathcal{Q}}_{{\rm L}\,l}^{x\,2},\,\widetilde{q}_{{\rm L}\,l}^{x}$ & $\bm{3}^x$ & $\bm{2}_l$ 
& $1/6$ & $-1$ & $1$ & $1/3$ \tabularnewline
$u^1_{{\rm R}\,x}$ & $\widetilde{u}^1_{{\rm R}\,x}$ & $\bm{3}_x$ & $\bm{1}$ & $-2/3$ & $0$ & $3$ & $-1/3$ \tabularnewline
$d^1_{{\rm R}\,x}$ & $\widetilde{d}^1_{{\rm R}\,x}$ & $\bm{3}_x$ & $\bm{1}$ & $1/3$ & $6$ & $3$ & $-1/3$ \tabularnewline
$u^{2,3}_{{\rm R}\,x}$ & $\widetilde{u}^{2,3}_{{\rm R}\,x}$ & $\bm{3}_x$ & $\bm{1}$ & $-2/3$ & $-4$ & $1$ & $-1/3$ \tabularnewline
$d^{2,3}_{{\rm R}\,x}$ & $\widetilde{d}^{2,3}_{{\rm R}\,x}$ & $\bm{3}_x$ & $\bm{1}$ & $1/3$ & $2$ & $1$ & $-1/3$ \tabularnewline
$D_{\rm L}^{x\,1}$ & $\widetilde{D}_{\rm L}^{x\,1}$ & $\bm{3}^x$ & $\bm{1}$ & $-1/3$ & $2$ & $1$ & $1/3$ \tabularnewline
$D_{\rm L}^{x\,2},\, \mathcal{B}_{\rm L}^{x}$ & $\widetilde{D}_{\rm L}^{x\,2},\, \widetilde{\mathcal{B}}_{\rm L}^{x}$ & $\bm{3}^x$ & $\bm{1}$ 
& $-1/3$ & $-2$ & $-1$ & $1/3$ \tabularnewline
$D_{{\rm R}\,x}^{1}$ & $\widetilde{D}_{{\rm R}\,x}^{1}$ & $\bm{3}_x$ & $\bm{1}$ & $1/3$ & $2$ & $1$ & $-1/3$ \tabularnewline
$D_{{\rm R}\,x}^{2},\,\mathcal{B}_{{\rm R}\,x}$ & $\widetilde{D}_{{\rm R}\,x}^{2},\, \widetilde{\mathcal{B}}_{{\rm R}\,x}$ & $\bm{3}_x$ & $\bm{1}$ 
& $1/3$ & $-2$ & $-1$ & $-1/3$ \tabularnewline
\hline
$\widetilde{g}^a$ & $G_{\rm C}^{\mu\,a}$ & $\bm{8}^a$ & $\bm{1}$  & $0$ & $0$ & $0$ & $0$ \tabularnewline
$\mathcal{T}_{\rm L}^i$ & $G_{\rm L}^{\mu\,i}$ & $\bm{1}$      & $\bm{3}^i$ & $0$ & $0$ & $0$ & $0$ \tabularnewline
$\mathcal{T}_{\rm R}^{\pm}$ & $G_{\rm R}^{\mu\,\pm}$ & $\bm{1}$ & $\bm{1}$  & $\pm 2$ & $0$ & $0$ & $0$ \tabularnewline
$\mathcal{T}_{\rm R}^{0}$ & $G_{\rm R}^{\mu\,0}$ & $\bm{1}$ & $\bm{1}$  & $0$ & $0$ & $0$ & $0$ \tabularnewline
$\mathcal{S}_{\rm L,R}$ & $G_{\rm L,R}^{\mu\,8}$ & $\bm{1}$ & $\bm{1}$ & $0$ & $0$ & $0$ & $0$  \tabularnewline
\hline
$\mathcal{H}_{\rm F}^1$ & $\widetilde{\mathcal{H}}_{\rm F}^1$ & $\bm{1}$ & $\bm{1}$  & $0$ & $4$ & $2$ & $0$ \tabularnewline
$\mathcal{H}_{\rm F}^2$ & $\widetilde{\mathcal{H}}_{\rm F}^2$ & $\bm{1}$ & $\bm{1}$  & $0$ & $0$ & $0$ & $0$ \tabularnewline
\hline
\end{tabular}
}
\caption[ ]{\it Field content and quantum numbers after the $\widetilde{\nu}_{\rm R}^{1}$ and $\widetilde{\phi}^{2}$ VEVs as in Eq.~\eqref{eq:brknuphi}. The charge for $\U{T}$ is to be rescaled with a factor $-1/6$, and the charge for $\U{T'}$ with a factor $-1/\sqrt{3}$.}
\label{Table:EFT-content-snu1Phi}
\end{table}

\twocolumngrid

Finally, there are two complex doublets from the real part of $\widetilde{\Delta}^a_{\rm{L,R,F}}$, transforming as $\bm{2}_{-1}$ and $\bm{2}_{1}$
\begin{align}
\nonumber
\renewcommand\arraystretch{1.3}
&\widetilde{G}_{\rm A}\equiv \dfrac{1}{\sqrt{2}} \mleft(
\begin{array}{ccc}
-\rm{Re}[\widetilde{\Delta}^6_{\rm A}] - \i \rm{Re}[\widetilde{\Delta}^7_{\rm A}]\\
\rm{Re}[\widetilde{\Delta}^4_{\rm A}] + \i \rm{Re}[\widetilde{\Delta}^5_{\rm A}]
\end{array}
\mright) \;,\; 
\end{align}
\begin{align}
\renewcommand\arraystretch{1.3}
&\widetilde{G}_{\rm A}^*= \dfrac{1}{\sqrt{2}} \mleft(
\begin{array}{ccc}
-\rm{Re}[\widetilde{\Delta}^6_{\rm A}] + \i \rm{Re}[\widetilde{\Delta}^7_{\rm A}]\\
\rm{Re}[\widetilde{\Delta}^4_{\rm A}] - \i \rm{Re}[\widetilde{\Delta}^5_{\rm A}]
\end{array}
\mright) \;,\label{eq:Doublets}
\end{align}
and two complex doublets from the imaginary part of $\widetilde{\Delta}^a_{\rm{L,R,F}}$, transforming as $\bm{2}_{-1}$ and $\bm{2}_{1}$
\begin{align}
\nonumber
\renewcommand\arraystretch{1.3}
&\mathcal{H}_{\rm A}  \equiv \dfrac{1}{\sqrt{2}} \mleft(
\begin{array}{ccc}
-\rm{Im}[\widetilde{\Delta}^6_{\rm A}] - \i \rm{Im}[\widetilde{\Delta}^7_{\rm A}]\\
\rm{Im}[\widetilde{\Delta}^4_{\rm A}] + \i \rm{Im}[\widetilde{\Delta}^5_{\rm A}]
\end{array}
\mright) \;,\; \\
\renewcommand\arraystretch{1.3}
&\mathcal{H}_{\rm A}^*  = \dfrac{1}{\sqrt{2}} \mleft(
\begin{array}{ccc}
-\rm{Im}[\widetilde{\Delta}^6_{\rm A}] + \i \rm{Im}[\widetilde{\Delta}^7_{\rm A}]\\
\rm{Im}[\widetilde{\Delta}^4_{\rm A}] - \i \rm{Im}[\widetilde{\Delta}^5_{\rm A}]
\end{array}
\mright) \;,\label{eq:Doublets2}
\end{align}
respectively, where the subscript $-1$ stands for the doublet with negative $\mathrm{T}^8$ eigenvalue.

\subsubsection{Scalar mass spectrum} 
\label{Sec:VacStab}

It is possible to write the minimisation conditions in a convenient way by recasting the scalar masses. 
In particular, the fundamental scalar masses can be collectively written as
\begin{align}
\label{eq:27scalars}
m^2_{\widetilde{\varphi}_i} =m^2_{\bm{27}} + c_1^i A_{\rm G} v + c_2^i A_{\rm F} v_{\rm F} \,,
\end{align}
where $c^i_{1,2}$ are constants with index $i$ running over all fundamental scalar eigenstates. 
For simplicity, the soft SUSY-breaking parameters and the family breaking VEV can be redefined in terms of a dimensionless parameter times a common scale $v$ as follows
\begin{align}
\label{eq:RelateTov}
v_{\rm F}	=	\beta v\,,~  
m_{\bm{27}}^{2}	=	\alpha_{\bm{27}} v^{2}\,, ~
A_{\rm G}	=	\sigma_{\rm G}v\,,~
A_{\rm F}	=	\sigma_{\rm F}v\,,
\end{align}
where, in the limit of low-scale SUSY-breaking, $\alpha_{\bm{27}},\,\sigma_{\rm G},\,\sigma_{\rm F}\ll 1$ and $\beta \sim {\cal O}\left( 1 \right)$ 
such that both gauge and family SSBs occur simultaneously at the GUT scale. Eq.~\eqref{eq:RelateTov} allows one to rewrite the scalar masses in terms of the common scale $v$
\begin{align}
\label{eq:27recast}
m^2_{\widetilde{\varphi}_i} = v^2  \left( \alpha_{\bm{27}} + c^i_1 \sigma_{\rm G} + c^i_2 \beta \sigma_{\rm F} \right) 
\equiv v^2 \omega_{\widetilde{\varphi}_i}  \,,
\end{align}
such that $\omega_{\widetilde{\varphi}_i} \ll 1$.
As the expression for the fundamental scalar masses contains three independent parameters, we may characterize the entire spectrum by the following three definitions 
\begin{align}
\label{eq:ScalarPars}
\omega_{\widetilde{H}^{(3)}} \equiv \xi\,,~ \omega_{\widetilde{E}_{\rm L,R}^{(1,2)}} \equiv \delta\,,~\omega_{\widetilde{H}^{(1,2)}} \equiv \kappa\,,
\end{align}
where the dimensionless parameters $\xi$, $\delta$ and $\kappa$ can span the entire spectrum by laying in the interval of 0 to 1, as the common mass scale is chosen to be the largest scale in the model, i.e.~the GUT scale $v$. With this, we can recast the scalar mass terms in the resulting EFT as
\begin{equation}
\label{eq:RedefMasses}
\begin{aligned}
m_{\widetilde{H}^{(3)}}^{2} 	&=	v^{2}\xi\,, \\
m_{\widetilde{E}_{\rm L,R}^{(3)}}^{2}	&=	v^{2}\left(\delta+\xi-\kappa\right)\,, \\
m_{\widetilde{\phi}^{(3)}}^{2}	&=	v^{2}\left(2\delta+\xi-2\kappa\right)\,,  \\
m_{\widetilde{\mathcal{Q}}_{\rm L,R}^{(3)}}^{2}	&=	\tfrac{1}{3}v^{2}\left(\delta+3\xi-\kappa\right)\,, \\
m_{\widetilde{D}_{\rm L,R}^{(3)}}^{2}	&=	\tfrac{1}{3}v^{2}\left(4\delta+3\xi-4\kappa\right)\,, \\
\end{aligned} \qquad
\begin{aligned}
m_{\widetilde{H}^{(1,2)}}^{2}	&=	v^{2}\kappa\,, \\
m_{\widetilde{E}_{\rm L,R}^{(1,2)}}^{2} 	&=	v^{2}\delta\,, \\
m_{\widetilde{\phi}^{(1,2)}}^{2}	&=	v^{2}\left(2\delta-\kappa\right)\,, \\
m_{\widetilde{\mathcal{Q}}_{\rm L,R}^{(1,2)}}^{2}	&=	\tfrac{1}{3}v^{2}\left(\delta+2\kappa\right)\,, \\
m_{\widetilde{D}_{\rm L,R}^{(1,2)}}^{2}	&=	\tfrac{1}{3}v^{2}\left(4\delta-\kappa\right)\,.
\end{aligned}
\end{equation}
Using Eq.~\eqref{eq:RedefMasses} the general set of conditions necessary to set the positivity of the fundamental scalar mass spectrum reads
\begin{eqnarray} \nonumber
\kappa > 0\, &\wedge& \, \Big[ \left( \dfrac{\kappa}{2} \leq \delta \leq \kappa \, \wedge \, \xi > -2 \delta + 2 \kappa  \right)\, \\ 
&\vee& \, 
\left( \delta > \kappa \, \wedge \, \xi > 0 \right) \Big]\,.
\label{eq:LQQcond}
\end{eqnarray}
Following the same procedure, we may redefine the parameters of the adjoint sector in terms of the GUT SSB scale $v$ as follows
\begin{equation}
\label{eq:RedefAdj}
\begin{aligned}
b_{\bm{1}}	&=	\tau_{\bm{1}} v^{2}\,, \\
b_{\bm{78}}	&=	\tau_{\bm{78}} v^{2}\,, \\
\mu_{\bm{1}} &= \alpha_{\bm{1}} v\,,   \\
\mu_{\bm{78}} &= \alpha_{\bm{78}} v\,, \\
\end{aligned} \qquad
\begin{aligned}
A_{\bm{1}}	&=	\sigma_{\bm{1}} v\,, \\
A_{\bm{78}}	&=	\sigma_{\bm{78}} v\,, \\
C_{\bm{78}}	&=	\theta_{\bm{78}} v\,.
\end{aligned}
\end{equation}
Substituting Eqs.~\eqref{eq:RedefAdj} in Tab.~\ref{table:SpecAdj} and, similarly to Eq.~\eqref{eq:27recast}, choosing
\begin{align}
\nonumber
&\omega_{\tilde{\mathcal{T}}_{\rm F}} \equiv \eta_{\rm F} \,,~ \omega_{\mathcal{H}_{\rm F}} \equiv \rho_{\rm F} \,,
~\omega_{\tilde{\mathcal{T}}^{\prime}_{\rm F}} \equiv \eta^{\prime}_{\rm F}\,,~ \omega_{\tilde{\mathcal{T}}_{\rm L,R}} \equiv \eta\,, \\
&~~~~~ \omega_{\tilde{\mathcal{H}}_{\rm L,R}} \equiv \rho\,,~ \omega_{\tilde{\Delta}^{\prime}_{\rm C}} \equiv \vartheta\,,\label{eq:ScalarPars}
\end{align}
where now $\omega_{\tilde{\varphi}_i \neq {\Scale[0.7]{\mathcal{H}_{\rm F}}} } \sim O(1)$ since only $\mathcal{H}_{\rm F}$ does not 
contain large $\mathcal{F}$- and $\mathcal{D}$-term contributions. Solving the system of equations w.r.t $\sigma_{\bm{1}},\, \tau_{\bm{1}},\, 
\alpha_{\bm{1}},\,\sigma_{\bm{78}},\, \tau_{\bm{78}},\, \alpha_{\bm{78}}$ we obtain
\begin{align}
&m_{\tilde{\mathcal{T}}_{\rm F}}^{2}	 = \eta_{\rm F}  v^{2}\,, \nonumber \\
&m_{\tilde{\mathcal{T}}^{\prime}_{\rm F}}^{2}	 = \eta^{\prime}_{\rm F}  v^{2}\,, \nonumber \\
&m_{\tilde{\mathcal{S}}_{\rm F}}^{2}	 = \tfrac{1}{6} v^{2} \left(\beta^2 \lambda^2_{\bm{1}} - 2 \eta_{\rm F} \right)\,,   \nonumber \\
&m_{\tilde{\mathcal{S}}^{\prime}_{\rm F}}^{2} = \tfrac{1}{6} v^{2} \left(\beta^2 \lambda^2_{\bm{1}} - 2 \eta^{\prime}_{\rm F} + 8 \rho_{\rm F} \right)\,, \nonumber \\
&m_{\mathcal{H}_{\rm F}}^{2} = \rho_{\rm F}  v^{2}\,, \nonumber \\
&m_{\tilde{\Delta}_{\rm C}}^{2}	 = \tfrac{1}{12}  v^{2} \left( 4 \eta - \lambda^2_{\bm{78}} \right)\,, \label{eq:RecastMassAdj} \\
&m_{\tilde{\mathcal{T}}_{\rm L,R}}^{2} = \eta v^{2}\,, \nonumber  \\
&m_{\tilde{\mathcal{T}}^{\prime}_{\rm L,R}}^{2} = \tfrac{1}{4} v^2 \left( \lambda^2_{\bm{78}} + 6 g^2_{\rm U} + 12 \vartheta - 8 \rho \right)\,, \nonumber \\
&m_{\tilde{\mathcal{S}}_{\rm L,R}}^{2} = \tfrac{1}{6} v^2 \left( \lambda^2_{\bm{78}} - 2 \eta \right)\,, \nonumber \\
&m_{\tilde{\mathcal{S}}^{\prime}_{\rm L,R}}^{2} = \tfrac{1}{12} v^2 \left(\lambda^2_{\bm{78}} - 18 g^2_{\rm U} - 12 \vartheta + 24 \rho \right) \,, \nonumber \\
&m_{\mathcal{H}_{\rm L,R}}^{2} = \rho  v^{2}\,, \nonumber \\
&m_{\tilde{\Delta}^{\prime}_{\rm C}}^{2} = \vartheta  v^{2}\,. \nonumber 
\end{align}

The scalar field components of the gauge and family adjoint sectors are treated separately. Noting that $\rho_\mathrm{F} \ll 1$, the general stability condition 
for the masses of the family sector read
\begin{equation}
\label{eq:Fcond}
\rho_{\rm F} \geq 0 \, \wedge \,  \left( \eta^{\prime}_{\rm F} > 4 \rho_{\rm F} \, \wedge \, x > 2  \eta^{\prime}_{\rm F} - 8  \rho_{\rm F} \, \wedge \,  
\eta_{\rm F} < \dfrac{x}{2} \right) \,,
\end{equation}
where we have defined $\beta^2 \lambda^2_{\bm{1}} \equiv x > 0$. Finally, the positivity conditions for the gauge sector are
\begin{eqnarray} \nonumber
&& \eta > 0 \, \wedge \, 2 \eta < y < 4 \eta \, \wedge \, \vartheta > 0 \,\wedge \\
&& \dfrac{1}{24} \left( z - y + 12 \vartheta \right) < \rho < \dfrac{1}{8} \left( y + 6 z + 12 \vartheta \right)\,,\label{eq:LRcond}
\end{eqnarray}
where we have defined $ \lambda^2_{\bm{78}} \equiv y > 0$ and $g^2_{\rm U} \equiv z > 0$. When conditions \eqref{eq:LQQcond}, \eqref{eq:Fcond} 
and \eqref{eq:LRcond} are simultaneously satisfied, the tree-level vacuum of the SHUT model is stable.

\onecolumngrid

\begin{table}[htb]
\ra{1.4}
\centering
\begin{tabular}{@{}lll@{}}
\hline
\small{d.o.f.'s}&\qquad\small{$(mass)^2$}&\small{Scalar components}\tabularnewline
\hline
8		& $m^2_{\bm{27}} - \tfrac{1}{\sqrt{6}} \left( A_\mathrm{G} v  + 2  A_\mathrm{F} \vev{F} \right) $ & $\widetilde{\nu}_{\rm R}^{(3)}\,, 
\widetilde{e}_{\rm R}^{(3)}\,, \widetilde{\nu}_{\rm L}^{(3)}\,, \widetilde{e}_{\rm L}^{(3)}$ \tabularnewline
2		& $m^2_{\bm{27}} - \tfrac{1}{\sqrt{6}} \left( 4 A_\mathrm{G} v  +  2 A_\mathrm{F} \vev{F} \right) $ & $\widetilde{\phi}^{(3)}$ \tabularnewline
8		& $m^2_{\bm{27}} + \tfrac{1}{\sqrt{6}} \left( 2 A_\mathrm{G} v  - 2  A_\mathrm{F} \vev{F} \right) $ & $H_{11}^{(3)}\,, H_{21}^{(3)} \, , 
H_{12}^{(3)}\,, H_{22}^{(3)}$ \tabularnewline
4		& $m^2_{\bm{27}} - \tfrac{1}{\sqrt{6}} \left( 4 A_\mathrm{G} v  -  A_\mathrm{F} \vev{F} \right) $ & $\widetilde{\phi}^{(1,2)}$ \tabularnewline
16		& $m^2_{\bm{27}} - \tfrac{1}{\sqrt{6}} \left( A_\mathrm{G} v  -  A_\mathrm{F} \vev{F} \right) $ & $\widetilde{\nu}_{\rm R}^{(1,2)}\,, 
\widetilde{e}_{\rm R}^{(1,2)}\,, \widetilde{\nu}_{\rm L}^{(1,2)}\,, \widetilde{e}_{\rm L}^{(1,2)}$ \tabularnewline
16		& $m^2_{\bm{27}} + \tfrac{1}{\sqrt{6}} \left( 2 A_\mathrm{G} v  +  A_\mathrm{F} \vev{F} \right) $  & $H_{11}^{(1,2)}\,, H_{21}^{(1,2)} \, , 
H_{12}^{(1,2)}\,, H_{22}^{(1,2)}$ \tabularnewline
24		& $m^2_{\bm{27}} + \tfrac{1}{\sqrt{6}} \left( A_\mathrm{G} v  - 2  A_\mathrm{F} \vev{F} \right) $  & $\widetilde{u}_{\rm L}^{(3)}\,,
\widetilde{d}_{\rm L}^{(3)}\,,\widetilde{u}_{\rm R}^{(3)}\,,\widetilde{d}_{\rm R}^{(3)}$ \tabularnewline
12		& $m^2_{\bm{27}} - \tfrac{1}{\sqrt{6}}\left( 2 A_\mathrm{G} v  + 2  A_\mathrm{F} \vev{F} \right) $ & $\widetilde{D}_{\rm L}^{(3)}\,, 
\widetilde{D}_{\rm R}^{(3)}$ \tabularnewline
48		& $m^2_{\bm{27}} + \tfrac{1}{\sqrt{6}} \left( A_\mathrm{G} v  +  A_\mathrm{F}\vev{F} \right) $  & $\widetilde{u}_{\rm L}^{(1,2)}\,,
\widetilde{d}_{\rm L}^{(1,2)}\,,\widetilde{u}_{\rm R}^{(1,2)}\,,\widetilde{d}_{\rm R}^{(1,2)}$ \tabularnewline
24               	& $m^2_{\bm{27}} - \tfrac{1}{\sqrt{6}} \left( 2 A_\mathrm{G} v  -  A_\mathrm{F} \vev{F} \right) $      & $\widetilde{D}_{\rm L}^{(1,2)}\,, 
\widetilde{D}_{\rm R}^{(1,2)}$  \tabularnewline
\hline
\end{tabular}
\caption[ ]{\it Scalar masses squared in the SHUT model for fields in the fundamental (tri-triplet) representation of 
the $[\SU{3}{}]^3\times \SU{3}{F}$ symmetry.}
\label{table:SpecFund}
\end{table}
\begin{table}[htb]
\ra{1.4}
\centering
\begin{tabular}{@{}lll@{}}
\hline
\small{d.o.f.'s}&\qquad\small{$(mass)^2$}&\small{Label}\tabularnewline
\hline
12		& $0$ & $\widetilde{G}_{\rm L,R,F}$ \tabularnewline
3		& $\sqrt{\tfrac{3}{2}} \tfrac{\vev{F}}{2} \left( 3 \lambda_{\bm{1}} \mu_{\bm{1}} + A_{\bm{1}} \right) $ & $\widetilde{\mathcal{T}}_{\rm F}$ \tabularnewline
1		& $ \tfrac{\vev{F}}{12} \left(  2 \vev{F} \lambda^2_{\bm{1}} - 3\sqrt{6}\lambda_{\bm{1}} \mu_{\bm{1}} - \sqrt{6} A_{\bm{1}}  \right) $ 
& $\widetilde{\mathcal{S}}_{\rm F}$ \tabularnewline
1		& $ -2 b_{\bm{1}} + \tfrac{\vev{F}}{12} \left(  \sqrt{6} \lambda_{\bm{1}} \mu_{\bm{1}} + 3 \sqrt{6} A_{\bm{1}} \right) $ & $\widetilde{\mathcal{S}}_F^{\prime}$ \tabularnewline
4		& $ -2 b_{\bm{1}} + \tfrac{\vev{F}}{12} \left( 2 \sqrt{6} \lambda_{\bm{1}} \mu_{\bm{1}} - \vev{F} \lambda^2_{\bm{1}} + 2 \sqrt{6} A_{\bm{1}}\right)$ 
& $\mathcal{H}_{\rm F}$ \tabularnewline
3		& $ -2 b_{\bm{1}} + \tfrac{\vev{F}}{12} \left( 5 \sqrt{6} \lambda_{\bm{1}} \mu_{\bm{1}} + 2 \vev{F} \lambda^2_{\bm{1}} - \sqrt{6} A_{\bm{1}}\right)$  
& $\widetilde{\mathcal{T}}_{\rm F}^{\prime}$ \tabularnewline
6		& $ \sqrt{\tfrac{3}{2}} \tfrac{v}{2} \left( 3 \lambda_{\bm{78}} \mu_{\bm{78}} + A_{\bm{78}} + 3 C_{\bm{78}} \right)  $  & $\widetilde{\mathcal{T}}_{\rm L,R}$ \tabularnewline
8		& $ \tfrac{v}{12} \left( - v \lambda^2_{\bm{78}} + 3 \sqrt{6} \lambda_{\bm{78}}\mu_{\bm{78}} + \sqrt{6} A_{\bm{78}} + 3 \sqrt{6} C_{\bm{78}} \right) $ 
& $\rm{Re}[\widetilde{\Delta}^{1,\cdots,8}_{\rm C}]$ \tabularnewline
2		& $\tfrac{v}{12} \left( 2 v \lambda^2_{\bm{78}} - 3 \sqrt{6} \lambda_{\bm{78}}\mu_{\bm{78}} - \sqrt{6} A_{\bm{78}} - 3 \sqrt{6} C_{\bm{78}} \right) $  
& $\widetilde{\mathcal{S}}_{\rm L,R}$ \tabularnewline
2               	& $ - 2 b_{\bm{78}} + \tfrac{\sqrt{6}}{12} v \left( \lambda_{\bm{78}} \mu_{\bm{78}} + 3 A_{\bm{78}} + C_{\bm{78}} \right) $    
& $\widetilde{\mathcal{S}}_{\rm L,R}^{\prime}$  \tabularnewline
8              	& $ - 2 b_{\bm{78}} + \tfrac{3}{4} g^2_{\rm U} v^2 + \tfrac{v^2}{12}\lambda^2_{\bm{78}} + \tfrac{\sqrt{6}}{6} v \left(\lambda_{\bm{78}} 
\mu_{\bm{78}} + A_{\bm{78}} + C_{\bm{78}} \right) $    & $\mathcal{H}_{\rm L,R}$  \tabularnewline
8              	& $ - 2 b_{\bm{78}} - \tfrac{v^2}{12}\lambda^2_{\bm{78}} + \tfrac{\sqrt{6}}{12} v \left( 3 \lambda_{\bm{78}} \mu_{\bm{78}} + 
A_{\bm{78}} + 3 C_{\bm{78}} \right) $    & $ \rm{Im}[\widetilde{\Delta}^{1,\cdots,8}_{\rm C}] $  \tabularnewline
6              	& $ - 2 b_{\bm{78}} + \tfrac{v^2}{6}\lambda^2_{\bm{78}} + \tfrac{\sqrt{6}}{12} v \left( 5 \lambda_{\bm{78}} \mu_{\bm{78}} -  
A_{\bm{78}} + 5 C_{\bm{78}} \right) $    & $\widetilde{\mathcal{T}}_{\rm L,R}^{\prime}$  \tabularnewline
\hline
\end{tabular}
\caption[ ]{\it Scalar masses squared in the SHUT model for fields in the adjoint representations of the $\SU{3}{L,R,C,F}$ symmetries.}
\label{table:SpecAdj}
\end{table}
\twocolumngrid

\subsection{Fermion Masses}

The masses of the fermions that originate from the gauge-adjoint sector are somewhat more complicated. For the sake of simplicity, we use a shortened 
notation and show the exact expressions for the fermion masses squared in Tab.~\ref{table:FermiSpect}. 

In particular, we parametrize the octet masses by $X^{\bm{8}}_{\rm C}$, $Y^{\bm{8}}_{\rm C}$ and $Z^{\bm{8}}_{\rm C}$, where the number in the superscript 
denotes the representation under the symmetry labeled in the subscript. The explicit form of such parameters reads
\begin{align}
X^{\bm{8}}_{\rm C} =& 4 M^2_0 +2 M^{\prime 2}_0 + \mu^2_{\bm{78}}\,, \label{eq:gluinos1} \\
Y^{\bm{8}}_{\rm C} =& 4 M^{\prime 2}_0 \left( 2 M_0 + \mu_{\bm{78}} \right)^2\,, \label{eq:gluinos2} \\
Z^{\bm{8}}_{\rm C} =& \left( \mu^2_{\bm{78}} - 4 M^2_0 \right)^2\,. \label{eq:gluinos3}
\end{align}
The singlet and triplet fermion masses depend on the $X^{\bm{1,3}}_{\rm{L,R}}$, $Y^{\bm{1,3}}_{\rm{L,R}}$ and $Z^{\bm{1,3}}_{\rm{L,R}}$ parameters 
which are given by
\begin{eqnarray}
X^{\bm{1,3}}_{\rm{L,R}} &=& \Big[ 2 v^2 \lambda^2_{\bm{78}} \mp 4 \sqrt{6} v \lambda_{\bm{78}} \mu_{\bm{78}} + 12 \Big( 4 M^2_0 \nonumber\\
&+& 2 M^{\prime 2}_0 + \mu^2_{\bm{78}} \Big)\Big]\,,  \label{eq:singlinos1} 
\end{eqnarray}
\begin{eqnarray}
Y^{\bm{1,3}}_{\rm{L,R}} &=& \Big[ \pm 2 \sqrt{6}  v \lambda_{\bm{78}} \mu_{\bm{78}} -  v^2 \lambda^2_{\bm{78}} - 6 \Big( 4 M_0^2 \nonumber\\
&+& 2 M_0^{\prime 2} +  \mu^2_{\bm{78}}  \Big) \Big]^2\,, \label{eq:singlinos2} 
\end{eqnarray}
\begin{eqnarray}
Z^{\bm{1,3}}_{\rm{L,R}} &=& 192 \Big[ 3 M_0^{\prime 4} \pm 2 M_0 M_0^{\prime 2} \left( \sqrt{6} v \lambda_{\bm{78}} \mp 6 \mu_{\bm{78}} \right) \nonumber \\  
&+& 2 M_0^2 \left( v^2 \lambda^2_{\bm{78}} \mp 2 \sqrt{6} v \lambda_{\bm{78}} \mu_{\bm{78}} + 
6 \mu^2_{\bm{78}} \right) \Big]\,. \label{eq:singlinos3}
\end{eqnarray}

For the new doublet fermions, the mass eigenstates are written in terms of $X^{\bm{2}}_{\rm{L,R}}$, 
$Y^{\bm{2}}_{\rm{L,R}}$ and $Z^{\bm{2}}_{\rm{L,R}}$ which read
\begin{align}
 X^{\bm{2}}_{\rm{L,R}} =& 96 M_0^2 + 48 M_0^{\prime 2} + 36 v^2 g^2_{\rm U} + v^2 \lambda^2_{\bm{78}} \nonumber \\ 
 -& 4 \sqrt{6} v \lambda_{\bm{78}} \mu_{\bm{78}} + 24 \mu^2_{\bm{78}}\,, \label{eq:Higgsinos1} 
 \end{align} 
\begin{align}
Y^{\bm{2}}_{\rm{L,R}} =& v^4 \lambda^4_{\bm{78}} - 8 \sqrt{6} v^3 \lambda^3_{\bm{78}} \mu_{\bm{78}} + 24 v^2 \lambda^2_{\bm{78}} \Big( 4 M_0^{\prime 2} \nonumber \\
-& 8 M^2_0 + 3 v^2 g^2_{\rm U} + 6 \mu^2_{\bm{78}}\Big)\,, \label{eq:Higgsinos2} 
\end{align}
\begin{align}
Z^{\bm{2}}_{\rm{L,R}} =& 96 \Big\{ 6 \left[ 4 M^{\prime 2}_0 + \left( \mu_{\bm{78}} - 2 M_0 \right)^2 \right] \Big[ 3 v^2 g^2_{\rm U} \nonumber \\
+& \left( \mu_{\bm{78}} + 2 M_0 \right)^2\Big] + \sqrt{6} v \lambda_{\bm{78}} \Big( 6 v^2 g^2_{\rm U} M_0   \nonumber \\
-& 8 M_0 M^{\prime 2}_0 + 
8 M_0^2 \mu_{\bm{78}} - 4 M_0^{\prime 2} \mu_{\bm{78}} \nonumber \\
-& 3 v^2 g^2_{\rm U} \mu_{\bm{78}} - 2 \mu^3_{\bm{78}} \Big) \Big\}\,. \label{eq:Higgsinos3}
\end{align}

\onecolumngrid

\begin{table}[htb]
\ra{1.4}
\centering
\scalebox{1.0}{
\begin{tabular}{@{}lll@{}}
\hline
\small{\# of Weyl spinors}&\small{$(mass)^2$}&\small{Fermionic components}\tabularnewline
\hline
81		& $0$ & $\small{\phi^{(1,2,3)}\,,\widetilde{H}^{(1,2,3)}\,,E^{(1,2,3)}_{\rm L,R}\,,\mathcal{Q}^{(1,2,3)}_{\rm L,R}}\,,D^{(1,2,3)}_{\rm L,R}$ \tabularnewline
1		& $\tfrac{1}{6} \left( \vev{F}^2 \lambda^2_{\bm{1}} - 2 \sqrt{6} \vev{F} \lambda_{\bm{1}} \mu_{\bm{1}} + 6 \mu^2_{\bm{1}} \right)$ 
& $ \Delta^{8}_{\rm F}\equiv \mathcal{S}_{\rm F}$ \tabularnewline
3		& $ \tfrac{1}{6} \left( \vev{F}^2 \lambda^2_{\bm{1}} + 2 \sqrt{6} \vev{F} \lambda_{\bm{1}} \mu_{\bm{1}} + 6 \mu^2_{\bm{1}} \right) $ 
& $\Delta^{1,2,3}_{\rm F} \equiv \mathcal{T}_{\rm F}$ \tabularnewline
4		& $\tfrac{1}{24} \left( \vev{F}^2 \lambda^2_{\bm{1}} - 4 \sqrt{6} \vev{F} \lambda_{\bm{1}} \mu_{\bm{1}} + 24 \mu^2_{\bm{1}} \right)$ 
& $\Delta^{4,5,6,7}_{\rm F}\equiv \widetilde{\mathcal{H}}_{\rm F}$ \tabularnewline
8		& $\tfrac{1}{2} \left( X^{\bm{8}}_{\rm C} - \sqrt{Y^{\bm{8}}_{\rm C} + Z^{\bm{8}}_{\rm C}} \right)$  & $c_{\theta_{\bm{8}}} \widetilde{\lambda}^a_{\rm C} - 
s_{\theta_{\bm{8}}} \Delta^a_{\rm C} \equiv \widetilde{g}^a$ \tabularnewline
8		& $\tfrac{1}{2} \left( X^{\bm{8}}_{\rm  C} + \sqrt{Y^{\bm{8}}_{\rm C} + Z^{\bm{8}}_{\rm C}} \right)$  & $s_{\theta_{\bm{8}}} \widetilde{\lambda}^a_{\rm C} + 
c_{\theta_{\bm{8}}} \Delta^a_{\rm C} \equiv \widetilde{g}^{a}_{\perp}$ \tabularnewline
2		& $\tfrac{1}{24} \left( X^{\bm{1}}_{\rm L,R} - \sqrt{ Y^{\bm{1}}_{\rm L,R} + Z^{\bm{1}}_{\rm L,R}} \right)$  
& $c_{\theta_{\bm{1}}} \widetilde{\lambda}^8_{\rm L,R} - s_{\theta_{\bm{1}}} \Delta^8_{\rm L,R} \equiv \mathcal{S}_{\rm L,R}$ \tabularnewline
2		& $\tfrac{1}{24} \left( X^{\bm{1}}_{\rm L,R} + \sqrt{Y^{\bm{1}}_{\rm L,R} + Z^{\bm{1}}_{\rm L,R}} \right)$  
& $s_{\theta_{\bm{1}}} \widetilde{\lambda}^8_{\rm L,R} + c_{\theta_{\bm{1}}} \Delta^8_{\rm L,R} \equiv \mathcal{S}^{\perp}_{\rm L,R}$ \tabularnewline
6		& $\tfrac{1}{24} \left( X^{\bm{3}}_{\rm L,R} - \sqrt{ Y^{\bm{3}}_{\rm L,R} + Z^{\bm{3}}_{\rm L,R}} \right)$  
& $c_{\theta_{\bm{3}}} \widetilde{\lambda}^{1,2,3}_{\rm L,R} - s_{\theta_{\bm{3}}} \Delta^{1,2,3}_{\rm L,R} \equiv \mathcal{T}_{\rm L,R}$ \tabularnewline
6		& $\tfrac{1}{24} \left( X^{\bm{3}}_{\rm L,R} + \sqrt{Y^{\bm{3}}_{\rm L,R} + Z^{\bm{3}}_{\rm L,R}} \right)$  
& $s_{\theta_{\bm{3}}} \widetilde{\lambda}^{1,2,3}_{\rm L,R} + c_{\theta_{\bm{3}}} \Delta^{1,2,3}_{\rm L,R} \equiv \mathcal{T}^{\perp}_{\rm L,R}$ \tabularnewline
8		& $\tfrac{1}{48}\left( X^{\bm{2}}_{\rm L,R} - \sqrt{ Y^{\bm{2}}_{\rm L,R} + Z^{\bm{2}}_{\rm L,R}} \right)$  
& $\varrho_1 \Delta^{4,6}_{\rm L,R} + \varrho_2 \Delta^{5,7}_{\rm L,R} + \varrho_3 \widetilde{\lambda}^{4,6}_{\rm L,R} + 
\varrho_4 \widetilde{\lambda}^{5,7}_{\rm L,R} \equiv \widetilde{\mathcal{H}}^{1,2}_{\rm L,R}$ \tabularnewline
8		& $\tfrac{1}{48}\left( X^{\bm{2}}_{\rm L,R} + \sqrt{Y^{\bm{2}}_{\rm L,R} + Z^{\bm{2}}_{\rm L,R}} \right)$  
& $\overline{\varrho}_1 \Delta^{4,6}_{\rm L,R} + \overline{\varrho}_2 \Delta^{5,7}_{\rm L,R} + \overline{\varrho}_3 \widetilde{\lambda}^{4,6}_{\rm L,R} + 
\overline{\varrho}_4 \widetilde{\lambda}^{5,7}_{\rm L,R} \equiv \widetilde{\mathcal{H}}^{1,2\,\perp}_{\rm L,R}$ \tabularnewline
\hline
\end{tabular}
}
\caption[ ]{\it Fermion masses squared and left singular eigenvectors in the SHUT model. The $c_{\theta_{\bm{R}}}$ and $s_{\theta_{\bm{R}}}$ coefficients 
denote the cosine and sine of the $2 \times 2$ mixing angles for the representation $\bm{R}$. Here, $\varrho_{1,2,3,4}$ and $\overline{\varrho}_{1,2,3,4}$ 
are coefficients that parametrize a unitary mixing. The fermion masses, for a given irrep $\bm{R}$ and gauge group $\rm{L,R,C}$, are determined in terms of the 
$X^{\bm{R}}_{\rm  A}$, $Y^{\bm{R}}_{\rm  A}$ and $Z^{\bm{R}}_{\rm  A}$ coefficients, with explicit expressions given in 
Eqs.~\eqref{eq:gluinos1}-\eqref{eq:Higgsinos3}.}
\label{table:FermiSpect}
\end{table}

\twocolumngrid

Note that the doublets $\widetilde{\mathcal{H}}_{\rm A}$, which are the left-handed Weyl fermions defined to transform as $\bm{2}_{1}$, 
form mass terms of the form $m\widetilde{\mathcal{H}}_{\rm A} \widetilde{\mathcal{H}}^{\prime}_{\rm A}$ with $\widetilde{\mathcal{H}}^{\prime}_{\rm A}$ 
being also the left-handed Weyl fermions transforming as $\bm{2}_{-1}$. 


\subsection{Gauge boson masses}

The gauge bosons of the $\SU{3}{C}$ group remain massless and are identified with the SM gluons whereas the massive gauge bosons are generated 
upon the SSB of the $\SU{3}{L,R}$ symmetries. The covariant derivative of the GUT symmetry reads
\begin{align}
\bm{D}^{\mu} =& \del^{\mu} \id{L} \So{\otimes} \id{R} \So{\otimes} \id{C} - \i g_{\rm U} 
\sum^{8}_{a=1} \Big[ G^{\mu a}_{\rm L} \bm{T}_{\rm L}^a \So{\otimes} \id{R} \So{\otimes} \id{C}  \nonumber \\
+ &
G^{\mu a}_{\rm R} \bm{T}_{\rm R}^a \So{\otimes} \id{L} \So{\otimes} \id{C} + 
G^{\mu a}_{\rm C} \bm{T}_{\rm C}^a \So{\otimes} \id{L} \So{\otimes} \id{R}\Big] \,,
\label{eq:covDeriv}
\end{align}
where $G^{\mu a}_{\rm L}$ are the gauge fields of the $\SU{3}{L}$ symmetry which cyclically transform into $G^{\mu a}_{\rm R}$ and $G^{\mu a}_{\rm C}$ 
by means of $\mathbb{Z}_3$-permutations. Considering the gauge-breaking VEVs \mbox{$\left \langle \widetilde{\Delta}^c_{\rm L,R} \right 
\rangle = \delta^c_8\, v$}, the relevant kinetic terms that couple the vector and scalar fields evaluated in the vacuum of the theory are given by
\begin{align}
\label{eq:covDerivVEV2}
\left \lvert D^{\mu} \left \langle \widetilde{\Delta}^b_{\rm L,R} \right\rangle \right\rvert^2 = \dfrac{3}{4} g^2_{\rm U} v^2 
\sum^{7}_{a=4} \eta_{\mu \nu} G^{\mu a}_{\rm L,R} G^{\nu a}_{\rm L,R}\,.
\end{align}
Therefore, there are eight massive gauge bosons in the model which transform as complex $\bm{2}_{1}$ representations of $\SU{2}{L,R} \times \U{L,R}$ 
whose charge eigenstates read
\begin{align}
\label{eq:GaugeBosonDoublets}
\renewcommand\arraystretch{1.3}
\mathcal{G}^{\mu}_{\rm L,R}\equiv \dfrac{1}{\sqrt{2}} \mleft(
\begin{array}{ccc}
G^{\mu 5}_{\rm L,R} + \i G^{\mu 4}_{\rm L,R}\\
G^{\mu 7}_{\rm L,R} + \i G^{\mu 6}_{\rm L,R}
\end{array}
\mright)\;,
\end{align}
with mass $m^2_{\mathcal{G}} = \tfrac{3}{4} g^2_{\rm U}\, v^2 $. In addition to the unbroken colour sector, the remaining gauge bosons are also massless 
at the SHUT SSB scale.

\section{Gauge couplings: $\beta$-functions and matching conditions} 
\label{sec:betabeta}

In general, the one-loop $\beta$-function for a gauge coupling is given by \cite{Luo:2002ti}
\begin{eqnarray}
\label{bebe}
\beta(g_i)&=&-\frac{g_i^3}{(4\pi)^2}\Big(\frac{11}{3}C_2(G) - \frac{4}{3}\kappa S_2(F) \nonumber \\
&-& \frac{1}{3}S_2(S)\Big)\equiv\frac{b_ig_i^3}{(4\pi)^2},
\end{eqnarray}


where $\kappa=1/2$ for Weyl fermions, $C_2(G)=N$ is the Casimir index, $S_2(F)$ is the Dynkin index for a fermion and $S_2(S)$ is the Dynkin index for a complex scalar. The one-loop $\beta$-function for the gauge coupling of a $\mathrm{U}(1)$ theory reads 

\begin{equation}
\label{2}
\beta(\tilde{g}_i)=\frac{\tilde{g}_i^3}{12\pi^2}\left(\kappa\sum_f Q_f^2 + \frac{1}{4}\sum_s Q_s^2\right)\equiv\frac{b_i\tilde{g}_i^3}{(4\pi)^2}.
\end{equation}

where again $\kappa$ is equal to $1/2$ for Weyl fermions, and where $Q_f$ and $Q_s$ are, respectively, the charges for all fermions and scalars in the theory. 

Rewriting the gauge couplings in terms of the inverse of the structure constants, $\alpha^{-1}=4\pi/g^2$, the solutions of \eqref{bebe} and \eqref{2} reads

\begin{equation}
\label{eqeq}
\alpha_i^{-1}(\mu_2)= \alpha_i^{-1}(\mu_1) - \frac{b_i}{2\pi}\log\left(\frac{\mu_2}{\mu_1}\right),
\end{equation}

where the $b_i$-coefficients are dependent on the number of particles and respective charges of a given EFT. Below, we specify such information for each of the four regions and provide the corresponding results for the one-loop $\beta$-functions.

\subsection{Region $\RN{1}$}\label{subsecRegion1}

As discussed in Sec.~\ref{sec:massespres}, all components of the fundamental scalars and fermions remain in the spectrum after the breaking of the T-GUT symmetry. In this region, the fermion sector also contains two adjoint triplets, $\mathcal{T}_\mathrm{L,R}$, two adjoint singlets, $\mathcal{S}_\mathrm{L,R}$ and one adjoint octet in color $\widetilde{g}^a$. Here adjoint triplets/doublets/singlets refers to triplet/doublet/singlet representations coming from an $\mathrm{SU}(3)$ octet. Heavy states, with masses the size of the T-GUT scale, are marked with a symbol $\perp$ in Tab.~\ref{table:FermiSpect} of Appendix~\ref{sec:masses}, and are integrated out. For the adjoint doublets, on the other hand, there is no distinct hierarchy between $\tilde{\mathcal{H}}_{\rm L,R},\tilde{\mathcal{H}}^{\dagger}_{\rm L,R}$ and their heavy counterparts, and can hence all be excluded from the spectrum. 

With this, there is a total of 18 fermions and 18 scalars in the fundamental/anti-fundamental rep of $\mathrm{SU}(3)_\mathrm{C}$, 18 fermions and 18 scalars in the fundamental/anti-fundamental rep of $\mathrm{SU}(2)_\mathrm{L}$ and $\mathrm{SU}(2)_\mathrm{R}$, one fermion and no scalars in the adjoint rep of $\mathrm{SU}(3)_\mathrm{C}$ and one fermion and no scalars in the adjoint rep of $\mathrm{SU}(2)_\mathrm{L}$ and $\mathrm{SU}(2)_\mathrm{R}$, resulting in

\vspace{-2mm}

\begin{equation}
\label{factor1}
b^{\RN{1}}_{g_{\mathrm{C}}}=0 \hspace{2mm} \mathrm{and}\hspace{2mm} b^{\RN{1}}_{g_{\mathrm{L,R}}}=3,
\end{equation}

with $b_i$ defined as $\beta(g_i)\equiv\frac{b_ig_i^3}{(4\pi)^2}$. Here ${g_{\mathrm{C}}}$ is the gauge coupling for $\mathrm{SU}(3)_{\mathrm{C}}$ and $g_{\mathrm{L,R}}$ is the gauge coupling for $\mathrm{SU}(2)_\mathrm{L}\times\mathrm{SU}(2)_\mathrm{R}$. 

For the $\mathrm{U}(1)_{\mathrm{L}}\times\mathrm{U}(1)_{\mathrm{R}}$ coupling, $\widetilde{g}_\mathrm{L,R}$, the $\beta$-function is calculated using the charges in Tab.~\ref{Table:EFT-content} of Appendix~\ref{sec:Symmetries}. With this we obtain

\vspace{-2mm}

\begin{equation}
\label{factor}
b^{\RN{1}}_{\widetilde{g}_\mathrm{L,R}}=9.
\end{equation}

\subsection{Region $\RN{2}$}\label{subsecRegion22}

In region $\RN{2}$, the adjoint scalars are integrated out, in addition to $D_\mathrm{L,R}$ in the second and third generation, which are the only fermions able to form a Dirac mass at this stage. When it comes to the fundamental scalars, there are no clear hierarchies in the spectrum, so here we will instead present the possible extreme values.  

As apparent from Eq.~\eqref{bebe} and \eqref{2}, the extreme values for each $b$ occur for the minimal- and maximal number of scalars, respectively. The maximal $b$-values are hence obtained when keeping all fundamental scalars, while the minimal $b$-values correspond to keeping only $H^f$, $\widetilde{E}_\mathrm{R}^f$ and $\widetilde{\phi}^f$. The latter scenario cannot be further reduced, as $H^f$ is required to remain as it contains the minimal amount of Higgs $\mathrm{SU}(2)_\mathrm{L}$-doublets required for Cabbibo mixing at tree-level ($H_u^{1,2}$ and $H_d^2$), while $\widetilde{E}_\mathrm{R}^f$ and $\widetilde{\phi}^f$ are required as they are involved in the breaking scheme down to the SM. 

With this, the $b$-values lie in the following intervals

\begin{align}
\nonumber
&-\frac{19}{3}\leq b^{\RN{2}}_{g_\mathrm{C}}\leq-\frac{10}{3},\;\;\;\; -\frac{2}{3} \leq b^{\RN{2}}_{g_\mathrm{L}}\leq\frac{5}{3},\\
&-\frac{1}{3} \leq b^{\RN{2}}_{g_\mathrm{R}}\leq\frac{5}{3},\;\;\;\;\frac{31}{3}\leq b^{\RN{2}}_{\widetilde{g}_\mathrm{L+R}}\leq\frac{46}{3},
\label{equationRegion2}
\end{align}

where hence the upper bound corresponds to the maximal field content and the lower bound to the minimal field content. 

\subsection{Region $\RN{3}$}\label{subsecRegion3}

In region $\RN{3}$, the fermion spectrum remains the same, while for the scalar sector we once again investigate the extreme values. The maximal field content is still to keep all fundamental scalars, while for the minimal field content we may now remove $\widetilde{E}_\mathrm{L}^2$, as $\mathrm{SU}(2)_\mathrm{F}$ is broken and only $\widetilde{E}_\mathrm{L}^1$ is involved in the breaking scheme down to the SM. 

With this, all $b$-values are identical to those in region $\RN{2}$, apart from the lower bound of $b_{\widetilde{g}_\mathrm{L+R}}$

\begin{align}
\nonumber
&-\frac{19}{3}\leq b^{\RN{3}}_{g_\mathrm{C}}\leq-\frac{10}{3},\;\;\;\; -\frac{2}{3} \leq b^{\RN{3}}_{g_\mathrm{L}}\leq\frac{5}{3},\\
&-\frac{1}{3} \leq b^{\RN{3}}_{g_\mathrm{R}}\leq\frac{5}{3},\;\;\;\;\frac{59}{6}\leq b^{\RN{3}}_{\widetilde{g}_\mathrm{L+R}}\leq\frac{46}{3},  \label{equationRegion3}
\end{align}

where again the upper bound corresponds to the maximal field content and the lower bound to the minimal field content. 

\subsection{Region $\RN{4}$}\label{subsecRegion2}

In region $\RN{4}$, the minimal field content corresponds to integrating out all scalars apart from three Higgs doublets, e.g. $H_u^{1,2}$ and $H_d^2$ and the field responsible for breaking the $\mathrm{U}(1)_\mathrm{T}$ symmetry, e.g.~$\widetilde{\phi}^1$. A minimum of two Higgs doublets are required to remain in order for all SM particles to gain a mass, while a third is needed for getting the appropriate Cabbibo mixing at tree level, as discussed in Sec.~\ref{quarkquark}. 

Among the fermions, ${D}_\mathrm{L}^{1,2,3}$ $\mathcal{D}_\mathrm{R}^{1,2,3}$, $\nu_\mathrm{R}^{1,2,3}$, $\phi^{1,2,3}$ and all Higgsinos are integrated out, as they can form massive states without the Higgs VEV. This can be seen from Tab.~\ref{Table:EFT-content-snu1Phi} of Appendix~\ref{sec:Symmetries}  (with $\mathrm{U}(1)_\mathrm{T'}$ broken). The remainder of the fundamental fermions are kept in the spectrum. Regarding the adjoints, both the octets $\widetilde{g}^a$, and the triplets, $\mathcal{T}_\mathrm{L}^i,\mathcal{T}_\mathrm{R}^{\pm}$ are integrated out, resulting in

\vspace{-2mm}

\begin{equation}
\label{mm1}
b^{\RN{4}}_{g_\mathrm{C}}=-7 \hspace{2mm} \mathrm{and}\hspace{2mm} b^{\RN{4}}_{g_\mathrm{L}}=-\frac{17}{6},
\end{equation}

where $g_\mathrm{L}$ is the gauge coupling for $\mathrm{SU}(2)_\mathrm{L}$. 

For $\mathrm{U}(1)_\mathrm{Y}$, the charges in Tab.~\ref{Table:EFT-content-snu1Phi} of Appendix~\ref{sec:Symmetries}, results in

\vspace{-2mm}

\begin{equation}
\label{mm2}
b^{\RN{4}}_{\widetilde{g}_\mathrm{Y}}=\frac{43}{6}
\end{equation}

where $\widetilde{g}_\mathrm{Y}$ is the gauge coupling for $\mathrm{U}(1)_\mathrm{Y}$.

\subsection{Matching conditions}\label{subsecMatching}

The gauge couplings unification condition at the GUT scale reads

\begin{equation}
\label{MGUT}
\alpha^{-1}_{\widetilde{g}_\mathrm{L,R}}(v)=\alpha^{-1}_{{g}_\mathrm{L,R}}(v)=\alpha^{-1}_{g}(v),
\end{equation}

with the charges in Tab.~\ref{Table:EFT-content} of Appendix~\ref{sec:Symmetries}.

At the soft scale, the gauge coupling matching conditions are obtained by finding the gauge boson mass eigenstates after the VEVs $\langle\widetilde{\phi}^2\rangle$, $\langle\widetilde{\phi}^3\rangle$ and $\langle\widetilde{\nu}_\mathrm{R}^1\rangle$, respectively, by expanding our old basis in terms of the new one, e.g. $\{G_\mathrm{R}^3,B_\mathrm{L},B_\mathrm{R}\}$ in terms of $\{B_\mathrm{L+R}, ...\}$\footnote{Here, $G_\mathrm{R}^3$ is the gauge boson corresponding to the third generator of $\mathrm{SU}(2)_\mathrm{R}$, $B_\mathrm{L,R}$ the gauge bosons for $\mathrm{U}(1)_\mathrm{L,R}$ and $B_\mathrm{L+R}$ the gauge boson for $\mathrm{U}(1)_\mathrm{L+R}$.}.  With this we have

\begin{equation}
{\alpha}^{-1}_{\widetilde{g}_\mathrm{L+R}}(\langle\widetilde{\phi}^3\rangle)={\alpha}^{-1}_{\widetilde{g}_\mathrm{L}}(\langle\widetilde{\phi}^3\rangle) + {\alpha}^{-1}_{\widetilde{g}_\mathrm{R}}(\langle\widetilde{\phi}^3\rangle),
\end{equation}

at the $\langle\widetilde{\phi}^3\rangle$ scale, and  

\begin{equation}
{\alpha}^{-1}_{\widetilde{g}_\mathrm{Y}}(\langle\widetilde{\nu}_\mathrm{R}^1\rangle)={\alpha}^{-1}_{g_\mathrm{R}}(\langle\widetilde{\nu}_\mathrm{R}^1\rangle) + \frac{1}{3}{\alpha}^{-1}_{\widetilde{g}_\mathrm{L+R}}(\langle\widetilde{\nu}_\mathrm{R}^1\rangle),
\end{equation}

at the $\langle\widetilde{\nu}_\mathrm{R}^1\rangle$ scale, while the matching at the $\langle\widetilde{\phi}^2\rangle$ scale is trivial, ${\alpha}^{-1}_{\widetilde{g}_\mathrm{L+R}}(\langle\widetilde{\phi}^2\rangle)={\alpha}^{-1}_{\widetilde{g}_\mathrm{L+R}}(\langle\widetilde{\phi}^2\rangle)$.  

Finally, at the $Z$-boson mass scale, the matching conditions between the electromagnetic coupling, the hypercharge coupling and the $\mathrm{SU}(2)_\mathrm{L}$ coupling are already well-known

\begin{equation}
\label{betti}
{\alpha}^{-1}_{\widetilde{g}_\mathrm{Y}} = {\cos^2\theta_\mathrm{W}}{\alpha}^{-1}_{\mathrm{EM}}\;\;\;\;\mathrm{and}\;\;\;\; {\alpha}^{-1}_{\widetilde{g}_L} = {\sin^2\theta_\mathrm{W}}{\alpha}^{-1}_{\mathrm{EM}},
\end{equation}

\noindent where $\theta_\mathrm{W}$ is the weak mixing angle, $\sin^2(\theta_\mathrm{W})\sim 0.2312$ \cite{Patrignani:2016xqp}.

\section{Lagrangian of the LR-symmetric effective theory} 
\label{sec:FullEffectiveLs}

The field content of the EFT is derived from the mass spectrum after the T-GUT symmetry breaking. As a general rule, the light fields, i.e.~those with 
a mass scale much smaller than the GUT scale $v$, are kept in the EFT spectrum whereas those with masses of the same order of magnitude 
as $v$ are integrated out.

The light field components and their group transformations under the LR-symmetry obtained after $v$ and $v_{\rm F}$ VEVs (see Eq.~\eqref{eq:brk2}) 
are shown in Tab.~\ref{Table:EFT-content}, where we use the notation given in Eq.~\eqref{1}.

\subsection{The scalar potential of the LR-symmetric effective model}

The scalar potential of the effective LR-symmetric theory generated after the T-GUT breaking can be summarized by
\begin{align}
V_{\rm LR} = V_2 + V_3 + V_4\,,
\end{align}
where $V_2$, $V_3$ and $V_4$ denote the quadratic, cubic and quartic scalar self-interactions, respectively. For simplicity, we will suppress colour 
indices in $V_{\rm LR}$ and, for all those terms that can be written from LR-parity transformations on the fields, we will show them 
within square brackets as $\widehat{\mathcal{P}}_{\rm LR} [\cdots ]$. Note that here we use this notation for both the cases of invariance 
or not under LR-parity. For instance, while for the LR-parity symmetric case we should preserve the couplings, for 
the LR-parity broken case we should also read $m\to\bar{m},\, A\to \bar{A},\, \lambda \to \bar{\lambda}$ whenever 
LR-parity transformation is applied.

We start by writing the scalar mass terms,
\begin{align*}
\begin{split}
V_2  = & {m_H^2}H^{\ast\,r}_{f\,l} H^{f\,l}_{r} 
+m_h^2 h^{\ast\,r}_{l} h^l_{r}
+ m_{\phi}^2 \widetilde{\phi}^\ast_{f} \widetilde{\phi}^f
+ m_{\varphi}^2 \widetilde{\varphi}^\ast \widetilde{\varphi} 
\\
+& m_{\Delta}^2 \widetilde{\mathcal{H}}_{{\rm F} f}^{\ast} 
\widetilde{\mathcal{H}}_{\rm F}^f 
+ \widehat{\mathcal{P}}_{\rm LR} \left[ m_{E}^2\widetilde{E}_{{\rm L}\;f\,l}^\ast \widetilde{E}_{\rm L}^{f\,l}
+ m_{\mathcal{E}}^2\widetilde{\mathcal{E}}_{{\rm L}\;l}^\ast \widetilde{\mathcal{E}}_{\rm L}^{l}
\right . \\
+ &\left. m_{\mathcal{Q}}^2\widetilde{\mathcal{Q}}_{{\rm L}\,f}^{\ast\,l} 
\widetilde{\mathcal{Q}}_{{\rm L}\,l}^{f} 
+ m_{q}^2\widetilde{q}_{\rm L}^{\ast\,l} \widetilde{q}_{{\rm L}\,l}
+ m_{D}^2\widetilde{D}_{{\rm L}\;f}^\ast \widetilde{D}_{\rm L}^{f}
+ m_{\mathcal{B}}^2 \widetilde{\mathcal{B}}_{\rm L}^\ast \widetilde{\mathcal{B}}_{\rm L} \right] 
\end{split}
\end{align*}
whereas the trilinear interactions are expressed as
\begin{align}
\begin{split}
V_3  = & \varepsilon_{f f'}\Big\{ \widehat{\mathcal{P}}_{\rm LR} \Big[
 A_{1} \widetilde{\mathcal{Q}}_{{\rm R}}^{f\,r} h^{l}_{r} \widetilde{\mathcal{Q}}_{{\rm L}\,l}^{f'}
+A_{2} \widetilde{D}_{\rm R}^f \widetilde{\varphi} \widetilde{D}_{\rm L}^{f'} \Big]
\\&
+ \widehat{\mathcal{P}}_{\rm LR} \Big[A_{3} \widetilde{q}_{\rm R}^{r} H^{f\,l}_{r} \widetilde{\mathcal{Q}}_{{\rm L}\,l}^{f'}
+ A_4 \widetilde{\mathcal{B}}_{\rm R} \widetilde{\phi}^{f'} \widetilde{D}_{\rm L}^{f} 
\\&
+  A_{5} \widetilde{\mathcal{B}}_{\rm R}  \widetilde{\mathcal{Q}}_{{\rm L}\,l}^{f} \widetilde{E}_{\rm L}^{f'\,l}
+ A_{6} \widetilde{D}_{\rm R}^{f} \widetilde{\mathcal{Q}}_{{\rm L}\,l}^{f'} \widetilde{\mathcal{E}}_{\rm L}^{\,l}
\\&
+A_{7}\widetilde{D}_{\rm R}^f \widetilde{q}_{{\rm L}\,l} \widetilde{E}_{\rm L}^{f'\,l} + {\rm c.c.}
\Big] \Big\}
\end{split}
\end{align}

Due to a large number of possible contractions of four scalar fields in the effective LR-symmetric model, we will employ a condensed 
notation to express the scalar quartic self-interactions. We describe below the five possible types of terms. 

For the first type, which we denote ``sc1'', we consider terms with \emph{one} reoccurring index, where we define 
the reoccurring index as an index possessed by all the four fields. For such a combination there are three possible contractions, 
out of which two of them are linearly independent. In particular, we have
\begin{align}
V_{\rm sc1} \supset
&\lambda_{k_1}\widetilde{D}_{{\rm L}\,x\,f'}^\ast\widetilde{D}_{\rm L}^{x\,f'} H^{\ast\,r}_{f\,l} H^{f\,l}_{r}
+
\lambda_{k_2}\widetilde{D}_{{\rm L}\,x\,f'}^\ast\widetilde{D}_{\rm L}^{x\,f} H^{\ast\,r}_{f\,l} H^{f'\,l}_r
\nonumber \\
\equiv
&\lam{k_1}{k_2} \widetilde{D}_{{\rm L}\,f'}^\ast\widetilde{D}_{\rm L}^{f'} H^{\ast\,r}_{f\,l} H^{f\,l}_{r}\,,
\label{sc1}
\end{align}
where colour indices are suppressed in the condensed form. 

For terms with \emph{two} reoccurring indices, denoted as ``sc2'', no matter if they are $\SU{2}{}$ indices or $\SU{3}{}$ indices\footnote{The two types coincide 
since for $\SU{2}{}$ the three combinations reduce down to two, using that $\varepsilon_{ij}\varepsilon^{kl}=\delta_i^k\delta_j^l-\delta_i^l\delta_j^k$, while 
for $\SU{3}{}$ there are only two possible contractions to begin with, and no Levi-Civita tensor to impose a reduction.}, there are four linearly independent contractions 
that read
\begin{align}
&V_{\rm sc2} \supset \nonumber \\
&\lambda_{n_1}\widetilde{E}_{{\rm L}\;l'\,f'}^\ast \widetilde{E}_{\rm L}^{l'\,f'} \widetilde{\mathcal{\mathcal{Q}}}_{{\rm L}\,x\,f}^{\ast\,l} 
\widetilde{\mathcal{\mathcal{Q}}}_{{\rm L}\,l}^{x\,f}
+
\lambda_{n_2}\widetilde{E}_{{\rm L}\;l'\,f'}^\ast \widetilde{E}_{\rm L}^{l'\,f}\widetilde{\mathcal{\mathcal{Q}}}_{{\rm L}\,x\,f}^{\ast\,l} 
\widetilde{\mathcal{\mathcal{Q}}}_{{\rm L}\,l}^{x\,f'}
\nonumber \\
&+
\lambda_{n_3}\widetilde{E}_{{\rm L}\;l'\,f'}^\ast \widetilde{E}_{\rm L}^{l\,f'}\widetilde{\mathcal{\mathcal{Q}}}_{{\rm L}\,x\,f}^{\ast\,l'} 
\widetilde{\mathcal{\mathcal{Q}}}_{{\rm L}\,l}^{x\,f}
+
\lambda_{n_4}\widetilde{E}_{{\rm L}\;l'\,f'}^\ast \widetilde{E}_{\rm L}^{l\,f}\widetilde{\mathcal{\mathcal{Q}}}_{{\rm L}\,x\,f}^{\ast\,l'} 
\widetilde{\mathcal{\mathcal{Q}}}_{{\rm L}\,l}^{x\,f'}
\nonumber \\
&\equiv
\lam{n_1}{n_4} \widetilde{E}_{{\rm L}\;l'\,f}^\ast \widetilde{E}_{\rm L}^{l'\,f} \widetilde{\mathcal{\mathcal{Q}}}_{{\rm L}\,f'}^{\ast\,l} 
\widetilde{\mathcal{\mathcal{Q}}}_{{\rm L}\,l}^{\,f'} \,.
\label{sc2}
\end{align}

The third type involves terms with \emph{two} reoccurring indices (either $\SU{2}{}$ or $\SU{3}{}$ indices) but \emph{identical} fields. We denote this case 
as ``sc3'' and observe that there are only two linearly independent terms of the form
\begin{align}
&V_{\rm sc3} \supset \nonumber \\
&\lambda_{j_1}\widetilde{D}_{{\rm L}\;x'\,f'}^\ast \widetilde{D}_{\rm L}^{x'\,f'} \widetilde{D}_{{\rm L}\;x\,f}^\ast \widetilde{D}_{\rm L}^{x\,f}
+
\lambda_{j_2}\widetilde{D}_{{\rm L}\;x'\,f'}^\ast \widetilde{D}_{\rm L}^{x\,f'} \widetilde{D}_{{\rm L}\;x\,f}^\ast \widetilde{D}_{\rm L}^{x'\,f}
\nonumber \\
& \equiv \lam{j_1}{j_2} \widetilde{D}_{{\rm L}\,f'}^\ast \widetilde{D}_{\rm L}^{f'} \widetilde{D}_{{\rm L}\;f}^\ast \widetilde{D}_{\rm L}^{f}\,,
\label{sc3}
\end{align}
where colour contractions are once again implicit. 

For terms with \emph{three} reoccurring indices and identical fields, labeled as ``sc4'', there are four linearly independent combinations that 
we write as
\begin{align}
V_{\rm sc4} \supset
&\lambda_{m_1}H^{\ast\,r'}_{f'\,l'} H^{f'\,l'}_{r'} H^{\ast\,r}_{f\,l} H^{f\,l}_r
+
\lambda_{m_2}H^{\ast\,r'}_{f'\,l'} H^{f'\,l'}_r H^{\ast\,r}_{f\,l} H^{f\,l}_{r'}
\nonumber \\
+&
\lambda_{m_3}H^{\ast\,r'}_{f'\,l'} H^{f\,l'}_{r'} H^{\ast\,r}_{f\,l} H^{f'\,l}_{r}
+
\lambda_{m_4}H^{\ast\,r'}_{f'\,l'} H^{f'\,l}_{r'} H^{\ast\,r}_{f\,l} H^{f\,l'}_{r}
\nonumber \\
\equiv &\lam{m_1}{m_4} H^{\ast\,r'}_{f'\,l'} H^{f'\,l'}_{r'}H^{\ast\,r}_{f\,l} H^{f\,l}_r
\label{sc4}
\end{align}
Note that the case with three reoccurring indices and different fields does not exist and the only case with one reoccurring index and identical fields is 
the one involving the gauge singlet $\phi^f$. 

Finally, the fifth type (``sc5'') involves terms without reoccurring indices or terms with one reoccurring index but four identical fields such as
\begin{align}
\label{sc5}
V_{\rm sc5} \supset \la{i} h^{\ast\,r}_{l} h^l_{r} \widetilde{\phi}^\ast_{f} \widetilde{\phi}^f + \la{j} \widetilde{\phi}^\ast_{f'}\widetilde{\phi}^{f'}    
\widetilde{\phi}^\ast_{f}\widetilde{\phi}^f \,.
\end{align}
Note that, for ease of notation, we assume that combinatorial factors were absorbed by various $\la{i}$ and $\lam{i}{j}$.

We will then consider five different scenarios organized according to the type of index contractions as described in detail in Eqs.~\eqref{sc1}, 
\eqref{sc2}, \eqref{sc3}, \eqref{sc4} and \eqref{sc5}:
\begin{equation}
V_4 = V_{\rm sc1} + V_{\rm sc2}+ V_{\rm sc3} + V_{\rm sc4} + V_{\rm sc5}\,.
\end{equation}
The first contribution reads

\onecolumngrid

\begin{align}
\begin{split}
&V_{\rm sc1} =   
\lam{1}{2} \widetilde{q}_{\rm L}^{\ast\,l} \widetilde{q}_{{\rm L}\,l} \widetilde{q}_{{\rm R}\,r}^{\ast} \widetilde{q}_{\rm R}^{r} 
+ \lam{3}{4} \widetilde{\mathcal{B}}_{\rm L}^\ast \widetilde{\mathcal{B}}_{\rm L} \widetilde{\mathcal{B}}_{\rm R}^\ast \widetilde{\mathcal{B}}_{\rm R} 
+\lam{5}{6} H^{\ast\,r}_{f'\,l} H^{f'\,l}_r \widetilde{\phi}^\ast_{f}\widetilde{\phi}^f  
+\lam{7}{8} \widetilde{E}_{{\rm L}\;f'\,l}^\ast \widetilde{E}_{\rm L}^{f'\,l} \widetilde{E}_{{\rm R}\;f}^{\ast\,r} \widetilde{E}_{{\rm R}\,r}^{f} 
+\widehat{\mathcal{P}}_{\rm LR} \left[ \lam{9}{10} \widetilde{q}_{\rm L}^{\ast\,l} \widetilde{q}_{{\rm L}\,l} 
\widetilde{\mathcal{Q}}_{{\rm R}\,f\,r}^{\ast} \widetilde{\mathcal{Q}}_{{\rm R}}^{f\,r}  
\right. \\
&
\left. 
+\lam{11}{12} \widetilde{\mathcal{B}}_{\rm L}^\ast \widetilde{\mathcal{B}}_{\rm L} \widetilde{D}_{{\rm R}\;f}^\ast \widetilde{D}_{\rm R}^{f}
+\lam{13}{14}\widetilde{q}_{\rm L}^{\ast\,l} \widetilde{q}_{{\rm L}\,l}\widetilde{D}_{{\rm L}\;f}^\ast \widetilde{D}_{\rm L}^{f} 
+\lam{15}{16}\widetilde{\mathcal{Q}}_{{\rm L}\;f}^{\ast\,l} \widetilde{\mathcal{Q}}_{{\rm L}\,l}^{f} \widetilde{\mathcal{B}}_{\rm L}^\ast \widetilde{\mathcal{B}}_{\rm L} 
+ \lam{17}{18} \widetilde{q}_{\rm L}^{\ast\,l} \widetilde{q}_{{\rm L}\,l}\widetilde{\mathcal{B}}_{\rm L}^\ast \widetilde{\mathcal{B}}_{\rm L}
+\lam{19}{20}\widetilde{q}_{\rm L}^{\ast\,l} \widetilde{q}_{{\rm L}\,l} \widetilde{D}_{{\rm R}\;f}^\ast \widetilde{D}_{\rm R}^{f}
 \right. \\
&
\left.
+\lam{21}{22}\widetilde{\mathcal{Q}}_{{\rm L}\;f}^{\ast\,l} \widetilde{\mathcal{Q}}_{{\rm L}\,l}^{f} \widetilde{\mathcal{B}}_{\rm R}^\ast \widetilde{\mathcal{B}}_{\rm R}
+ \lam{23}{24} \widetilde{\mathcal{Q}}_{{\rm L}\;f}^{\ast\,l'} \widetilde{\mathcal{Q}}_{{\rm L}\,l'}^{f}  h^{\ast\,r}_{l} h^l_{r}
+\lam{25}{26} \widetilde{q}_{\rm L}^{\ast\,l} \widetilde{q}_{{\rm L}\,l}\widetilde{\mathcal{B}}_{\rm R}^\ast \widetilde{\mathcal{B}}_{\rm R}
+\lam{27}{28} \widetilde{q}_{\rm L}^{\ast\,l'} \widetilde{q}_{{\rm L}\,l'} H^{\ast\,r}_{f\,l} H^{f\,l}_{r}  
+ \lam{29}{30} \widetilde{q}_{\rm L}^{\ast\,l'} \widetilde{q}_{{\rm L}\,l'} h^{\ast\,r}_{l} h^l_{r}
\right. \\
&
\left. 
+ \lam{31}{32} \widetilde{D}_{{\rm L}\,f'}^\ast \widetilde{D}_{\rm L}^{f'} H^{\ast\,r}_{f\,l} H^{f\,l}_{r}
+ \lam{33}{34} \widetilde{q}_{\rm L}^{\ast\,l} \widetilde{q}_{{\rm L}\,l} \widetilde{E}_{{\rm L}\;f\,l'}^\ast \widetilde{E}_{\rm L}^{f\,l'} 
+ \lam{35}{36} \widetilde{\mathcal{Q}}_{{\rm L}\,f}^{\ast\,l} \widetilde{\mathcal{Q}}_{{\rm L}\,l}^{f} \widetilde{\mathcal{E}}_{{\rm L}\;l'}^\ast \widetilde{\mathcal{E}}_{\rm L}^{l'} 
+\lam{37}{38} \widetilde{q}_{\rm L}^{\ast\,l} \widetilde{q}_{{\rm L}\,l}  \widetilde{\mathcal{E}}_{{\rm L}\;l'}^\ast \widetilde{\mathcal{E}}_{\rm L}^{l'}
\right. \\
&
\left. 
+ \lam{39}{40} \widetilde{\mathcal{Q}}_{{\rm R}\;f'}^{\ast\;r} \widetilde{\mathcal{Q}}_{{\rm R}\;\,r}^{f'}  \widetilde{E}_{{\rm L}\;f\,l}^\ast \widetilde{E}_{\rm L}^{f\,l}   
+ \lam{41}{42 }\widetilde{D}_{{\rm L}\,f'}^\ast \widetilde{D}_{\rm L}^{f'} \widetilde{E}_{{\rm L}\;f\,l}^\ast \widetilde{E}_{\rm L}^{f\,l} 
+\lam{43}{44} \widetilde{D}_{{\rm R}\;f'}^\ast \widetilde{D}_{\rm R}^{f'}  \widetilde{E}_{{\rm L}\;f\,l}^\ast \widetilde{E}_{\rm L}^{f\,l} 
+ \lam{45}{46} \widetilde{\mathcal{Q}}_{{\rm L}\,f'}^{\ast\,l} \widetilde{\mathcal{Q}}_{{\rm L}\,l}^{f'} \widetilde{\phi}^\ast_{f}\widetilde{\phi}^f 
\right. \\
&
\left. 
+\lam{47}{48} \widetilde{D}_{{\rm L}\,f'}^\ast \widetilde{D}_{\rm L}^{f'}\widetilde{\phi}^\ast_{f}\widetilde{\phi}^f 
+ \lam{49}{50} h^{\ast\,r}_{l'} h^{l'}_{r}  \widetilde{E}_{{\rm L}\;f\,l}^\ast \widetilde{E}_{\rm L}^{f\,l}
+ \lam{51}{52}  \widetilde{D}_{{\rm L}\;f}^\ast \widetilde{D}_{\rm L}^{f} \widetilde{E}_{{\rm L}\;f\,l}^\ast \widetilde{E}_{\rm L}^{f\,l}  
+\lam{53}{54}  H^{\ast\,r}_{f\,l'} H^{f\,l'}_{r}  \widetilde{\mathcal{E}}_{{\rm L}\;l}^\ast \widetilde{\mathcal{E}}_{\rm L}^{l} 
\right. \\
&
\left. 
+\lam{55}{56} h^{\ast\,r}_{l'} h^{l'}_{r} \widetilde{\mathcal{E}}_{{\rm L}\;l}^\ast \widetilde{\mathcal{E}}_{\rm L}^{l}
+ \lam{57}{58} \widetilde{\mathcal{B}}_{\rm L}^\ast \widetilde{\mathcal{B}}_{\rm L}  \widetilde{D}_{{\rm L}\;f}^\ast \widetilde{D}_{\rm L}^{f} 
+\lam{59}{60} \widetilde{E}_{{\rm L}\;f'\,l}^\ast \widetilde{E}_{\rm L}^{f'\,l} \widetilde{\phi}^\ast_{f}\widetilde{\phi}^f 
+ \lam{61}{62} \widetilde{\phi}^f H^{f'\,l}_r  \widetilde{E}_{{\rm L}\;f'\,l}^\ast \widetilde{E}_{{\rm R}\;f}^{\ast\,r} 
\right. \\
&
\left. 
+ \left( \lam{63}{64} h^{\ast\,r}_{l'} H^{f\,l'}_{r}  \widetilde{E}_{{\rm L}\;f\,l}^\ast \widetilde{\mathcal{E}}_{\rm L}^{l}
+ \lam{65}{66}  h^{\ast\,r}_{l'} H^{f\,l'}_{r} \widetilde{\mathcal{Q}}_{{\rm L}\,f}^{\ast\,l} \widetilde{q}_{{\rm L}\,l} 
+ \lam{67}{68} \widetilde{E}_{{\rm L}\;f\,l'}^\ast \widetilde{\mathcal{E}}_{\rm L}^{l'} \widetilde{q}_{\rm L}^{\ast\,l} \widetilde{\mathcal{Q}}_{{\rm L}\,l}^{f}
+ \lam{69}{70} \widetilde{D}_{{\rm L}\,f'}^\ast \widetilde{\mathcal{Q}}_{{\rm L}\,l}^{f'} \widetilde{E}_{{\rm R}\;f}^{\ast\,r} H^{f\,l}_{r}
\right. \right. \\
&
\left. \left.
+ \lam{71}{72} \widetilde{D}_{{\rm L}\,f'}^\ast \widetilde{\mathcal{Q}}_{{\rm L}\,l}^{f'} \widetilde{E}_{\rm L}^{f\,l}  \widetilde{\phi}^\ast_{f}
+ \lam{73}{74} \widetilde{\mathcal{B}}_{\rm L} \widetilde{\mathcal{Q}}_{{\rm R}\;\,r}^{f} \widetilde{D}_{{\rm L}\;f}^\ast \widetilde{q}_{{\rm R}\,r}^{\ast}
+ {\rm c.c.} \right)  \right] + V_{\rm sc1}^{\rm gen}
 \,,
\end{split}
\end{align}

\twocolumngrid

with $V_{\rm sc1}^{\rm gen}$ corresponding to the interactions generated only after the matching procedure, i.e.~not directly obtained by expansion of the Lagrangian 
of the original theory, and given by
\begin{align*}
\begin{split}
&V_{\rm sc1}^{\rm gen} =   
\widehat{\mathcal{P}}_{\rm LR} \Big[
\delt{1}{2} \widetilde{\mathcal{Q}}_{{\rm L}\,f'}^{\ast\,l} \widetilde{\mathcal{Q}}_{{\rm L}\,l}^{f'} \widetilde{\mathcal{H}}_{{\rm F} f}^{\ast} \widetilde{\mathcal{H}}_{\rm F}^f
+\delt{3}{4}\widetilde{D}_{{\rm L}\,f'}^\ast \widetilde{D}_{\rm L}^{f'}\widetilde{\mathcal{H}}_{{\rm F} f}^{\ast} \widetilde{\mathcal{H}}_{\rm F}^f 
\\
&
+ \delt{5}{6} \widetilde{E}_{{\rm L}\;f'\,l}^\ast \widetilde{E}_{\rm L}^{f'\,l}  \widetilde{\mathcal{H}}_{{\rm F} f}^{\ast} \widetilde{\mathcal{H}}_{\rm F}^f
 \Big]
+ \delt{7}{8} H^{\ast\,r}_{f'\,l} H^{f'\,l}_r  \widetilde{\mathcal{H}}_{{\rm F} f}^{\ast} \widetilde{\mathcal{H}}_{\rm F}^f \,.
\end{split}
\end{align*}
The effective quartic interactions with two reoccurring indices are given by
\begin{align*}
\begin{split}
&V_{\rm sc2} =  
\lam{75}{78} \widetilde{\mathcal{Q}}_{{\rm L}\,f'}^{\ast\,l} \widetilde{\mathcal{Q}}_{{\rm L}\,l}^{f'} \widetilde{\mathcal{Q}}_{{\rm R}\,f\,r}^{\ast} 
\widetilde{\mathcal{Q}}_{{\rm R}}^{f\,r}
+\lam{79}{82} \widetilde{D}_{{\rm L}\,f'}^\ast \widetilde{D}_{\rm L}^{f'} \widetilde{D}_{{\rm R}\;f}^\ast \widetilde{D}_{\rm R}^{f} 
\\
&
+\lam{83}{86} h^{\ast\,r'}_{l'} h^{l'}_{r'} H^{\ast\,r}_{f\,l} H^{f\,l}_{r}
+ \widehat{\mathcal{P}}_{\rm LR} \left[ \lam{75}{78}  \widetilde{\mathcal{Q}}_{{\rm L}\;f'}^{\ast\,l} \widetilde{\mathcal{Q}}_{{\rm L}\,l}^{f'} 
\widetilde{D}_{{\rm L}\;f}^\ast \widetilde{D}_{\rm L}^{f} 
\right. \\
&
\left. 
+ \lam{87}{90}\widetilde{\mathcal{Q}}_{{\rm L}\;f'}^{\ast\,l} \widetilde{\mathcal{Q}}_{{\rm L}\,l}^{f'} \widetilde{D}_{{\rm R}\;f}^\ast \widetilde{D}_{\rm R}^{f}
+ \lam{91}{94} \widetilde{q}_{\rm L}^{\ast\,l'} \widetilde{q}_{{\rm L}\,l'} \widetilde{\mathcal{Q}}_{{\rm L}\,f}^{\ast\,l} \widetilde{\mathcal{Q}}_{{\rm L}\,l}^{f} 
\right. \\
&
\left. 
+\lam{95}{98} \widetilde{\mathcal{Q}}_{{\rm L}\;f'}^{\ast\,l'} \widetilde{\mathcal{Q}}_{{\rm L}\,l'}^{f'} H^{\ast\,r}_{f\,l} H^{f\,l}_{r}
+ \lam{99}{102}\widetilde{\mathcal{Q}}_{{\rm L}\,f'}^{\ast\,l} \widetilde{\mathcal{Q}}_{{\rm L}\,l}^{f'} \widetilde{E}_{{\rm L}\;f\,l'}^\ast 
\widetilde{E}_{\rm L}^{f\,l'} 
\right. \\
&
\left. 
+\lam{103}{106}  H^{\ast\,r}_{f'\,l'} H^{f'\,l'}_{r} \widetilde{E}_{{\rm L}\;f\,l}^\ast \widetilde{E}_{\rm L}^{f\,l} \right]\,.
\end{split}
\end{align*}
The third contribution, which accounts for identical multiplets and two reoccurring indices, has the form
\begin{align*}
\begin{split}
&V_{\rm sc3} =   
\lam{107}{108} h^{\ast\,r'}_{l'} h^{l'}_{r'} h^{\ast\,r}_{l} h^l_{r} + \widehat{\mathcal{P}}_{\rm LR} \left[ \lam{101}{102}\widetilde{q}_{\rm L}^{\ast\,l'} 
\widetilde{q}_{{\rm L}\,l'} \widetilde{q}_{\rm L}^{\ast\,l} \widetilde{q}_{{\rm L}\,l} \right. \\
&
\left. 
+ \lam{109}{110} \widetilde{D}_{{\rm L}\,f'}^\ast \widetilde{D}_{\rm L}^{f'} \widetilde{D}_{{\rm L}\;f}^\ast \widetilde{D}_{\rm L}^{f}
\right. \\
&
\left.
+\lam{111}{112}\widetilde{E}_{{\rm L}\;f'\,l'}^\ast \widetilde{E}_{\rm L}^{f'\,l'} \widetilde{E}_{{\rm L}\;f\,l}^\ast \widetilde{E}_{\rm L}^{f\,l} \right] \,,
\end{split}
\end{align*}
while the forth scenario, where identical fields with three reoccurring indices are considered, reads
\begin{align*}
\begin{split}
&V_{\rm sc4} =   
\lam{113}{116} H^{\ast\,r'}_{f'\,l'} H^{f'\,l'}_{r'} H^{\ast\,r}_{f\,l} H^{f\,l}_{r} \\
+ & 
\widehat{\mathcal{P}}_{\rm LR} \left[ \lam{117}{120} 
\widetilde{\mathcal{Q}}_{{\rm L}\;f'}^{\ast\,l'} \widetilde{\mathcal{Q}}_{{\rm L}\,l'}^{f'} \widetilde{\mathcal{Q}}_{{\rm L}\,f}^{\ast\,l} \widetilde{\mathcal{Q}}_{{\rm L}\,l}^{f} \right]\,.
\end{split}
\end{align*}
Finally, for those terms that contain only one independent type of contraction we have

\onecolumngrid

\begin{align*}
\begin{split}
& V_{\rm sc5} =   
\la{121} h^{\ast\,r}_{l} h^l_{r} \widetilde{\phi}^\ast_{f}\widetilde{\phi}^f 
+ \la{122} H^{\ast\,r}_{f\,l} H^{f\,l}_{r} \widetilde{\varphi}^\ast \widetilde{\varphi} 
+ \la{123} h^{\ast\,r}_{l} h^l_{r}  \widetilde{\varphi}^\ast \widetilde{\varphi}
+ \la{124}  \widetilde{\mathcal{E}}_{{\rm L}\;l}^\ast \widetilde{\mathcal{E}}_{\rm L}^{l} 
\widetilde{\mathcal{E}}_{\rm R}^{\ast\,r} \widetilde{\mathcal{E}}_{{\rm R}\,r} 
+\la{125} \widetilde{\phi}^\ast_{f'}\widetilde{\phi}^{f'}    \widetilde{\phi}^\ast_{f}\widetilde{\phi}^f 
+ \la{126}  \widetilde{\phi}^\ast_{f}\widetilde{\phi}^f \widetilde{\varphi}^\ast \widetilde{\varphi}  
\nonumber \\
&
+ \la{127} \widetilde{\varphi}^\ast \widetilde{\varphi}\,\widetilde{\varphi}^\ast \widetilde{\varphi} 
+ \widehat{\mathcal{P}}_{\rm LR} \left[  
 \la{128} \widetilde{\varphi} h^l_{r} \widetilde{\mathcal{E}}_{{\rm L}\;l}^\ast \widetilde{\mathcal{E}}_{\rm R}^{\ast\,r}
+ \la{129} \widetilde{D}_{{\rm L}\;f}^\ast \widetilde{\mathcal{B}}_{\rm L} \widetilde{\mathcal{B}}_{\rm R}^\ast \widetilde{D}_{\rm R}^{f}
+ \la{130} \widetilde{\mathcal{Q}}_{{\rm L}\,f}^{\ast\,l} \widetilde{q}_{{\rm L}\,l} \widetilde{q}_{{\rm R}\,r}^{\ast} \widetilde{\mathcal{Q}}_{{\rm R}}^{f\,r} 
+ \la{131}  \widetilde{\mathcal{B}}_{\rm L}^\ast \widetilde{\mathcal{B}}_{\rm L} \widetilde{\mathcal{B}}_{\rm L}^\ast \widetilde{\mathcal{B}}_{\rm L}
\right. \nonumber \\
&
\left.
+ \la{132} \widetilde{\mathcal{B}}_{\rm L}^\ast \widetilde{\mathcal{B}}_{\rm L}  H^{\ast\,r}_{f\,l} H^{f\,l}_{r}
+ \la{133} \widetilde{D}_{{\rm L}\;f}^\ast \widetilde{D}_{\rm L}^{f} h^{\ast\,r}_{l} h^l_{r} 
+ \la{134}  \widetilde{\mathcal{B}}_{\rm L}^\ast \widetilde{\mathcal{B}}_{\rm L} h^{\ast\,r}_{l} h^l_{r}
+ \la{135} \widetilde{q}_{{\rm R}\,r}^{\ast} \widetilde{q}_{\rm R}^{r}  \widetilde{E}_{{\rm L}\;f\,l}^\ast \widetilde{E}_{\rm L}^{f\,l}
+ \la{136}  \widetilde{\mathcal{Q}}_{{\rm R}\,f\,r}^{\ast} \widetilde{\mathcal{Q}}_{{\rm R}}^{f\,r}  \widetilde{\mathcal{E}}_{{\rm L}\;l}^\ast \widetilde{\mathcal{E}}_{\rm L}^{l} 
\right. \nonumber \\
&
\left. 
+ \la{137} \widetilde{q}_{{\rm R}\,r}^{\ast} \widetilde{q}_{\rm R}^{r} \widetilde{\mathcal{E}}_{{\rm L}\;l}^\ast \widetilde{\mathcal{E}}_{\rm L}^{l} 
+ \la{138}  \widetilde{D}_{{\rm L}\;f}^\ast \widetilde{D}_{\rm L}^{f} \widetilde{E}_{{\rm L}\;f\,l}^\ast \widetilde{E}_{\rm L}^{f\,l} 
+ \la{139}  \widetilde{D}_{{\rm L}\;f}^\ast \widetilde{D}_{\rm L}^{f} \widetilde{\mathcal{E}}_{{\rm L}\;l}^\ast \widetilde{\mathcal{E}}_{\rm L}^{l} 
+ \la{140} \widetilde{\mathcal{B}}_{\rm L}^\ast \widetilde{\mathcal{B}}_{\rm L}  \widetilde{\mathcal{E}}_{{\rm L}\;l}^\ast \widetilde{\mathcal{E}}_{\rm L}^{l}
+ \la{141} \widetilde{\mathcal{B}}_{\rm R}^\ast \widetilde{\mathcal{B}}_{\rm R}  \widetilde{E}_{{\rm L}\;f\,l}^\ast \widetilde{E}_{\rm L}^{f\,l}
\right. \nonumber \\
&
\left. 
+ \la{142}  \widetilde{D}_{{\rm R}\;f}^\ast \widetilde{D}_{\rm R}^{f} \widetilde{\mathcal{E}}_{{\rm L}\;l}^\ast \widetilde{\mathcal{E}}_{\rm L}^{l} 
+ \la{143}  \widetilde{\mathcal{B}}_{\rm R}^\ast \widetilde{\mathcal{B}}_{\rm R}  \widetilde{\mathcal{E}}_{{\rm L}\;l}^\ast \widetilde{\mathcal{E}}_{\rm L}^{l}
+ \la{144} \widetilde{q}_{\rm L}^{\ast\,l} \widetilde{q}_{{\rm L}\,l} \widetilde{\phi}^\ast_{f}\widetilde{\phi}^f  
+ \la{145} \widetilde{\mathcal{Q}}_{{\rm L}\,f}^{\ast\,l} \widetilde{\mathcal{Q}}_{{\rm L}\,l}^{f} \widetilde{\varphi}^\ast \widetilde{\varphi} 
+ \la{146} \widetilde{q}_{\rm L}^{\ast\,l} \widetilde{q}_{{\rm L}\,l}  \widetilde{\varphi}^\ast \widetilde{\varphi}
+ \la{147}\widetilde{\mathcal{B}}_{\rm L}^\ast \widetilde{\mathcal{B}}_{\rm L} \widetilde{\phi}^\ast_{f}\widetilde{\phi}^f 
\right. \nonumber \\
&
\left. 
+ \la{148} \widetilde{D}_{{\rm L}\;f}^\ast \widetilde{D}_{\rm L}^{f} \widetilde{\varphi}^\ast \widetilde{\varphi} 
+ \la{149}\widetilde{\mathcal{B}}_{\rm L}^\ast \widetilde{\mathcal{B}}_{\rm L}  \widetilde{\varphi}^\ast \widetilde{\varphi} 
+ \la{150} \widetilde{\mathcal{E}}_{{\rm L}\;l'}^\ast \widetilde{\mathcal{E}}_{\rm L}^{l'}  \widetilde{E}_{{\rm L}\;f\,l}^\ast \widetilde{E}_{\rm L}^{f\,l}
+ \la{151} \widetilde{\mathcal{E}}_{{\rm L}\;l'}^\ast \widetilde{\mathcal{E}}_{\rm L}^{l'} \widetilde{\mathcal{E}}_{{\rm L}\;l}^\ast \widetilde{\mathcal{E}}_{\rm L}^{l} 
+ \la{152} \widetilde{\mathcal{E}}_{{\rm L}\;l}^\ast \widetilde{\mathcal{E}}_{\rm L}^{l} \widetilde{E}_{{\rm R}\;f}^{\ast\,r} \widetilde{E}_{{\rm R}\,r}^{f}
\right. \nonumber \\
&
\left. 
+\la{153} \widetilde{\mathcal{E}}_{{\rm L}\;l}^\ast \widetilde{\mathcal{E}}_{\rm L}^{l} \widetilde{\phi}^\ast_{f}\widetilde{\phi}^f 
+\la{154} \widetilde{E}_{{\rm L}\;f\,l}^\ast \widetilde{E}_{\rm L}^{f\,l} \widetilde{\varphi}^\ast \widetilde{\varphi} 
+\la{155} \widetilde{\mathcal{E}}_{{\rm L}\;l}^\ast \widetilde{\mathcal{E}}_{\rm L}^{l} \widetilde{\varphi}^\ast \widetilde{\varphi}
+\left( 
 \la{156} \widetilde{E}_{{\rm L}\;f\,l}^\ast \widetilde{\mathcal{E}}_{\rm L}^{l} \widetilde{\varphi}^\ast \widetilde{\phi}^f
+ \la{157}  \widetilde{E}_{{\rm L}\;f\,l}^\ast \widetilde{\phi}^f h^l_{r} \widetilde{\mathcal{E}}_{\rm R}^{\ast\,r}
\right. \right. \nonumber\\
&
\left. \left. 
+ \la{158} H^{f\,l}_{r} \widetilde{E}_{{\rm R}\;f}^{\ast\,r} \widetilde{\mathcal{E}}_{{\rm L}\;l}^\ast \widetilde{\varphi}
+ \la{159} \widetilde{E}_{{\rm L}\;f\,l}^\ast \widetilde{\mathcal{E}}_{\rm L}^{l} \widetilde{\mathcal{B}}_{\rm L}^\ast \widetilde{D}_{\rm L}^{f}
+ \la{160} \widetilde{\varphi}^\ast \widetilde{\phi}^f \widetilde{D}_{{\rm L}\;f}^\ast \widetilde{\mathcal{B}}_{\rm L} 
+ \la{161} \widetilde{D}_{{\rm L}\;f}^\ast \widetilde{\mathcal{B}}_{\rm L} \widetilde{q}_{\rm L}^{\ast\,l} \widetilde{\mathcal{Q}}_{{\rm L}\,l}^{f} 
+ \la{162} \widetilde{\mathcal{B}}_{\rm L}^\ast \widetilde{q}_{{\rm L}\,l} \widetilde{E}_{{\rm R}\;f}^{\ast\,r} H^{f\,l}_{r}
\right. \right.\nonumber \\
&
\left. \left. 
+ \la{163} \widetilde{\mathcal{B}}_{\rm L}^\ast \widetilde{\mathcal{Q}}_{{\rm L}\,l}^{f} \widetilde{E}_{{\rm R}\;f}^{\ast\,r} h^{l}_{r}
+ \la{164} \widetilde{\mathcal{B}}_{\rm L}^\ast \widetilde{q}_{{\rm L}\,l} \widetilde{\mathcal{E}}_{\rm R}^{\ast\,r} h^{l}_{r}
+ \la{165} \widetilde{\mathcal{B}}_{\rm L}^\ast \widetilde{q}_{{\rm L}\,l} \widetilde{E}_{\rm L}^{f\,l} \widetilde{\phi}^\ast_{f}
+ \la{166} \widetilde{\mathcal{B}}_{\rm L}^\ast \widetilde{\mathcal{Q}}_{{\rm L}\,l}^{f} \widetilde{\mathcal{E}}_{\rm L}^{l} \widetilde{\phi}^\ast_{f} 
+ \la{167} \widetilde{\mathcal{B}}_{\rm L}^\ast \widetilde{q}_{{\rm L}\,l} \widetilde{\mathcal{E}}_{\rm L}^{l} \widetilde{\varphi}^\ast 
\right. \right. \nonumber\\
&
\left. \left. 
+ \la{168} \widetilde{\mathcal{B}}_{\rm L}^\ast \widetilde{D}_{\rm L}^{f} \widetilde{E}_{{\rm R}\;f}^{\ast\,r} \widetilde{\mathcal{E}}_{{\rm R}\,r}
+ \la{169} \widetilde{D}_{{\rm L}\, f}^{\ast} \widetilde{q}_{{\rm L}\,l} \widetilde{\mathcal{E}}_{\rm R}^{\ast\,r} H^{f\,l}_{r}
+ \la{170} \widetilde{D}_{{\rm L}\, f}^{\ast} \widetilde{\mathcal{Q}}_{{\rm L}\,l}^{f} \widetilde{\mathcal{E}}_{\rm R}^{\ast\,r} h^{l}_{r}
+ \la{171} \widetilde{D}_{{\rm L}\, f}^{\ast} \widetilde{q}_{{\rm L}\,l} \widetilde{E}_{\rm L}^{f\,l} \widetilde{\varphi}^\ast
\right. \right. \nonumber \\
&
\left. \left. 
+\la{172} \widetilde{D}_{{\rm L}\, f}^{\ast} \widetilde{\mathcal{Q}}_{{\rm L}\,l}^{f} \widetilde{\mathcal{E}}_{\rm L}^{l} \widetilde{\varphi}^\ast
+ {\rm c.c.} \right)\right]  
+ V_{\rm sc5}^{\rm gen}\,.
\end{split}
\end{align*}

\twocolumngrid

Here, the terms generated after the breaking are
\begin{align*}
\begin{split}
&V_{\rm sc5}^{\rm gen} =   
\la{173} h^{\ast\,r}_{l} H^{f\,l}_{r} \widetilde{\phi}^\ast_{f} \widetilde{\varphi}
+ \la{174} \widetilde{E}_{{\rm L}\;f\,l}^\ast \widetilde{\mathcal{E}}_{\rm L}^{l} \widetilde{\mathcal{E}}_{\rm R}^{\ast\,r} \widetilde{E}_{{\rm R}\,r}^{f}
\\
&
+ \de{9}  h^{\ast\,r}_{l} h^l_{r}  \widetilde{\mathcal{H}}_{{\rm F} f}^{\ast} \widetilde{\mathcal{H}}_{\rm F}^f 
+ \de{10} \widetilde{\mathcal{H}}_{{\rm F} f'}^{\ast} \widetilde{\mathcal{H}}_{\rm F}^{f '} \widetilde{\mathcal{H}}_{{\rm F} f}^{\ast} \widetilde{\mathcal{H}}_{\rm F}^f  
\\
&
+ \de{11} \widetilde{\varphi}^\ast \widetilde{\varphi} \widetilde{\mathcal{H}}_{{\rm F} f}^{\ast} \widetilde{\mathcal{H}}_{\rm F}^f 
+ \de{12} \widetilde{\phi}^\ast_{f'}\widetilde{\phi}^{f'} \widetilde{\mathcal{H}}_{{\rm F} f}^{\ast} \widetilde{\mathcal{H}}_{\rm F}^f 
\\
&
+ \widehat{\mathcal{P}}_{\rm LR} \left[
 \la{175} h^{\ast\,r}_{l} H^{f\,l}_{r} \widetilde{D}_{{\rm L}\;f}^\ast \widetilde{\mathcal{B}}_{\rm L} 
+ \la{176} \widetilde{\varphi}^\ast \widetilde{\phi}^f \widetilde{\mathcal{Q}}_{{\rm L}\,f}^{\ast\,l} \widetilde{q}_{{\rm L}\,l}
\right. \\
&
\left.
+ \la{177} \widetilde{\mathcal{B}}_{\rm L}^\ast \widetilde{D}_{\rm L}^{f} \widetilde{E}_{{\rm L}\,f\,l}^\ast \widetilde{\mathcal{E}}_{\rm L}^{l}
+ \la{178} \widetilde{E}_{{\rm L}\;f\,l}^\ast \widetilde{\mathcal{E}}_{\rm L}^{l} \widetilde{q}_{{\rm R}\,r}^{\ast} \widetilde{\mathcal{Q}}_{{\rm R}}^{f\,r}
\right. \\
&
\left.
+ \de{13} \widetilde{q}_{\rm L}^{\ast\,l} \widetilde{q}_{{\rm L}\,l} \widetilde{\mathcal{H}}_{{\rm F} f}^{\ast} \widetilde{\mathcal{H}}_{\rm F}^f
+ \de{14} \widetilde{\mathcal{B}}_{\rm L}^\ast \widetilde{\mathcal{B}}_{\rm L} \widetilde{\mathcal{H}}_{{\rm F} f}^{\ast} \widetilde{\mathcal{H}}_{\rm F}^f
\right. \\
&
\left.
+ \de{15} \widetilde{\mathcal{E}}_{{\rm L}\;l}^\ast \widetilde{\mathcal{E}}_{\rm L}^{l} \widetilde{\mathcal{H}}_{{\rm F} f}^{\ast} \widetilde{\mathcal{H}}_{\rm F}^f
 \right]\,.
\end{split}
\end{align*}
%

\subsection{The fermion sector of the LR-symmetric EFT}

The part of the Lagrangian of the effective LR-symmetric theory that involves purely quadratic fermion interactions as well as the Yukawa terms reads 
\begin{align}
\mathcal{L}_{\rm fermi} = \mathcal{L}_{\rm M} + \mathcal{L}_{\rm Yuk}\,.
\label{eq:Lfermi-LR}
\end{align}  
For the mass terms we have
\begin{align}
\begin{aligned}
\label{LM}
&\mathcal{L}_{\rm M}  =
 \widehat{\mathcal{P}}_{\rm LR} \[\tfrac{1}{2} m_{S_{\rm L}} S_{\rm L} S_{\rm L} 
+ \tfrac{1}{2} m_{\mathcal{T}_{\rm L}} \mathcal{T}^i_{\rm L} \mathcal{T}^i_{\rm L} + {\rm c.c.} \] \\ 
+&  \widehat{\mathcal{P}}_{\rm LR} \[ \tfrac{1}{2} m_{\tilde{g}} \tilde{g}^a \tilde{g}^a + m_{{\rm LR}} S_{\rm L} S_{\rm R}
+  \tfrac{1}{2} m_{\mathcal{H}} \mathcal{H}_{{\rm F} f}^{\ast} \mathcal{H}_{\rm F}^f \]\,,
\end{aligned}
\end{align}
while for the Yukawa ones we write for convenience,
\begin{align}
\mathcal{L}_{\rm Yuk} = \mathcal{L}_{\rm 3c} + \mathcal{L}_{\rm 2c} + \mathcal{L}_{\rm 1c} + \mathcal{L}_{\mathcal{S}} + 
\mathcal{L}_{\mathcal{T}} + \mathcal{L}_{\rm \widetilde{g}}\,,
\end{align}
where the first three terms, which involve only the fields from the fundamental representations of the trinification group, denote three, 
two and one $\SU{2}{}$ contractions, respectively, whereas the last ones describe the Yukawa interactions of the singlet $\mathcal{S}$, 
triplet $\mathcal{T}$ and octet $\widetilde{g}^a$ fermions. 

The terms with three $\SU{2}{}$ contractions are given by
\begin{align}
\begin{split}
\label{L3c}
&\mathcal{L}_{\rm 3c}  = 
\varepsilon_{f f'} \left(\widehat{\mathcal{P}}_{\rm LR} \[ \y1 {\mathcal{Q}}_{\rm R}^{f\,r} h^{l}_{r} \mathcal{Q}_{{\rm L}\,l}^{f'}\]
+ \widehat{\mathcal{P}}_{\rm LR} \left[\y2 \widetilde{q}_{\rm R}^{r}  \widetilde{H}^{f\,l}_{r} \mathcal{Q}_{{\rm L}\,l}^{f'}
\right. \right.  \\
&
\left.  \left.
+ \y3 \widetilde{\mathcal{Q}}_{\rm R}^{f\,r}\, \widetilde{h}^{l}_{r}  \mathcal{Q}_{{\rm L}\,l}^{f'} 
+ \y4 q_{\rm R}^{\,r} H^{f\,l}_{r} \mathcal{Q}_{{\rm L}\,l}^{f'} + {\rm c.c.} \right] \right) \,,
\end{split}
\end{align}
those with two $\SU{2}{}$ contractions are written as
\begin{align}
\begin{split}
\label{L2c}
&\mathcal{L}_{\rm 2c}  = 
\varepsilon_{f f'} \widehat{\mathcal{P}}_{\rm LR} \left[\y{5} \widetilde{\mathcal{B}}_{\rm R} \mathcal{Q}_{{\rm L}\,l}^{f} E_{\rm L}^{f'\,l}
+\y{6} \widetilde{D}_{\rm R}^{f}  \mathcal{Q}_{{\rm L}\,l}^{f'} \mathcal{E}_{\rm L}^{l}
\right.  \\
&
\left.  
+\y{7} \widetilde{D}_{\rm R}^{f} q_{{\rm L}\,l} E_{\rm L}^{f'\,l}
+\y{8}\mathcal{B}_{\rm R} \widetilde{\mathcal{Q}}_{{\rm L}\,l}^{f} E_{\rm L}^{f'\,l}
 \right.  \\
&
\left.  
+\y{9} D_{\rm R}^{f}  \widetilde{\mathcal{Q}}_{{\rm L}\,l}^{f'} \mathcal{E}_{\rm L}^{l}
+\y{10} D_{\rm R}^{f} \widetilde{q}_{{\rm L}\,l} E_{\rm L}^{f'\,l}
+\y{11} \mathcal{B}_{\rm R}  \mathcal{Q}_{{\rm L}\,l}^{f} \widetilde{E}_{\rm L}^{f'\,l} 
\right.  \\
&
\left. 
+\y{12} D_{\rm R}^{f} \mathcal{Q}_{{\rm L}\,l}^{f'} (\widetilde{\mathcal{E}}_{\rm L})^{l}
+\y{13} D_{\rm R}^{f} q_{{\rm L}\,l} \widetilde{E}_{\rm L}^{f'\,l} + {\rm c.c.} \right]\,,
\end{split}
\end{align}
and for those with one $\SU{2}{}$ contraction we have
\begin{align}
\begin{split}
\label{L1c}
&\mathcal{L}_{\rm 1c}  = 
\varepsilon_{f f'}  \left( \widehat{\mathcal{P}}_{\rm LR} \[
\y{14} D_{\rm R}^f \widetilde{\varphi} D_{\rm L}^{f'}\]
+ \widehat{\mathcal{P}}_{\rm LR} \left[
\y{15} \widetilde{\mathcal{B}}_{\rm R} \phi^f  D_{\rm L}^{f'}
\right. \right. \\
&
\left. \left. 
+\y{16} \widetilde{D}_{\rm R}^{f} \phi^{f'} \mathcal{B}_{\rm L} 
+\y{17}  \widetilde{D}_{\rm R}^{f} \varphi  D_{\rm L}^{f'}
+\y{18} \mathcal{B}_{\rm R}  \widetilde{\phi}^f D_{\rm L}^{f'} + {\rm c.c.} \right] \right) \,.
\end{split}
\end{align}
The part of the Lagrangian involving the singlets $\mathcal{S}_{\rm L,R}$ reads
\begin{align}
\begin{split}
\label{LS}
&\mathcal{L}_{\mathcal{S}}  = 
 \widehat{\mathcal{P}}_{\rm LR} \left[  
\y{19} \widetilde{\mathcal{Q}}^{\ast\,l}_{{\rm L}\,f}\mathcal{S}_{\rm L} {\mathcal{Q}}_{{\rm L}\,l}^{f}
+\y{20} \widetilde{q}^{\ast\,l}_{\rm L}\mathcal{S}_{\rm L} {q}_{{\rm L}\,l}
+\y{21} \widetilde{D}_{{\rm L}\,f}^\ast \mathcal{S}_{\rm L} {D}_{\rm L}^{f}
\right.  \\
&
\left. 
+\y{22} \widetilde{\mathcal{B}}_{\rm L}^\ast \mathcal{S}_{\rm L} {\mathcal{B}}_{\rm L}
+\y{23} H^{\ast \,r}_{f\,l} \mathcal{S}_{\rm L} \widetilde{H}^{f\,l}_{r}
+\y{24} h^{\ast \,r}_{l}\mathcal{S}_{\rm L} \widetilde{h}^{l}_{r}
\right. \\
&
\left. 
+\y{25} \widetilde{E}_{{\rm L}\,f\,l}^\ast \mathcal{S}_{\rm L} {E}_{\rm L}^{f\,l}
+\y{26}  \widetilde{\mathcal{E}}_{{\rm L}\,l}^\ast \mathcal{S}_{\rm L}{\mathcal{E}}_{\rm L}^{\,l}
+\y{27}{\widetilde{\phi}^\ast_f} \mathcal{S}_{\rm L} {\phi}^{f}
\right. \\
&
\left. 
+\y{28}{\widetilde{\varphi}^\ast} \mathcal{S}_{\rm L} {\varphi}
+y \mathcal{H}_{{\rm F} f}^{\ast} \mathcal{S}_{\rm L} \mathcal{H}_{\rm F}^f
+\y{29} \widetilde{E}_{{\rm L}\,f\,l}^\ast \mathcal{S}_{\rm R} {E}_{\rm L}^{f\,l} 
\right. \\
&
\left. 
+\y{30} \widetilde{\mathcal{B}}_{\rm L}^\ast \mathcal{S}_{\rm R} {\mathcal{B}}_{\rm L}
+\y{31} \widetilde{D}_{{\rm L}\,f}^\ast \mathcal{S}_{\rm R} {D}_{\rm L}^{f}
+\y{32} \widetilde{\mathcal{Q}}^{\ast\,l}_{{\rm L}\,f}\mathcal{S}_{\rm R} {\mathcal{Q}}_{{\rm L}\,l}^{f}
\right. \\
&
\left. 
+\y{33} \widetilde{q}^{\ast\,l}_{\rm L}\mathcal{S}_{\rm R} {q}_{{\rm L}\,l}
+\y{34}  \widetilde{\mathcal{E}}_{{\rm L}\,l}^\ast \mathcal{S}_{\rm R}{\mathcal{E}}_{\rm L}^{\,l}
+ {\rm c.c.}
\right]\,,
\end{split}
\end{align}
while those interactions that couple to $\mathcal{T}^i_{\rm L,R}$ read
\begin{align}
\begin{split}
\label{LT}
&\mathcal{L}_{\mathcal{T}}  = 
\widehat{\mathcal{P}}_{\rm LR} \left[ \left({\sigma_i}\right)^{l}_{l'}   \left(
\y{35}  \widetilde{\mathcal{Q}}^{\ast\,l'}_{{\rm L}\,f}\mathcal{T}^i_L {\mathcal{Q}}_{{\rm L}\,l}^{f}
+\y{36} \widetilde{q}^{\ast \,l'}_{\rm L} \mathcal{T}^i_L{q}_{{\rm L}\,l}
\right. \right. \\
&
\left. \left.
+\y{37} H^{\ast \,r}_{f\,l} \mathcal{T}^i_L \widetilde{H}^{f\,l'}_{r}
+\y{38} h^{\ast \,r}_{l} \mathcal{T}^i_L \widetilde{h}^{l'}_{r}
\right. \right. \\
&
\left. \left.
+\y{39} \widetilde{E}_{{\rm L}\,f\,l}^\ast \mathcal{T}^i_L {E}_{\rm L}^{f\,l'}
+\y{40} \widetilde{\mathcal{E}}_{{\rm L}\,l}^\ast \mathcal{T}^i_L {\mathcal{E}}_{\rm L}^{l'}
+ {\rm c.c.}
\right) \right]\,.
\end{split}
\end{align}
Finally, the Yukawa interactions involving gluinos are given by
\begin{align}
\begin{split}
\label{Lglu}
&\mathcal{L}_{\widetilde{g}}  = 
\widehat{\mathcal{P}}_{\rm LR} \left[     
\y{41} \widetilde{\mathcal{Q}}^{\ast \,l}_{{\rm L}\,f}{\bm{T}^a}\widetilde{g}^a \mathcal{Q}_{{\rm L}\,l}^{f}
+\y{42} \widetilde{q}^{\ast \,l}_{\rm L} {\bm{T}^a}\widetilde{g}^a{q}_{{\rm L}\,l}
\right. \\
&
\left.
+\y{43} \widetilde{D}_{{\rm L}\,f}^\ast {\bm{T}^a}\widetilde{g}^a {D}_{\rm L}^{f}
+\y{44} \widetilde{\mathcal{B}}_{\rm L}^\ast  {\bm{T}^a}\widetilde{g}^a{\mathcal{B}}_{\rm L}
+ {\rm c.c.}
 \right] \,.
\end{split}
\end{align}
%

\subsection{The gauge sector of the LR-symmetric EFT}

In this section, we consider interactions involving the gauge bosons of the effective SHUT-LR model. For ease of reading, 
we separate those into the gauge-scalar (gs), gauge-fermion (gf) and pure-gauge (pg) interaction types,
\begin{align}
\mathcal{L}_{\rm gauge} = \mathcal{L}_{\rm gs} + \mathcal{L}_{\rm gf} + \mathcal{L}_{\rm pg}\,,
\end{align}
where Eqs.~\eqref{CovDer-1}, \eqref{CovDer-2} and \eqref{Fmn} of appendix \ref{sec:Cov} can be employed to write
\begin{align}
\begin{split}
\label{Lg}
&\mathcal{L}_{\rm gs}  = 
\( \bm{D}_{\mu} \widetilde{\varphi} \)^{\ast}  \( \bm{D}^{\mu} \widetilde{\varphi} \)
+ \( \bm{D}_{\mu} \widetilde{\phi} \)^{\ast}_f \( \bm{D}^{\mu} \widetilde{\phi} \)^f
\\
&
+ \( \bm{{D}}_{\mu} h \)^{\dagger\,r}_{l} \( \bm{{D}}^{\mu} h \)^{l}_{r}
+ \( \bm{{D}}_{\mu} H \)^{\dagger\,r}_{f\,l} \( \bm{{D}}^{\mu} H \)^{f\,l}_{r}
\\
&
+ \eta_{\mu \nu} \widehat{\mathcal{P}}_{\rm LR} \[ 
\(\bm{D}^{\nu} \widetilde{\mathcal{E}}_{\rm L} \)^{\dagger}_{l} \(\bm{D}^{\mu}\widetilde{\mathcal{E}}_{\rm L} \)^{l} 
+ \(\bm{D}^{\nu} \widetilde{E}_{\rm L} \)^{\dagger}_{f\,l} \(\bm{D}^{\mu} \widetilde{E}_{\rm L} \)^{f\,l}
\right.
\\
&
\left.
+ \(\bm{D}^{\nu} \widetilde{q}_{\rm L} \)^{\dagger\,l} \(\bm{D}^{\mu}\widetilde{q}_{\rm L} \)_{l}  
+ \(\bm{D}^{\nu} \widetilde{\mathcal{Q}}_{\rm L} \)^{\dagger\,l}_{f} \(\bm{D}^{\mu}\widetilde{\mathcal{Q}}_{\rm L} \)^{f}_{l}
\right.
\\
&
\left.
+ \(\bm{D}^{\nu}  \widetilde{\mathcal{B}}_{\rm L} \)^{\dagger} \(\bm{D}^{\mu} \widetilde{\mathcal{B}}_{\rm L} \)
+ \(\bm{D}^{\nu}\widetilde{D}_{\rm L} \)^{\dagger}_{f} \(\bm{D}^{\mu} \widetilde{D}_{\rm L} \)^{f}  \] 
\\
&\mathcal{L}_{\rm gf}  = 
\i \varphi^{\dagger} \overline{\sigma}_{\mu} \bm{D}^{\mu} \varphi
+ \i \phi^{\dagger}_f \overline{\sigma}_{\mu} \( \bm{D}^{\mu} \phi\)^f
+ \i \widetilde{h}^{\dagger\,r}_{l} \overline{\sigma}_{\mu} \( \bm{{D}}^{\mu} \widetilde{h} \)^{l}_{r} 
\\
&
+ \i \widetilde{H}^{\dagger\,r}_{f\,l} \overline{\sigma}_{\mu} \( \bm{{D}}^{\mu} \widetilde{H} \)^{f\,l}_{r} 
+ \widehat{\mathcal{P}}_{\rm LR} \[ \i \mathcal{E}^{\dagger}_{{\rm L}\,l} \overline{\sigma}_{\mu} \(\bm{D}^{\mu}\mathcal{E}_{\rm L} \)^{l} 
\right. \\
&
\left.
+ \i E^{\dagger}_{{\rm L}\,f\,l} \overline{\sigma}_{\mu} \(\bm{D}^{\mu} E_{\rm L} \)^{f\,l}
+ \i q^{\dagger\,l}_{\rm L} \overline{\sigma}_{\mu} \(\bm{D}^{\mu} q_{\rm L} \)_{l}  
\right. \\
&
\left.
+ \i \mathcal{Q}^{\dagger\,l}_{{\rm L}\,f} \overline{\sigma}_{\mu} \(\bm{D}^{\mu} \mathcal{Q}_{\rm L} \)^{f}_{l}
+ \i \mathcal{B}^{\dagger}_{\rm L} \overline{\sigma}_{\mu} \bm{D}^{\mu} \mathcal{B}_{\rm L}
+ \i D^{\dagger}_{{\rm L}\,f} \overline{\sigma}_{\mu} \(\bm{D}^{\mu} D_{\rm L} \)^{f} \] 
\\
&
+ \sum_{\rm A = L,R} \[ \i \mathcal{S}^\dagger_A\overline{\sigma}_\mu\partial^\mu\mathcal{S}_A 
+ \i \mathcal{T}^{i\dagger}_A \overline{\sigma}_\mu \( \bm{D}^{\mu} \mathcal{T}_A  \)^i  \]
\\
&
+ \i \widetilde{g}^{a\,\dagger} \overline{\sigma}_\mu \( \bm{D}^{\mu} \widetilde{g}  \)^a  + {\rm c.c.}
\\
&
\mathcal{L}_{\rm pg}  =
- \dfrac{1}{4} \[ \sum_{\rm A = L,R} \( B^{\mu \nu}_{\Sr{0.7}{A}} B_{\Sr{0.7}{A} \mu \nu}
+ F^{\mu \nu\,i}_{\Sr{0.7}{A}} F^{i}_{\Sr{0.7}{A} \mu \nu} \) 
\right. \\
&
\left.
+ G^{\mu \nu\,a} G^{a}_{\mu \nu}
+ B^{\mu \nu}_{\Sr{0.7}{L}} B_{\Sr{0.7}{R} \mu \nu}  \]\,.
\end{split}
\end{align}

\subsubsection{Covariant derivatives and field strengths}  
\label{sec:Cov}

The covariant derivatives of the LR-symmetric effective model can be written in a compact matrix form as follows
\begin{equation}
\label{CovDer-1}
\begin{aligned}
&\bm{D}^{\mu} \left(H,h\right) = \left( \id{L} \So{\otimes} \id{R}\del^{\mu} - \i \g{\rm L} A^{\mu\, i}_{\rm L}\bm{\tau}^i \So{\otimes} \id{R} - \i \g{\rm R} A^{\mu\,i}_{\rm R}\bm{ \tau}^i \So{\otimes} \id{L} 
\right.
\\
&
\left.
+ \i \g{L}^{\prime} \q{\rm L} B^{\mu}_{\rm L} \id{L} \So{\otimes} \id{R} 
+ \;\i \g{R}^{\prime} \q{R} B^{\mu}_{\rm R} \id{L} \So{\otimes} \id{R}\right) \left(H,h\right) \,,
\\
&
\widehat{\mathcal{P}}_{\rm LR} \left[\bm{D}^{\mu}  \left(E_L,\mathcal{E}_L\right)\right]= \widehat{\mathcal{P}}_{\rm LR} 
\left[\left( \id{L}\del^{\mu} - \i \g{\rm L} A^{\mu\, i}_{\rm L}\bm{ \tau}^i 
\right.\right.
\\
&
\left.\left.
+ \i \g{L}^{\prime} \q{\rm L} B^{\mu}_{\rm L} \id{L} + 
\i \g{R}^{\prime} \q{R} B^{\mu}_{\rm R}\id{L} \right) \left(E_L,\mathcal{E}_L\right)\right]
\\
&\bm{D}^{\mu}  \left(\phi,\varphi\right) =\left(\del^{\mu} + \i \g{L}^{\prime} \q{\rm L} B^{\mu}_{\rm L}  + 
\i \g{R}^{\prime} \q{R} B^{\mu}_{\rm R} \right)  \left(\phi,\varphi\right)  \,,
\\
&
\widehat{\mathcal{P}}_{\rm LR} \left[\bm{D}^{\mu}\left(Q_L,q_L\right) \right]=\widehat{\mathcal{P}}_{\rm LR} 
\left[\left(  \id{C}\So{\otimes} \id{L} \del^{\mu}  - \i \g{C} G^{\mu \, a}_{\rm C} \bm{T}^a \So{\otimes} \id{L}   
\right. \right.
\\
&\left. \left.
- \i \g{\rm L} A^{\mu\, i}_{\rm L}\bm{ \tau}^i \So{\otimes} \id{C} 
+ \i \g{L}^{\prime} \q{\rm L} B^{\mu}_{\rm L}\id{C} \So{\otimes} \id{L}\right) \left(Q_L,q_L\right) \right]\,,
\\
&\widehat{\mathcal{P}}_{\rm LR} \left[\bm{D}^{\mu}\left(D_L,\mathcal{B}_L\right)\right] =\widehat{\mathcal{P}}_{\rm LR} 
\left[\left( \id{C}  \del^{\mu} - \i \g{C} G^{\mu \, a}_{\rm C} \bm{T}^{a} 
\right. \right.
\\
&\left. \left.
+ \i \g{L}^{\prime} \q{\rm L} B^{\mu}_{\rm L} \id{C} \right) \left(D_L,\mathcal{B}_L\right)\right] \,,
\end{aligned} 
\end{equation}
\begin{equation}
\label{CovDer-2}
\begin{aligned}
&\bm{D}^{\mu}\;\mathcal{T}_A =\left(  \ida{L,R} \del^{\mu} - \i \g{\rm L,R} A^{\mu\, i}_{\rm L,R}\bm{\tau}_{\Scale[0.6]{\rm adj}}^i \right) \mathcal{T}_A
\\
&\bm{D}^{\mu}\;\widetilde{g} =\left(  \ida{C} \del^{\mu} - \i \g{\rm C} G^{\mu\, a}_{\rm C}\bm{T}_{\Scale[0.6]{\rm adj}}^a\right) \widetilde{g} \,,
\end{aligned} 
\end{equation}
where summation is assumed over each pair of repeated indices, $\q{\rm A}$ is the $\U{A}$ hypercharge and $\id{A}$ and $\ida{A}$ are the identity matrices 
with the same dimensions of the fundamental and adjoint representations, respectively. The field strength tensors of the $\U{A}$, $\SU{2}{A}$ and $\SU{3}{C}$ 
gauge symmetries are given by
\begin{equation}
\label{Fmn}
\begin{aligned}
B^{\mu \nu}_{\Sr{0.7}{A}} &= \del^{\mu} B_{\rm A}^{\nu} - \del^{\nu} B_{\rm A}^{\mu} \\
F^{\mu \nu\,i}_{\Sr{0.7}{A}} &= \del^{\mu} A_{\rm A}^{\nu\,i} - \del^{\nu} A_{\rm A}^{\mu\,i} +
\g{A} \varepsilon^{ijk} A_{\rm A}^{\mu\,j} A_{\rm A}^{\nu\,k} \\
G^{\mu \nu\,a} &= \del^{\mu} G_{\rm C}^{\nu\,a} - \del^{\nu} G_{\rm C}^{\mu\,a} +
\g{C} f^{abc} G_{\rm C}^{\mu\,b} G_{\rm C}^{\nu\,c} \,.
\end{aligned} 
\end{equation}

\vspace{0.5cm}

\subsubsection{Abelian $D$-terms}  
\label{sec:ADterms}

The $\U{L,R}$ $D$-terms of the LR-symmetric theory read
\begin{align*}
\begin{split}
&\mathcal{D}_{\mathrm{L}} = \frac{1}{\left(1-\frac{\chi^2}{4}\right)}
\left[-\frac{1}{2}\chi\left( X_{\mathrm{R}} - \kappa\right) + 
X_{\mathrm{L}} + \kappa \right] \,,
\\
&\mathcal{D}_{\mathrm{R}} = \frac{1}{\left(1-\frac{\chi^2}{4}\right)}
\left[-\frac{1}{2}\chi\left( X_{\mathrm{L}} + \kappa\right) + 
X_{\mathrm{R}} - \kappa \right] \,,
\\
& X_{\mathrm{L}} =  H^{\ast\,r}_{f\,l} H^{l\,f}_{r} 
-2
\widetilde{\phi}_{f}^\ast\widetilde{\phi}^{f}
+
\widetilde{E}_{{\rm L}\;f\,l}^\ast \widetilde{E}_{\rm L}^{f\,l}
-2
\widetilde{E}_{{\rm R}\;f}^{\ast\,r} \widetilde{E}_{{\rm R}\,r}^{f}
\\
&\qquad \qquad \quad 
-\,
\widetilde{\mathcal{Q}}_{{\rm L}\,f}^{\ast\,l}\widetilde{\mathcal{Q}}_{{\rm L}\,l}^{f}
+2
\widetilde{D}_{{\rm L}\;f}^\ast\widetilde{D}_{\rm L}^{f} \,,
\\
& X_{\mathrm{R}} =
-H^{\ast\,r}_{f\,l} H^{l,f}_{r} 
+2
\widetilde{\phi}_{f}^\ast\widetilde{\phi}^{f}
+2
\widetilde{E}_{{\rm L}\;f\,l}^\ast \widetilde{E}_{\rm L}^{f\,l}
-
\widetilde{E}_{{\rm R}\;f}^{\ast\,r} \widetilde{E}_{{\rm R}\,r}^{f}
\\
&\qquad \qquad \quad 
+\,
\widetilde{\mathcal{Q}}_{{\rm R}\,f\,r}^{\ast}\widetilde{\mathcal{Q}}_{{\rm R}}^{f\,r}
-2
\widetilde{D}_{{\rm R}\;f}^\ast\widetilde{D}_{\rm R}^{f} \,,
\end{split}
\end{align*}
with $f=1,2,3$.

\newpage

\bibliography{bib}

\end{document}